\documentclass[aps,pra,reprint,superscriptaddress,notitlepage,showpacs,floatfix,twocolumn,longbibliography]{revtex4-1}

\usepackage{graphicx,graphics,epsfig,subfigure,times,bm,bbm,amssymb,amsmath,amsfonts,mathrsfs}
\usepackage[scr=boondoxo,scrscaled=1.05]{mathalfa}
\usepackage[matrix,frame,arrow]{xypic}
\usepackage[pdfstartview=FitH]{hyperref}
\usepackage[pdftex]{color}
\usepackage{dsfont}
\usepackage[export]{adjustbox}
\usepackage[table]{xcolor}

\newcommand{\beq}{\begin{equation}}
\newcommand{\eneq}{\end{equation}}
\newcommand{\beqnn}{\begin{equation*}}
\newcommand{\eneqnn}{\end{equation*}}
\newcommand{\beqy}{\begin{eqnarray}}
\newcommand{\eneqy}{\end{eqnarray}}
\newcommand{\beqynn}{\begin{eqnarray*}}
\newcommand{\eneqynn}{\end{eqnarray*}}

\newcommand{\ket}[1]{ | #1 \rangle  }

\newcommand{\expect}[1]{\langle #1 \rangle}

\newcommand{\erf}[1]{Eq. (\ref{#1})}
\def\Xint#1{\mathchoice
   {\XXint\displaystyle\textstyle{#1}}%
   {\XXint\textstyle\scriptstyle{#1}}%
   {\XXint\scriptstyle\scriptscriptstyle{#1}}%
   {\XXint\scriptscriptstyle\scriptscriptstyle{#1}}%
   \!\int}
\def\XXint#1#2#3{{\setbox0=\hbox{$#1{#2#3}{\int}$}
     \vcenter{\hbox{$#2#3$}}\kern-.5\wd0}}
\def\ddashint{\Xint=}

\newcommand{\bes} {\begin{subequations}}
\newcommand{\ees} {\end{subequations}}
	\newcommand{\bea} {\begin{eqnarray}}
	\newcommand{\eea} {\end{eqnarray}}

\newcommand{\ignore}[1]{}

\begin{document}

\title{Optimally band-limited spectroscopy of control noise using a qubit sensor}

\author{Leigh M. Norris}
\affiliation{\mbox{Department of Physics and Astronomy, Dartmouth 
College, 6127 Wilder Laboratory, Hanover, NH 03755, USA}}

\author{Dennis Lucarelli}
\affiliation{\mbox{Johns Hopkins University, Applied Physics Laboratory,
11100 Johns Hopkins Road, Laurel, MD 20723, USA}}

\author{Virginia M. Frey}
\affiliation{ARC Centre for Engineered Quantum Systems, School of Physics, 
The University of Sydney, NSW 2006 Australia \& \\
National Measurement Institute, West Lindfield NSW 2070 Australia}

\author{Sandeep Mavadia} 
\affiliation{ARC Centre for Engineered Quantum Systems, School of Physics, 
The University of Sydney, NSW 2006 Australia \& \\
National Measurement Institute, West Lindfield NSW 2070 Australia}

\author{Michael J. Biercuk}
\affiliation{ARC Centre for Engineered Quantum Systems, School of Physics, 
The University of Sydney, NSW 2006 Australia \& \\
National Measurement Institute, West Lindfield NSW 2070 Australia}

\author{Lorenza Viola}
\affiliation{\mbox{Department of Physics and Astronomy, Dartmouth 
College, 6127 Wilder Laboratory, Hanover, NH 03755, USA}}

\begin{abstract}
Classical control noise is ubiquitous in qubit devices, making its accurate spectral characterization 
essential for designing optimized error suppression strategies at the physical level. Here, we focus on multiplicative 
Gaussian amplitude control noise on a driven qubit sensor and show that sensing protocols using optimally band-limited 
Slepian modulation offer substantial benefit in realistic scenarios.
Special emphasis is given to laying out the theoretical framework necessary for extending {\em non-parametric 
multitaper spectral estimation} to the quantum setting by highlighting key points of contact and 
differences with respect to the classical formulation. In particular, we introduce and analyze two approaches
(adaptive vs. single-setting) to quantum multitaper estimation, and show how they provide a practical means to 
both identify fine spectral features not otherwise detectable by existing protocols and to obtain reliable prior estimates 
for use in subsequent parametric estimation, including high-resolution Bayesian techniques. 
We quantitatively characterize the performance of both single- and multitaper Slepian estimation protocols 
by numerically reconstructing representative spectral densities, and demonstrate their advantage over 
dynamical-decoupling noise spectroscopy approaches in reducing bias from spectral leakage as well as in 
compensating for aliasing effects while maintaining a desired sampling resolution.
\end{abstract}

\date{\today}
\maketitle

\section{Introduction}

Spectral estimation is a central task in classical statistical signal processing, with widespread applications across  
many areas of the physical and engineering sciences \cite{percival1993spectral}.  The simplest case involves an unknown 
random signal described by a stationary Gaussian process, having a frequency content completely characterized by the 
power spectral density $S(\omega)$ (PSD, or ``spectrum''). Given access to a finite record of measurement data, 
the goal of spectral estimation is to reconstruct the PSD. While a variety of techniques for spectral estimation have 
been developed, an important distinction is whether they are {\em parametric} or {\em non-parametric} in nature. In 
the former case, a specific functional form for $S(\omega)$ is assumed, which reduces the 
problem to estimating unknown functional parameters. Non-parametric approaches, in contrast, aim at spectral 
reconstruction through appropriate sampling in frequency space. Though parametric methods are known to offer 
more accurate estimates if the data closely agrees with the assumed model, non-parametric approaches are 
clearly preferable if minimal or no {\em a priori} knowledge is available.  

In 1982, Thomson introduced {\em multitaper spectral estimation} \cite{thomson_multitaper}, a non-parametric 
approach that provides multiple independent spectral estimates from a single set of time-domain samples. This is made possible 
by leveraging the orthogonality and {\em optimal spectral concentration} properties of a 
particular set of discrete-time functions known as the ``discrete prolate spheroidal sequences'' (DPSS) \cite{Slepian1961,Landau1961a,Landau1961b,Slepian1964,Papoulis1972,Slepian1978}. The discovery of the DPSS by Slepian and co-workers was prompted by a fundamental question originally posed by Shannon \cite{verdu}: To what extent can functions, which are confined to a finite bandwidth, be also concentrated in the time domain? As it turns out, among all sequences with the same total duration, DPSS maximize the fraction of spectral power within a target 
sensing band compared to the total spectral power. Due in part to the unique characteristics of the DPSS, multitaper spectral estimation offers superior performance in terms of bandwidth, bias and variance efficiency \cite{Bronez,babadi}, in addition to having the ability to handle spectra with complex structure.  As such, it is nowadays regarded as the non-parametric method of choice \cite{percival1993spectral}, and is widely employed in applications ranging from from radar and seismic data analysis to electroencephalography,  speech processing, geosciences and climatology. 

In the quantum setting, it has long been appreciated that a controlled quantum system (a qubit in the simplest case) can serve as a 
``spectrometer'' or sensor of its own noisy environment \cite{Aash,Faoro,Young2012}. In response to continued progress in quantum 
technologies and high-fidelity quantum information processing (QIP), however, interest in developing {\em quantum spectral estimation} 
or ``quantum noise spectroscopy'' (QNS) techniques has heightened in recent years.  While experimental coherent-control 
capabilities have reached superb levels of accuracy with two-qubit gate fidelities now exceeding 99\% \cite{wineland,petta}, 
it has also become increasingly clear that the manner in which spatial noise correlations decay is crucial in determining the ultimate 
feasibility of full-fledged fault-tolerant QIP architectures \cite{NgPreskill2009,Mucciolo2013}. Even in near-term 
devices which are expected to operate without quantum error correction or possibly with variational error-correcting 
algorithms \cite{Sergio,Peter}, accurate characterization of the noisy environment experienced by qubits remains a prerequisite 
for achieving {\em noise-optimized control design and actuation} at the physical layer. 

Among non-parametric approaches, QNS via dynamical decoupling (DD) has attracted significant theoretical attention and 
seen broad adoption across experimental QIP. Here, one exploits the fact that DD, implemented via 
(nearly) instantaneous pulses or piecewise-constant (``flat-top'') control waveforms, creates a ``band-pass'' filter in Fourier 
space, which suppresses all noise outside a select range of frequencies known as the ``passband". The value of the PSD within 
the passband is then estimated from measurements of the sensor following DD-driven evolution. Changing the evolution 
time $T$ and/or other parameters of the DD sequence adjusts the position of the passband, allowing the PSD to be sampled at multiple locations in the frequency domain~\cite{BiercukJPB2011}. For a qubit sensor subject to pure dephasing noise, there are more sophisticated versions of QNS via DD that utilize sequence repetition. In the long-time limit, repeating a fixed DD sequence of duration $T$ produces a filter function (FF) that approximates a {\em frequency comb} with peaks centered at multiples of a fundamental harmonic, that is, $\omega_h \equiv h (2\pi/T)$, $h \in {\mathbb Z}$. Measuring the qubit decay rate for 
different DD sequences results in a linear inversion problem, whose solution yields a sampling of the PSD,  
$\{S(\omega_h)\}$, at a subset of the harmonic frequencies. From a theory standpoint, recent advances include the design 
of single-qubit QNS protocols for high-order spectra arising from a class of {\em non-Gaussian} stationary dephasing environments \cite{Norris_Spectroscopy}, and two-qubit QNS protocols for estimation of both classical and quantum cross-correlation and 
self-spectra in {\em multiqubit} Gaussian dephasing settings \cite{Cywinski2016,Paz2017}. In parallel, experimental implementations 
have been reported for single-qubit sensors ranging from spins in NMR \cite{Alvarez_Spectroscopy} and semiconductor quantum dots 
\cite{Morello2014,Malinowski2017} to superconducting qubits \cite{Bylander2011}, NV centers \cite{Jelezko2015}, 
trapped ions \cite{Ozeri2013} and atomic ensembles \cite{Davidson2016}. 

Despite the above advances, DD QNS methods suffer from significant shortcomings. In comb-based approaches, care is needed to avoid ill-conditioning and numerical instabilities stemming from the need for linear inversion.  Furthermore, even in the idealized limit where the comb approximation is taken to be exact, only sufficiently smooth spectra are amenable to sampling -- which prevents sensing of fine spectral structure or ``spur-like'' components.  Lastly, deviations from the frequency-comb approximation 
may significantly complicate the spectral reconstruction procedure from input data \cite{LukasAccuracy}. Such deviations inevitably arise in reality, due to the necessarily finite duration of the control pulses and/or finite evolution time. On a more fundamental level, a key issue is that the band-pass filters generated by DD do not guarantee optimal noise blocking at frequencies outside the intended passband: the presence of FF harmonics in the Fourier domain outside the passband can lead to spurious contributions and bias -- so-called {\em spectral leakage} -- an issue that has already manifested in experimental settings of interest \cite{degen_retract,degen_followup}. 

In previous work \cite{Frey2017}, we introduced and experimentally validated an approach to quantum spectral estimation that is designed to incorporate bandwidth constraints from the outset and thus optimally reduce spectral leakage, while also avoiding comb approximations and linear inversion. This was accomplished by utilizing DPSS functions to define the envelopes of applied control fields, which enabled the first proof-of-principle adaptive multitaper estimation of an engineered Gaussian amplitude-noise spectrum using trapped $^{171}\text{Yb}^+$ ions, along with a first application of Bayesian estimation tools to the task of QNS. Here, we continue our investigation of Slepian-based QNS protocols with a twofold motivation.  On the one hand, we provide a complete discussion of the theoretical underpinning for Ref. \cite{Frey2017}, by keeping our focus on the simplest setting of {\em classical, Gaussian multiplicative amplitude control noise} on a single-qubit sensor. On the other hand, we substantially expand our analysis by establishing a number of new results and by providing a quantitative comparative account of the impact of leakage and aliasing effects in spectral reconstruction. The content of the paper and our main contributions may be summarized as follows.

In Sec. \ref{sec::settings} and Sec. \ref{sec::background} we describe the relevant control-theoretic setting and the 
necessary background from classical spectral estimation theory, respectively.  While no new results are included, 
Sec. \ref{sec::settings} presents, in particular, an expanded discussion of the three-axis measurement strategy required to selectively 
sense amplitude control noise in the presence of additive dephasing, along with a more precise definition of the estimator 
used in \cite{Frey2017}. A qualitative understanding of the spectral leakage problem is also included 
in the context of standard DD QNS employing Carr-Purcell-Meiboom-Gill (CPMG) sequences, in preparation for introducing 
optimally band-limited DPSS functions and classical multitaper estimation in Sec. \ref{sec::background}.
Altogether, we make it clear that there is an analogy between quantum ``passband estimators'' and classical ``tapered estimates,'' 
which allows classical concepts like tapering, multitaper, leakage, spectral concentration and aliasing to be imported into the quantum setting.

Sec. \ref{sec::MultiQNS} is the core theory section of the paper.  Here, we expand significantly the description of multitaper 
QNS given in \cite{Frey2017}, by both highlighting important differences between the classical and quantum 
estimation settings and by supplying detailed derivations of broadband and local biases. 
In addition, a number of new results are presented, which include: (i) a strategy for shifting the FF passband (so-called {\em CS modulation}) that both removes unwanted cross terms in existing modulation schemes and preserves, unlike single-sideband modulation, the completeness properties of the DPSS so crucial for multitaper estimation; 
(ii) a discussion of {\em aliasing} and the bandwidth limitations it imposes on the spectral estimate whenever the underlying control waveform is digital;
(iii) an alternative {\em single-setting} implementation of multitaper QNS closer in spirit to the classical one, in that a single noise realization is tapered with multiple DPSS before measurement, as opposed to \cite{Frey2017} in which each Slepian order corresponds to a different measurement setting;
(iv) a streamlined approach for combining DPSS modulation with Bayesian analysis \cite{BayesianQNS}, which benefits from the introduction of an interpolated estimate based on the Fisher information as an empirical prior.

In Sec. \ref{sec::results}, we numerically implement and analyze Slepian QNS protocols in concrete 
illustrative scenarios. In particular, we examine how spectral leakage can significantly bias spectral estimates, with the extent of the error depending on the structure of the underlying spectrum. Even for estimation protocols that employ a single DPSS taper, we demonstrate substantial leakage suppression over conventional DD-based schemes.
For QNS protocols in general, we show that the range of spectral reconstruction is limited by aliasing that occurs beyond the Nyquist frequency. Moreover, without proper compensation, aliasing can represent a significant source of error.
In this respect, a main advantage that DPSS approaches offer in comparison to comb-based QNS stems from the possibility of {\em independently} tuning both the Nyquist frequency and the sampling resolution, extending the range of spectral reconstruction under realistic constraints. Finally, we demonstrate that a key application of multitaper QNS is the detection of fine spectral components, which are often encountered in realistic signals. 
The initial multitaper estimate can then serve as a prior in a subsequent Bayesian analysis aimed to producing a 
{\em high-resolution estimate} of  fine spectral features. While the choice of system and control parameter in our 
simulations is tailored to the trapped-ion platform of \cite{Frey2017}, our methods 
and main conclusions are applicable to spectral estimation of multiplicative control noise in a variety of qubit devices 
supporting similar control capabilities. 

We conclude in Sec. \ref{sec::outlook} by highlighting a number of directions for future investigation, whereas we include in 
Appendixes \ref{app::TimeConcen} and \ref{app::fisher} a brief discussion of a dual spectral concentration problem in the 
time domain, and additional technical detail on the evaluation of the Fisher information, respectively. 


\section{System and control setting}
\label{sec::settings}

\subsection{Time domain: Noise, control, and measurement}
\label{sec::time}

We consider a qubit sensor that is subject to both open-loop control and time-dependent, 
temporally correlated noise. We assume that the qubit may be controlled via modulation of 
the amplitude $\Omega(t)$ and phase $\phi(t)$ of an external driving field, as available in a variety of 
qubit technologies -- in particular, trapped ions \cite{Soare2014bath,SoareNatPhys2014}. 
In a frame co-rotating with the carrier frequency, at resonance with the qubit transition frequency 
and under the rotating wave approximation, the ideal driven dynamics (in units $\hbar=1$) is described by 
the Hamiltonian
\begin{equation}
\label{eq::Hc}
H_{c}(t) = \frac{\Omega(t)}{2} \left[\cos\phi(t)\sigma_x + \sin\phi(t)\sigma_y \right] . 
\end{equation}
Noise causes the qubit dynamics to deviate from the ideal evolution generated by $H_{c}(t)$.  
This noise may be due to both ambient (control-independent) sources and control imperfections, 
and can be generated by classical or quantum-mechanical degrees of freedom. Here, we focus on a 
semiclassical limit in which all the relevant noise sources may be modeled through a ``universal noise 
Hamiltonian'' of the form
\begin{equation*}
H_N(t) = \vec{\xi}(t) \cdot \vec{\sigma} = \sum_{u=x,y,z} \xi_u(t) \sigma_u , 
\end{equation*}
where each noise component $\xi_u(t)$ corresponds to a stationary, Gaussian, zero-mean process.
While in principle each $\xi_u(t)$ could take contributions from both ambient and control noise, 
following \cite{Ball2014,Frey2017} we focus on a simple yet realistic scenario where ambient noise is solely 
contributed by {\em additive dephasing} and control noise solely arise from fluctuations in one control 
variable, resulting in {\em multiplicative amplitude noise}. The relevant noise Hamiltonian then specializes to 
\begin{equation}
H_N(t) = \beta_z(t) \sigma_z + \beta_{\Omega}(t) H_{c}(t), 
\label{eq::HN}
\end{equation}
that is, $\xi_z(t)=\beta_z(t)$, $\xi_x(t)= \beta_{\Omega}(t) \frac{\Omega(t)}{2} \cos\phi(t)$, 
$\xi_y(t)= \beta_{\Omega}(t) \frac{\Omega(t)}{2} \sin\phi(t)$, and the 
two stochastic processes $\beta_z(t)$ and $\beta_\Omega(t)$ are taken to be independent.
The total Hamiltonian in the rotating frame, describing the effects of both 
the noise and the applied control, is then given by
\begin{align}
\!\!H(t) =  H_{c}(t)+H_N(t)= \beta_z(t)\sigma_z + [1 + \beta_{\Omega}(t)]  H_{c}(t) .
\label{eq::Ham}
\end{align}
Note that a formally similar Hamiltonian describes a singlet-triplet qubit in a double quantum dot, 
simultaneously exposed to hyperfine-induced additive dephasing and multiplicative noise in the exchange control due 
to charge fluctuations. A main difference in this case is that Hamiltonian \eqref{eq::Ham} applies directly in the physical frame 
and $\beta_z(t) \equiv \beta_0 +\delta\beta_z(t)$ has a non-zero mean to account for the known static magnetic field 
gradient $\beta_0$ between the two dots  \cite{aDCG}.

The dynamical contribution of the noise can be isolated by further transforming into the toggling frame, that is, the interaction picture with respect to $H_c(t)$. This is described by the canonical transformation $\tilde{\rho}(t) \equiv U_c(t)^\dagger \rho(t) U_c(t)$, where $U_c(t)$ and $\rho(t)$ are the propagator generated by the ideal control Hamiltonian in Eq. \eqref{eq::Hc} and the qubit density operator in the physical (rotating) frame, respectively. In the toggling frame, the effective qubit Hamiltonian, or ``error Hamiltonian" \cite{KavehDCG,PazFFF}, may be expressed in the form 
\begin{align}
\label{eq::toggle}
\!\!\!\tilde{H}(t) = U_{c}(t)^{\dagger} H_N(t) U_{c}(t) \equiv 
\!\!\sum_{u,v =x,y,z} Y_{uv}(t) \beta_u(t) \sigma_v , 
\end{align}
where the  {\em control matrix} $Y(t) \equiv \{Y_{uv}(t)\}\in\mathbb{R}^{3\times 3}$ depends only upon the control variables 
$\{\Omega(t), \phi(t)\}$ and $ \beta_x(t) =\beta_y(t) \equiv {\beta_\Omega (t)}/{2}$. 
Let the matrix $R(t) \equiv \{R_{uv}(t)\}\in \text{SO}(3)$ be defined by letting $R_{uv}(t) \equiv \frac{1}{2} \text{Tr} 
\{U_c(t)^\dagger \sigma_u U_c(t)\sigma_v\}$ \cite{Ball2014}.  Then, the control matrix takes the explicit form
\begin{eqnarray*}
&&Y_{xv}(t) = \Omega(t) \cos \phi(t) R_{xv}(t), \\
&&Y_{yv}(t) = \Omega(t) \sin \phi(t) R_{xv}(t), \\
&&Y_{zv}(t) = R_{zv}(t).
\end{eqnarray*}   

Accordingly, evolution in the toggling frame is described by the error propagator $\tilde{U}(t) = \mathcal{T}_+\!\exp\!\left[- \mathrm{i} \! \int_0^t \mathrm{d}s \tilde{H} (s)\right]$,  which is related to the 
rotating-frame propagator by $U(t)=U_c(t)\tilde{U}(t)$. Following \cite{GreenNJP2013}, it is convenient to 
parametrize $\tilde{U}(t)$ in terms of the real, time-dependent ``error vector'' $\mathbf{a}(t) \equiv [a_x(t), a_y(t), a_z(t)]$, such that
\begin{equation}
\tilde{U}(t) = \exp[-\mathrm{i} \mathbf{a}(t) \cdot\boldsymbol{\sigma}] \equiv e^{-i\sum_{\ell =1}^\infty \mathbf{a}^{(\ell)}(t) 
\cdot\boldsymbol{\sigma} }, 
\label{eq::Magnus}
\end{equation}
where a perturbative Magnus expansion is introduced in the second equality. In general, it is impossible to obtain a closed-form expression for $\,\mathbf{a}(t)$.  Here, we consider a regime where the noise is sufficiently weak and the time scale is small 
enough that only the leading (first) order term of Eq. \eqref{eq::Magnus} enters the 
 evolution of the observables of interest [see also Eq. \eqref{eq::PzActual} below]. 
In this case, the error vector is given by 
$\mathbf{a}(t)\approx\mathbf{a}^{(1)}(t) \equiv \{ a_v^{(1)}(t), v=x,y,z \}$, with components 
\begin{align}
\label{eq::ErrV1}
\hspace*{-3mm} a_v^{(1)}(t) =\frac{1}{2} \mbox{Tr}\Big[\sigma_v \!\int_0^t \!\!\text{d}s\,\tilde{H}(s) \Big]\! = 
\sum_{u= x,y,z} \int_0^t \!\!\text{d}s \,Y_{u v}(s) \beta_u(s).
\end{align}

Proper design of the control Hamiltonian in \erf{eq::Hc} is essential for QNS. Throughout this paper, 
we take the control to be bounded, with the amplitude of the driving field being restricted by the maximum Rabi rate, 
$0 \leq \Omega(t)\leq \Omega_{\text{max}} $, and the phase attaining the values $\phi(t)\in\{0,\pi\}$. Since switching the phase between $0$ and $\pi$ is equivalent to changing the sign of $\Omega(t)$, for simplicity we shall take $\phi(t)\equiv 0$ henceforth and allow the amplitude to assume negative values, $-\Omega_{\text{max}}\leq \Omega(t)\leq \Omega_{\text{max}}$. The control propagator under 
{\em pure amplitude modulation} simplifies to 
$$U_c(t)= \exp[-i \Theta(t) \sigma_x/2], \quad \Theta (T)\equiv \int_0^T\text{d}t\,\Omega(t).$$
Explicit calculation then yields a control matrix whose only non-vanishing entries are 
\begin{equation}
\!Y_{xx}(t) = {\Omega(t)}, \;Y_{zz}(t) =  \cos\Theta(t), \;Y_{zy}(t) = \sin\Theta(t), 
\label{eq::AM}
\end{equation}
and leading-order error-vector components given by
\begin{align}
a_x^{(1)} (t)&= \frac{1}{2}\int_0^t \text{d}s \,\Omega(s)\beta_\Omega(s) , 
\label{eq::ax}\\
a_y^{(1)} (t)&= \int_0^t \text{d}s \sin\Theta(s)  \beta_z(s) ,  \label{eq::ay} \\
a_z^{(1)} (t)&= \int_0^t \text{d}s \cos\Theta(s) \beta_z(s). \label{eq::az}
\end{align}
Thus, $a_x^{(1)} (t)$ 
depends solely on the amplitude noise, whereas the dephasing noise enters both $a_y^{(1)} (t)$ and $a_z^{(1)} (t)$.
As long as the first term of the Magnus expansion in \erf{eq::Magnus} makes the dominant contribution to the dynamics, the $x$-component of the error vector is the relevant quantity for sensing amplitude control noise, which is the main focus of this work.

While in principle any quantity depending on $a_x^{(1)}(t)$ may be used to {\em selectively} gain information about the 
amplitude noise (that is, free from the effects of dephasing), specific choices are dictated by the measurement capabilities 
of an experimental implementation. Here, we work under the standard assumption that projective qubit measurements may 
be performed along the three axes of the Bloch's sphere (possibly with the aid of fast $\pi/2$ rotations if only 
measurements in the computational $z$ basis are directly
available, as in \cite{Frey2017}). Second-order moments of the error vector components play then a natural role, as they 
relate directly to outcome probabilities.  Consider, specifically, the expected squared magnitude 
\begin{align}
\mathcal{S}(T)\equiv\expect{[a_x^{(1)}(T)]^2}_{\beta_\Omega}, \quad T>0,
\label{eq::S}
\end{align}
where $\expect{\cdot}_{\beta_\Omega}$ denotes the ensemble average over realizations of $\beta_\Omega(t)$. 
If the qubit is initially prepared in $\ket{\!\uparrow_z}$ and evolves under the simultaneous influence of noise and control 
[Eq. \eqref{eq::Ham}], the noise-averaged probability of measuring it in $\ket{\!\uparrow_z}$ at time $T$ in the toggling frame 
is given by
\begin{align}
& {\mathbb P}(\uparrow_z,T)   = \langle\, |\langle \uparrow_z| \tilde{U}(T) | \uparrow_z \rangle |^2 
\,\rangle_{\beta_\Omega,\beta_z} \notag \\
& \!= \langle \cos^2 ||\mathbf{a}(T)|| + (a_z(T) / ||\mathbf{a}(T)||)^2 \sin^2 ||\mathbf{a}(T)|| \,\rangle_{\beta_\Omega,\beta_z} \notag \\
& \!= \!\Big\langle \frac{1}{2} + \frac{a_z(T)^2 + [a_x(T)^2 + a_y(T)^2] \cos(2||\mathbf{a}(T)|| )}{2 ||\mathbf{a}(T)||^2 } 
\Big\rangle_{\beta_\Omega,\beta_z}\notag \\
& \!\approx 1 - \langle [a_x^{(1)}(T)]^2 \rangle_{\beta_\Omega} - \langle[a_y^{(1)}(T)]^2\rangle_{\beta_z}, 
\label{eq::PzActual}
\end{align}
where in the last line we have expanded the cosine term and truncated the expression to second order in $T$. The analogous average survival probabilities for a qubit initially prepared 
in $\ket{\uparrow_x}$ and $\ket{\uparrow_y}$ are, respectively,
\begin{align}
{\mathbb P}(\uparrow_x,T) & \approx 1 - \langle [a_y^{(1)}(T)]^2 \rangle_{\beta_z}  - \langle [a_z^{(1)}(T)]^2 \rangle_{\beta_z} ,  
\label{eq::PxActual}\\
{\mathbb P}(\uparrow_y,T) & \approx 1 - \langle [a_x^{(1)}(T)]^2 \rangle_{\beta_\Omega}  - \langle [a_z^{(1)}(T)]^2 \rangle_{\beta_z}.  
\label{eq::PyActual}
\end{align}
While measurements cannot be implemented directly in the toggling frame, the above expressions still yield the desired  
probabilities in the physical frame provided that a fast rotation $R_x=U_c(T)^\dagger$ is implemented at $t=T$ immediately following 
evolution under $H(t)$ to effectively undo the rotation enacted by the ideal control \cite{Frey2017}.

If, in practice, a total of $M$ measurements are made along a fixed axis $i\in\{x,y,z\}$, each corresponding to a different (independent) 
noise realization, the estimated survival probability is $\hat{\mathbb P}(\uparrow_i,T)=\frac{1}{M}\sum_{m=1}^M i_m$, where $i_m$ 
is a random variable taking the values 1 (0) if the qubit is measured in $\ket{\uparrow_i}$ $\big( \ket{\downarrow_i}\big)$. 
From Eqs. (\ref{eq::PzActual})-(\ref{eq::PyActual}), we can estimate $\mathcal{S}(T)$ by
\begin{align}
\hat{\mathcal{S}}(T)
=\frac{1}{2} \Big[1 + \hat{\mathbb P}(\uparrow_x,T) - \hat{\mathbb P}(\uparrow_y,T) - \hat{\mathbb P}(\uparrow_z,T)\Big].
\label{eq::AxEst}
\end{align}
Formally, $\expect{\hat{\mathcal{S}}(T)}_\mathcal{M}=\mathcal{S}(T)$, where $\expect{\cdot}_\mathcal{M}$ denotes expectation 
in the limit $M\rightarrow\infty$ and $\mathcal{S}(T)$ is given in Eq. \eqref{eq::S}. Measurements of $\hat{\mathcal{S}}(t)$ for different 
controlled evolutions and sufficiently large $M$ will form the basis of the QNS procedure.

\subsection{Frequency domain: Noise spectra and filter functions}
\label{sec::FFs}

As we aim to determine the spectral properties of the amplitude noise affecting the sensor,  we work primarily in the frequency domain. Under the assumptions of Gaussianity and independence, the amplitude and dephasing noise are each fully characterized by their respective PSD. Stationarity, namely, invariance of statistical properties under time translation, allows us to write the noise autocorrelation function as 
$$\expect{\beta_u(t_1)\beta_u(t_2)}_{\beta_u}=\expect{\beta_u(\tau)\beta_u(0)}_{\beta_u}, \quad u\in\{\Omega,z\},$$ 
where $\tau \equiv t_1-t_2$ is the lag time. From the Wiener-Khinchin theorem, the PSD may be obtained as the Fourier transform of the autocorrelation function \cite{ClerkRMP}, 
\begin{equation*}
S_{u}(\omega)=\frac{1}{2\pi}\int_{-\infty}^\infty \mathrm{d}\tau \,e^{-i\omega \tau}\expect{\beta_u(\tau)\beta_u(0)}_{\beta_u},
\quad u\in\{\Omega,z\}.
\end{equation*}
For real, scalar and stationary classical stochastic processes like $\beta_u(t)$, 
the PSD is a non-negative, even function.

In the frequency domain, the action of the external control is most conveniently described within a {\em transfer filter function} (FF) formalism \cite{Kofman2004,Biercuk_Filter,GreenNJP2013,Ball2014,PazFFF}. In particular, the functions that describe, to the leading order, the filtering of the noise terms $\beta_u(t) \sigma_v$ in the toggling-frame Hamiltonian of Eq. \eqref{eq::toggle} are directly 
related to the first-order {\em fundamental FFs} $F^{(1)}_{uv}(\omega,t)$ \cite{PazFFF}, 
which are defined in terms of the Fourier transform of the control matrix elements, 
$$F^{(1)}_{uv}(\omega,t) \equiv \int_0^t \text{d}s \, Y_{uv}(s) e^{i \omega s}, \quad u,v \in \{x,y,z\}.$$
Fundamental FFs may be used as building blocks to evaluate how the qubit's response to amplitude and dephasing noise is modified by the applied control.  For instance, the expected squared magnitude of the first-order error vector $\mathbf{a}^{(1)}(t)$, which determines the leading contribution to the operational fidelity loss, can be seen to depend upon overlap integrals between the PSDs and the associated FFs \cite{GreenNJP2013,PazFFF}.  Using the explicit form of the control matrix for 
amplitude modulation in Eq. \eqref{eq::AM}, and noting that, since $ \beta_x(t) =\beta_y(t) = {\beta_\Omega (t)}/{2}$, 
 the corresponding spectra obey $S_x(\omega)=S_y(\omega)=S_\Omega(\omega)/4$, 
one may show that 
\begin{equation*}
\langle||\mathbf{a}^{(1)}(t)||^2\rangle_{\beta_\Omega,\beta_z} \equiv \frac{1}{2\pi} 
\sum_{u = \Omega, z} \int_{-\infty}^{\infty} \!\mathrm{d}\omega \, S_u(\omega) F_u( \omega, t ), 
\end{equation*}
where the effective amplitude and dephasing first-order FFs (or ``filters'' for short) are given by 
\begin{eqnarray}
&& F_\Omega(\omega,t) \equiv \frac{1}{4} \,| F^{(1)}_{xx} (\omega,t)|^2, \\
&& F_z(\omega, t) \equiv | F^{(1)}_{zz} (\omega,t)|^2 +  | F^{(1)}_{zy} (\omega,t)|^2. 
\end{eqnarray}

Because we specialize here to spectroscopy of amplitude noise, our focus is the {\em amplitude filter} entering the signal $\mathcal{S}(T)$, 
corresponding to the $u=\Omega$ term above, namely \cite{One}, 
\begin{eqnarray}
\label{eq::errx}
\mathcal{S}(T) &=& \frac{1}{2\pi} \int_{-\infty}^{\infty}\! \mathrm{d}\,\omega\, S_{\Omega}(\omega) F_{\Omega}(\omega,T),\\
F_{\Omega}(\omega,T) &=& \Big| \frac{1}{2}\int_0^T\!\!\mathrm{d}s\, e^{i\omega s}\Omega(s) \Big|^2.
\label{eq::errx1}
\end{eqnarray}
Assume now that, as in Ref. \cite{Frey2017}, the amplitude modulation is 
{\em piecewise-constant}, with sampling interval $\Delta t >0$, 
\begin{align}
\label{eq::piecwiseOmega}
\Omega(t)\equiv \Omega_n,\;\; t\in [n\Delta t,\, (n+1)\Delta t] , \, n=0,\ldots, N-1.
\end{align}
Over a total duration $T=N\Delta t$, this control generates a rotation of the qubit about $\sigma_x$ by 
angle $\Theta (T)=\Delta t\sum_{n=0}^{N-1}\Omega_n$.
The corresponding amplitude fundamental FF becomes
\begin{align*}
F_{xx}^{(1)}(\omega,T) &=\frac{1}{2}\sum_{n=0}^{N-1}\int_{n\Delta t}^{(n+1)\Delta t}\!\!\!\!\mathrm{d}s\,e^{i\omega s}\Omega_n\\
&=\frac{1}{2}\,e^{i\omega(N-1)\Delta t/2}\bigg(\int_{0}^{\Delta t}\!\!\!\!\mathrm{d}s\,e^{i\omega s}\bigg)\tilde{\Omega}(\omega),
\end{align*}
where $\tilde{\Omega}(\omega)$ denotes the discrete-time Fourier transform (DTFT) of the sequence $\{\Omega_n\}$,
defined by 
$$\tilde{\Omega}(\omega)\equiv \sum_{n=0}^{N-1}\Omega_n \, e^{i\omega[n-(N-1)/2]\Delta t}.$$
The amplitude filter appearing in $\mathcal{S}(T)$ thus simplifies to
\begin{align}
\label{eq::amplitudeFF}
\!\!\!F_{\Omega}(\omega,T)\!=  \!\frac{\text{sin}^2(\omega\Delta t/2)}{\omega^2}\big|\tilde{\Omega}(\omega)\big|^2 \! \equiv 
s(\omega, \Delta t) \,\big |\tilde{\Omega}(\omega)\big|^2,
\end{align}
where the envelope function $s(\omega, \Delta t)$ is a consequence of the finite duration of the sampling interval $\Delta t$. 

It is worth noting that, by Parseval's theorem, the total energy (as quantified by the $L^2$-norm) of the time-domain 
``signal" $Y_{uv}(t)$ is preserved under Fourier transform, implying 
\begin{equation*}
\int_0^T \!\!\mathrm{d}t \, Y_{uv}(t)^2 = \frac{1}{2\pi}\int_{-\infty}^\infty \!\!\!\! \mathrm{d}\omega\, |F^{(1)}_{uv}(\omega,T)|^2, \;\: u,v \in \{x,y,z\}.
\label{eq::parseval0}
\end{equation*} 
Accordingly, the area underneath the amplitude filter obeys
\begin{eqnarray}
\int_0^T \!\mathrm{d}t \,\Omega(t)^2 = \frac{2}{\pi} \int_{-\infty}^\infty \! \mathrm{d}\omega\, F_\Omega (\omega,T).
\label{eq::parseval}
\end{eqnarray}

\subsection{Quantum noise spectroscopy: A first look}
\label{sec::QNS1}

Broadly speaking, QNS protocols characterize noise by using a quantum system as a probe of its environment. The basic steps in this procedure are (1) prepare the sensor in a known state; (2) let it evolve under both noise and a known control sequence; (3) measure an observable; (4) repeat and gather measurement statistics for different controlled evolutions and observables; (5) process the acquired data to produce an estimate of the noise PSD. In the frequency domain, subjecting the quantum sensor to different controlled evolutions is equivalent to modifying its spectral response with different FFs. Appropriate FF design is thus essential to access information about the noise PSD from the measurement statistics. 

In the setting under consideration, Eqs. \eqref{eq::errx}-\eqref{eq::errx1} show explicitly how the applied control modifies the sensor's frequency response to amplitude noise.  By selecting particular control waveforms for $\Omega(t)$, we can take advantage of this relationship to estimate $S_\Omega(\omega)$ at specific frequencies. Suppose we choose a control waveform that produces a band-pass filter centered at $\omega_s$, i.e.,  $F_{\Omega}(\omega,T)$ is large in some frequency interval, or \emph{passband}, 
$(\omega_s\!-\!\Delta b,\omega_s\!+\!\Delta b)\subset (0,\infty)$ and small outside. In \erf{eq::errx}, the effect of such a filter is to attenuate the PSD when $\omega\notin (\omega_s\!-\!\Delta b,\omega_s\!+\!\Delta b)$, making the qubit most sensitive to noise at frequencies within the passband. If $F_{\Omega}(\omega,T)=0$ for $\omega\notin (\omega_s - \Delta b,\omega_s+ \Delta b)$, 
the filter is said to be {\em band-limited}. For small enough $\Delta b$, the target PSD is approximately flat within the passband, i.e., $S_\Omega(\omega)\approx S_\Omega(\omega_s)$ for $\omega\in (\omega_s\!-\!\Delta b,\omega_s\!+\!\Delta b)$. When the filter is band-limited, 
\begin{align}
\label{eq::Ax}
\mathcal{S}(T)\approx \frac{S_\Omega(\omega_s)}{\pi}\int_{\omega_s\!-\!\Delta b}^{\omega_s\!+\!\Delta b}\mathrm{d}\omega F_{\Omega}(\omega,T), \quad  \omega_s > \Delta b.
\end{align}
Since $F_{\Omega}(\omega,T)$ is known, 
knowledge of $\mathcal{S}(T)$ suffices to determine $S_\Omega(\omega_s)$. 
The estimate of the PSD 
is then
\begin{align}
\label{eq::BandpassEst}
\hat{S}_\Omega(\omega_s)&= \frac{\hat{\mathcal{S}}(T)}{ A^{(\omega_s)}},\;A^{(\omega_s)} \equiv \frac{1}{\pi} \int_{\omega_s-\Delta b}^{\omega_s+\Delta b}d\omega F_{\Omega}(\omega,T).
\end{align}
The expected value of this {\em passband estimator} depends on the overlap integral of 
$F_{\Omega}(\omega,T)$ and the actual PSD,
\begin{align}
\expect{\hat{S}_\Omega(\omega_s)}_{\mathcal{M}}&=  \frac{1}{ \pi A^{(\omega_s)} }
\int_{0}^{\infty} \mathrm{d}\omega\, F_{\Omega}(\omega,T) S_{\Omega}(\omega).
\label{eq::BandpassExpect}
\end{align}

\begin{figure}[t]
\includegraphics[scale=.76]{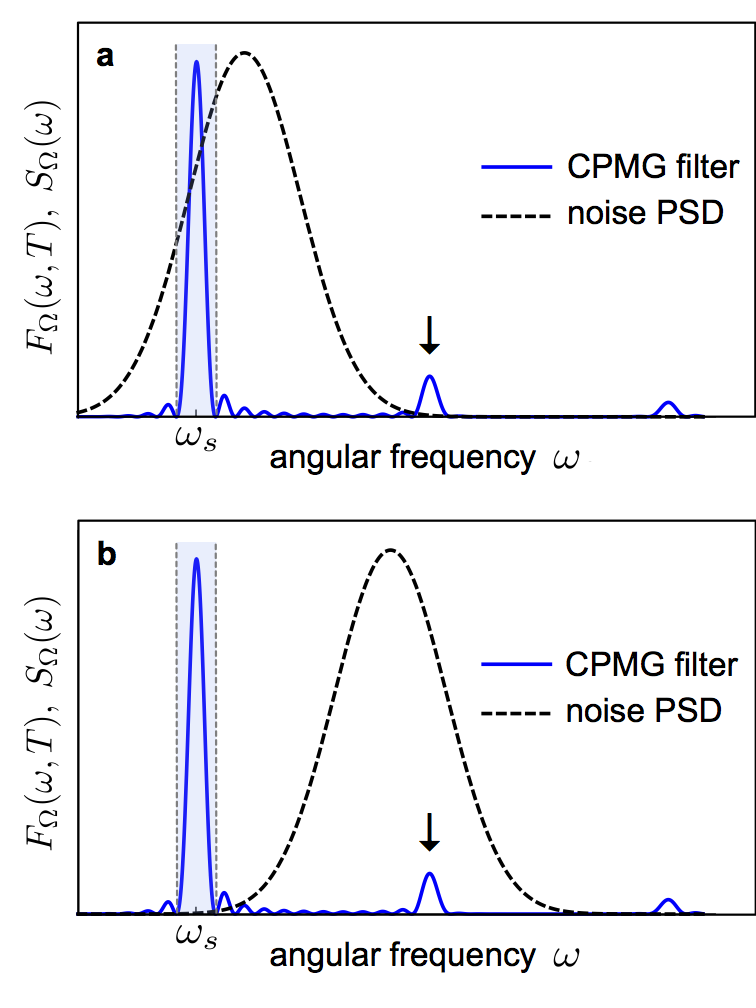}
\vspace*{-3mm}
\caption{(Color online) 
Effect of spectral leakage and bias on PSD estimation. The FF generated by a 12-pulse CPMG control sequence (blue solid line) is depicted in two spectral estimation scenarios, with the shaded area corresponding to the passband centered at $\omega_s$.  Leakage is manifest in the presence of a high frequency lobe, marked with an arrow. In (a), the PSD being estimated (black dashed line) has minimal spectral overlap with the high frequency lobe. Consequently, the leakage has little effect on the expected value of the estimate in \erf{eq::BandpassExpect}. In (b), the PSD (black dashed line) and the high frequency lobe have significant overlap. This causes the high frequency lobe to contribute to the overlap integral in \erf{eq::BandpassExpect}, resulting in a biased, over-estimated PSD at $\omega_s$. 
\label{fig::Leakage}}
\end{figure}

The strategy outlined above hinges on our capability to create narrow, (approximately) band-limited FFs. There is a fundamental barrier to how well this can be achieved, however. In realistic sensing applications, where data is acquired over a finite duration, a signal like 
$\Omega(t)$ is necessarily {\em time-limited}, i.e., nonzero only within a finite time interval. Since it is impossible for a signal to be simultaneously time-limited and band-limited in the frequency domain, $F_{\Omega}(\omega,T)$ will necessarily suffer from {\em leakage}, namely, non-zero spectral components outside the passband. Depending on both the extent and distribution of the leakage and on the PSD itself, the estimate of the PSD can be significantly biased. 

As a simple example, consider an $n$-pulse CPMG rotary spin echo (RSE) sequence, which is described 
by a flat-top control waveform in which a constant amplitude, 
$|\Omega(t)|=\Omega$, switches sign at times $t={T}/{2n},{3T}/{2n},\ldots,{(2n-1)T}/{2n}$ \cite{Two}.
The amplitude FFs created by such a RSE (more generally, by a {\em Walsh RSE} \cite{Ball2014}) have the same functional form as the 
dephasing FFs produced by CPMG sequences with instantaneous $\pi$-pulses, which have been used to characterize dephasing noise \cite{Alvarez_Spectroscopy,Bylander2011,BarGill2012}. The plots in Fig. \ref{fig::Leakage} depict the FF of a $n=12$ CPMG sequence, which is used to estimate two different PSDs at $\omega_s$. The FF  has a high frequency lobe outside the passband centered at $\omega_s$. If the PSD being estimated is nonzero in the region of the high frequency lobe, the integral in \erf{eq::BandpassExpect} can be significantly influenced by the value of the PSD outside the passband, biasing the estimate at $\omega_s$.

\section{Background from classical spectral estimation theory}
\label{sec::background}

\subsection{The spectral concentration problem}
\label{sec::sc}

Estimation of the PSD of an unknown signal from a set of discrete time samples is a central problem in classical spectral analysis. Consider a stationary, Gaussian stochastic process $x(t)$ with PSD $S_x(\omega)$. Suppose that $x(t)$ is measured at times $0,\Delta t,\ldots,(N-1)\Delta t$, producing a finite, discrete-time sequence $\{x_0,\ldots,x_{N-1}\}\equiv \{x_n\}$.  One of the earliest methods to estimate $S_x(\omega)$ was Schuster's periodogram \cite{percival1993spectral}, 
given by $\hat{S}_x^\text{p}(\omega_s) \equiv |\tilde{x}(\omega_s)|^2/N$, where $\tilde{x}(\omega_s)$ 
is the sample's DTFT. In the frequency domain, the expected value of $\hat{S}_x^\text{p}(\omega_s)$ is an overlap integral of the true spectrum and the so-called Fej\'{e}r kernel $F_N(\omega-\omega_s)$, that is, 
\begin{eqnarray*}
\!\!\!\expect{\hat{S}_x^\text{p}(\omega_s)}_x&=&\frac{1}{2\pi}\int_{-\omega_\text{N}}^{\omega_\text{N}} \!\! d\omega \,F_N(\omega-\omega_s)S_x(\omega),\;\;
\label{eq::periodogram} \\
F_N(\omega-\omega_s)&=&\frac{1}{N}\frac{\text{sin}^2[N(\omega-\omega_s)\Delta t/2]}{\text{sin}^2[(\omega-\omega_s)\Delta t/2]}.
\nonumber
\end{eqnarray*}
Note that the above integral extends over the \emph{principal domain}, $(-\omega_\text{N},\omega_\text{N}]$, defined by the {\em Nyquist frequency} $\omega_\text{N}\equiv\pi/\Delta t$. Like other estimators that depend on the squared magnitude of a DTFT, the periodogram is only capable of estimating the PSD within the principal domain, since $|\tilde{x}(\omega_s)|^2$ is periodic over $2\omega_\text{N}$. 

\begin{figure*}[th]
\centering
\includegraphics[scale=.85]{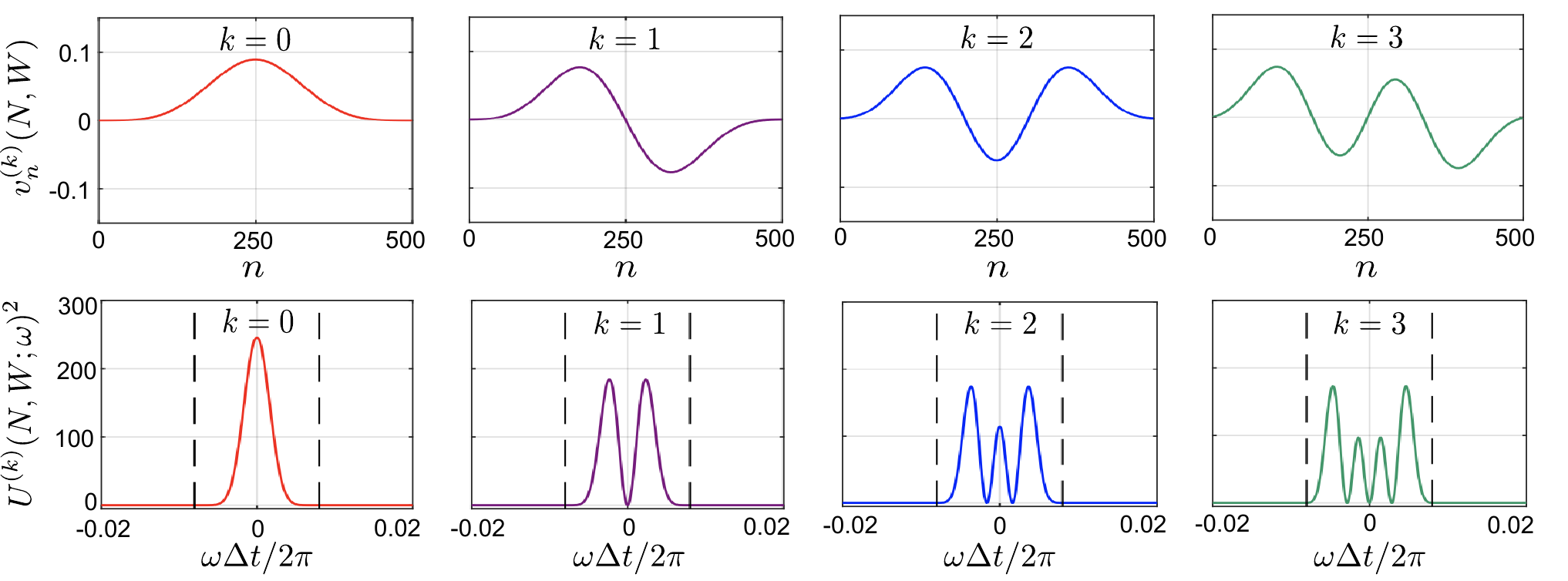}
\vspace*{-2mm}
\caption{(Color online) DPSS and DPSWF for $N=500$ and $W=4/N$. The DPSS $\{v_n^{(k)}(N,W)\}$ (top row) are plotted vs. $n$ for orders $k=0,\,1,\,2,\,3$ (left to right). The squared magnitudes of DPSWF $|U^{(k)} \!(N,W;\omega)|^2$ (bottom row) are plotted vs. angular frequency scaled by $\Delta t/(2\pi)$ for orders $k=0,\,1,\,2,\,3$ (left to right). The boundaries of the passbands at $\omega\Delta t/(2\pi)=\pm W$ are marked by dashed vertical lines.
\label{fig::SlepianPlots}}
\end{figure*}

Owing to the fact that the samples $ \{x_n\}$ are acquired over a finite time duration, the periodogram necessarily suffers from leakage bias. Leakage is evident in the shape of the Fej\'{e}r kernel, which serves as a classical analogue of the amplitude filter in \erf{eq::BandpassExpect}. Though the Fej\'{e}r kernel  approaches $2\pi\delta(\omega-\omega_s)$ as $N\rightarrow\infty$, it is an oscillatory function  for finite $N$, having numerous sidelobes outside the main passband that bias the estimate. By the 1970's, it was standard practice to suppress this bias with spectral windows or ``tapering functions'' \cite{percival1993spectral,Harris1978}. A taper is a discrete-time sequence $\{t_n\}$, with $\sum_{n=0}^{N-1}t_n^2=1$, which defines a modified DTFT of the sample, 
\begin{align*}
\tilde{x}_t(\omega_s) \equiv \sum_{n=0}^{N-1}t_ne^{i\omega_s[n-(N-1)/2]\Delta t}x_n.
\end{align*}
This produces the {\em tapered estimate} 
$\hat{S}_x^t(\omega_s) \equiv |\tilde{x}_t(\omega_s)|^2,$
whose expected value again takes a form similar to \erf{eq::BandpassExpect},
\begin{align}
\label{eq::expecttaper}
\expect{\hat{S}_x^t(\omega_s)}_x = \frac{1}{2\pi}\int_{-\omega_\text{N}}^{\omega_\text{N}} 
d\omega\, S_x(\omega) |\tilde{t}(\omega-\omega_s)|^2.
\end{align}

A natural question is what tapered estimate has the least bias due to leakage. That is, we seek the finite, discrete-time sequence $\{t_n\}$, of length $N$ and sampling time $\Delta t$, such that $|\tilde{t}(\omega)|^2$ is maximally concentrated in a chosen passband, $(-\Delta b, \Delta b)$, with minimal leakage outside. The {\em spectral concentration} of $|\tilde{t}(\omega)|^2$ in $(-\Delta b, \Delta b)$ is quantified by
\begin{align}
\label{eq::SpectralCon}
E_{\Delta b}[\{t_n\}] \equiv \frac{\int_{-\Delta b}^{\Delta b}d\omega |\tilde{t}(\omega)|^2}{\int_{-\omega_\text{N}}^{\omega_\text{N}} 
d\omega |\tilde{t}(\omega)|^2},
\end{align}
which is the ratio of energy contained in $(-\Delta b, \Delta b)$ to that in the principal domain. The task of finding the taper maximizing $E_{\Delta b}[\{t_n\}]$ is termed the ``spectral concentration problem''.

\subsection{The Slepian sequences and their Fourier transforms}
\label{sec::dpss}

The taper maximizing \erf{eq::SpectralCon} was first determined by Papoulis and Bertran \cite{Papoulis1972}. In the last of a series of influential papers that explored the extent to which time-limited signals could be band-limited (and vice-versa), Slepian showed that the solution of Papoulis and Bertran belonged to a family of discrete-time functions called the discrete prolate spheroidal sequences \cite{Slepian1961,Landau1961a,Landau1961b,Slepian1964,Slepian1978}. As Slepian was the first to systematically study their mathematical properties, the DPSS are nowadays often termed ``Slepians". For $n,k\in\{0,\ldots,N-1\}$ and {\em bandwidth parameter} $W\equiv\Delta b\Delta t/2\pi$, the $k$th-order DPSS $\{v_n^{(k)}(N,W)\}$ is a  real sequence of length $N$ solving the following Toeplitz matrix eigenvalue equation:
\begin{align*}
\sum_{m=0}^{N-1}\frac{\text{sin}\,2\pi W(n-m)}{\pi(n-m)}v_m^{(k)}(N,W) = \lambda_k(N,W)\,v_n^{(k)}(N,W).
\end{align*}
The DPSS are an orthonormal basis of $\mathbb{R}^N$, satisfying $\sum_{n=0}^{N-1}v_n^{(k)}(N,W)v_n^{(\ell)}(N,W)=\delta_{k,\ell}$. The order $k$ of a DPSS is determined by the size of its eigenvalue, where $1>\lambda_0(N,W)>\lambda_1(N,W)> \ldots >\lambda_{N-1}(N,W)>0$. The eigenvalue $\lambda_k$ also determines the spectral concentration of a DPSS in $B_{0}\equiv(-2\pi W/\Delta t,2\pi W/\Delta t)$, 
\begin{align}
\label{eq::SlepConcentration}
E_{2\pi W/\Delta t}[\{v_n^{(k)}(N,W)\}]=\lambda_k(N,W). 
\end{align}
The optimal taper is, thus, the DPSS of order $k=0$. Lower order DPSS, with $k< K\equiv \lfloor 2NW\rfloor$ (the so-called Shannon number), are the most spectrally concentrated in $B_{0}$. Representative DPSS are plotted in Fig. \ref{fig::SlepianPlots}(a).
Remarkably, the DPSS extended to $n\in\mathbb{Z}$  solve a time concentration problem analogous to \erf{eq::SpectralCon} (see Appendix \ref{app::TimeConcen}).

The spectral properties of the DPSS are captured by their DTFTs $\{ U^{(k)} \!(N,W;\omega)\}$, 
known as the {\em discrete prolate spheroidal wave functions} (DPSWF), 
\begin{align}
\label{eq::DPSWF}
\!U^{(k)} \!(N,W;\omega)\!=\epsilon_k\!\sum_{n=0}^{N-1}\!v_n^{(k)}\!(N,W)e^{i\omega[n-(N-1)/2]\Delta t},
\end{align}
where $\epsilon_k=1\,(i)$ for even (odd) $k$, ensuring that $U^{(k)} \!(N,W;\omega)$ 
are real functions. Each $U^{(k)} \!(N,W;\omega)$
has $k$ zeros in $B_0$ and $N-1$ zeros in the principal domain, and is odd or even depending on the parity of $k$.
For $k<K$, it follows from \erf{eq::SlepConcentration} that the DPSWF  are approximately band-limited to $B_{0}$, as shown in Fig. \ref{fig::SlepianPlots} for $k=0, 1, 2, 3$.

\begin{figure}[t]
\centering
\includegraphics[scale=.61]{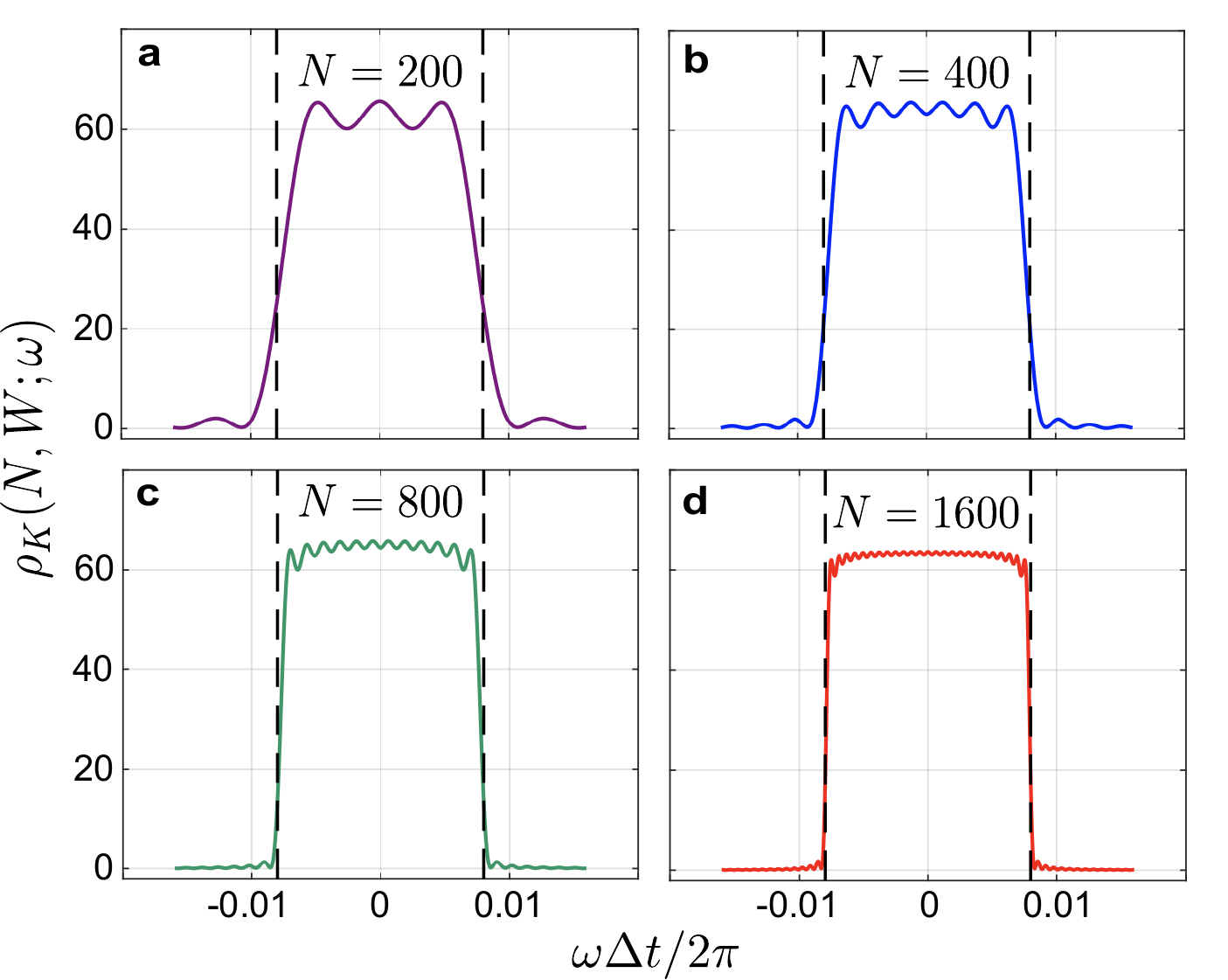}
\vspace*{-3mm}
\caption{(Color online)
Convergence to an ideal band-pass filter with increasing $N$.  The average squared magnitude of the first $K$
DPSWF, denoted by $\rho_K(N,W;\omega)$ [\erf{eq::complete}], for $W=0.008$ and (a) $N=200$ and $K=3$, (b) $N=400$ and $K=6$, (c) $N=800$ and $K=12$, and (d) $N=1600$ and $K=25$. Boundaries of the passband at $\omega\Delta t/(2\pi)=\pm W$ are marked with vertical dashed lines. With increasing $N$, $\rho_K(N,W;\omega)$ approximates an ideal (square, band-limited) filter with increasing accuracy. 
\label{fig::IdealFilter}}
\end{figure}

The orthogonality properties of the DPSWF are particularly relevant from the standpoint of spectral estimation. 
In fact, the DPSWFs are orthogonal in {\em both} $B_0$ and the principal domain, 
\begin{align*}
\frac{2\pi\delta_{k,\ell}}{\Delta t}=&\frac{1}{\lambda_k(N\!,\!W)}\!\int_{B_0}\!\!\text{d}\omega\, U^{(k)}\!(N\!,\!W\!;\omega)U^{(\ell)}\!(N\!,\!W\!;\omega)\\
=&\int_{-\omega_\text{N}}^{\omega_\text{N}}\!\!\!\text{d}\omega\, U^{(k)}\!(N,W;\omega)U^{(\ell)}\!(N,W;\omega).
\end{align*}
On account of their orthogonality and spectral concentration in $B_0$, the DPSS of orders $k<K$ form an {\em approximately complete} functional basis in $B_0$. As a consequence, Thomson observed that the sum of the squares of the first $K$ DPSWF approximates a square wave or an ideal band-pass filter 
\cite{thomson_multitaper},
\begin{align}
\label{eq::complete}
\rho_K(N,W;\omega)\equiv 
\frac{1}{K}\sum_{k=0}^{K-1} U^{(k)}\!(N,W;\omega)^2\approx\frac{1}{2W}\mathds{1}_{B_0}(\omega),
\end{align}
where $\mathds{1}_{B_0}(\omega)$ denotes the indicator function of $B_0$.
This correspondence becomes more accurate with increasing $N$, as illustrated in Fig. \ref{fig::IdealFilter}. 
Exact convergence in the $L_1$-norm as $N\rightarrow\infty$ was recently rigorously established in Ref. \cite{Abreu2016}. 

\subsection{Multitaper classical spectral estimation}
\label{sec::multitaper}

Irrespective of the tapering function, all tapered spectral estimates $\hat{S}_x^t(\omega_s)$ suffer from a serious drawback: for $N\gg1$, 
$\hat{S}_x^t(\omega_s)$ is approximately distributed as $\chi_2^2$, a chi-square with two degrees of freedom \cite{Brillinger1981}. Thus, the variance of $\hat{S}_x^t(\omega_s)$ does {\em not} decrease with the sample size $N$, meaning that $\hat{S}_x^t(\omega_s)$ is an {\em inconsistent} estimator. A variety of approaches have been proposed to remedy this problem, from the methods of Bartlett and Welch, which  involve dividing $\{x_n\}$ into multiple smaller samples, to convolutional smoothing of the tapered estimate \cite{percival1993spectral}. However, the consistency offered by these techniques comes at the expense of additional bias. In 1982, Thomson devised a solution that provided a consistent estimator while introducing less bias than other approaches  \cite{thomson_multitaper}. 

The key to multitaper estimation is to form multiple independent estimates with the same set of time-domain samples $\{x_n\}$.  Rather than tapering the samples with a single function, the multitaper approach utilizes a set of tapers formed by the first $K$ DPSS. In Thomson's terminology, each DPSS of order $k$ produces a unique ``eigenestimate,'' 
\begin{align}
\label{eq::ceigenest}
\hat{S}_x^{k}(\omega_s)&=\Big|\sum_{n=0}^{N-1}v_n^{(k)}(N,W)e^{-i\omega_s[n-(N-1)/2]\Delta t}x_n\Big|^2 ,
\end{align}
with expected value
\begin{align*}
\expect{\hat{S}_x^{k}(\omega_s)}_x=\int_{-\omega_\text{N}}^{\omega_\text{N}}
d\omega\, U^{(k)}(N,W;\omega_s-\omega)^2\,S_x(\omega).
\end{align*}
For white noise, the orthogonality of the DPSS and DPSWF implies that the eigenestimates are statistically independent. In the presence of colored noise, this holds to good approximation depending on the extent to which $S_x(\omega)$ is locally flat in $B_{\omega_s}\equiv(-2\pi W/\Delta t+\omega_s,2\pi W/\Delta t+\omega_s)$  \cite{thomson_multitaper}. 

The final multitaper estimate, $\hat{S}_x^\text{m}(\omega_s)$, is formed by combining the eigenestimates into a weighted average,
\begin{align*}
\hat{S}_x^\text{m}(\omega_s) \equiv \sum_{k=0}^{K-1}d_k(\omega_s)\hat{S}_x^k(\omega_s), \quad 
 \sum_{k=0}^{K-1}d_k(\omega_s)=1.
\end{align*}
Since it is the sum of $K$ approximately independent $\chi_2^2$ distributed random variables, $\hat{S}_x^\text{m}(\omega_s)$ is well described by a chi-squared distribution with $2K$ degrees of freedom. If the weights are taken to be $d_k(\omega_s)=1/K$ for all $k$, the variance of $\hat{S}_x^\text{m}(\omega_s)$ decreases with $N$, scaling as $1/K=1/\lfloor2WN\rfloor$. The expected value of the multitaper estimate is 
\begin{align*}
\expect{\hat{S}_x^{\text{m}}(\omega_s)}_x=\int_{-\omega_\text{N}}^{\omega_\text{N}}
d\omega \,\rho_K(N,W; \omega-\omega_s)\,S_x(\omega).
\end{align*}
Since, by Eq. \eqref{eq::complete}, $\rho_K(N,W; \omega-\omega_s)$ converges to an ideal band-pass filter as $N$ increases, 
$\expect{\hat{S}_x^{\text{m}}(\omega_s)}_x$ asymptotically approaches $\frac{\Delta t}{4\pi W}\int_{B_{\omega_s}}\!\!d\omega\, S_x(\omega)$, the average of the PSD in the passband. For finite $N$, however, $\rho_K(N,W; \omega-\omega_s)$ is not ideal, 
despite the spectral concentration of the DPSS tapers. To combat the effects of leakage, 
Thomson introduced an adaptive procedure to determine the $d_k(\omega_s)$, which down-weights more biased eigenestimates \cite{thomson_multitaper}. This technique, which we will revisit in Sec. \ref{sec::adaptive}, produces a variance that decreases with 
$N$, thereby ensuring consistency, while at the same time minimizing leakage bias.

\section{Multitaper Quantum Spectral Estimation}
 \label{sec::MultiQNS}

\subsection{Quantum vs. classical estimation settings}

The FF formalism provides a natural starting point to adapt multitaper spectral estimation to the quantum setting. 
Unsurprisingly, however, this entails significant differences in both the formulation of the spectral estimation problem 
and in the implementation of multitaper itself. As a consequence, we ultimately find that multitaper
techniques serve different purposes in quantum vs. classical estimation tasks. While the latter will be concretely illustrated in 
Sec. \ref{sec::results}, we highlight here the key conceptual features responsible for these differences. 

First, and perhaps most notably, the classical approach relies on a collection of discrete-time samples $\{ x_n\}$ of the noise 
process. Convolving these noise samples with a tapering function and taking the squared modulus, as described in Sec. \ref{sec::sc},
determines the tapered estimate $|\tilde{x}_t(\omega_s)|^2$, which is the basic building block of the classical multitaper. The quantum analogue of the tapered estimate is the passband estimator in \erf{eq::BandpassEst}, which depends on $a_x^{(1)} (T)\propto\int_0^T \text{d}t \,\Omega(t)\beta_\Omega(t)$. The amplitude control waveform $\Omega(t)$, thus, serves as a ``continuous taper" of the amplitude noise $\beta_\Omega(t)$. Note that tapering takes place {\em during} the controlled evolution of the qubit, \emph{prior} to measurement. In contrast, tapering in the classical case is a form of {\em post-processing} on the experimental data.

The manner in which data is acquired has also significant implications for multitaper QNS. Rather than a set of discrete time-domain samples, spectral estimation in the quantum setting is based on the qubit measurement outcomes $\{i_1,\ldots,i_M\}\equiv\{i_m\}$, 
where $i_m=1\,(0)$ as described in Sec. \ref{sec::time}. Classically, multitaper estimation is largely motivated by the fact that the 
variance of the tapered estimate does not decrease with the sample size $N$.  Multitapering has a different utility in the quantum 
setting, however, since the variance of the passband estimator does decrease with $M$. This is seen by noting that 
$\hat{S}_\Omega(\omega_s)\propto[1 + \hat{\mathbb P}(\uparrow_x, t) - \hat{\mathbb P}(\uparrow_y,t) - \hat{\mathbb P}(\uparrow_z,t)],$ where each $\hat{\mathbb P}(\uparrow_i,t)$ is the sample mean of $\{i_m\}$. The $i_m$ are i.i.d. random variables, which implies that 
the variances of the sample means, and consequently $\hat{S}_\Omega(\omega_s)$, scale as $1/M$. Despite its well-behaved 
variance, the quantum passband estimator is nonetheless inconsistent, as consistency also requires $\hat{S}_\Omega(\omega_s)$ to be 
unbiased as $M\rightarrow\infty$. The topic of bias will be discussed in detail in Sec. \ref{sec::adaptive}. Recall for now that 
leakage bias is present for any finite evolution time $T$. Because leakage is independent of $M$, it  
persists in the asymptotic limit.

Unlike the classical tapered estimate, which is $\chi_2^2$-distributed for $N\gg1$, the quantum passband estimator is normally distributed for $M\gg1$. Note that measuring the qubit in $\ket{\!\uparrow_i}$ or $\ket{\!\downarrow_i}$ at time $T$ is the outcome of a Bernoulli trial with variance ${\mathbb P}(\uparrow_i,T)[1-{\mathbb P}(\uparrow_i,T)]$, $i\in\{x,y,z\}$. Dependent on the experimental implementation, noise in the measurement apparatus may dominate over these statistical fluctuations.  To account for either possibility, we denote the variance of an individual outcome $i_m$ (or measurement shot) by $\sigma_{\uparrow_i}^2$. As a consequence of the central limit theorem when $M\gg1$, the sample mean $\hat{\mathbb P}(\uparrow_i,T)$ is normally distributed with variance 
$\sigma_{\uparrow_i}^2/M$.  From Eqs. (\ref{eq::BandpassEst}) and (\ref{eq::AxEst}), $\hat{S}_\Omega(\omega_s)$ is thus a sum of normally distributed random variables when $M \gg1$ and is itself normally distributed with variance
\begin{align}
\label{eq::VarEst}
\text{var}[\hat{S}_\Omega(\omega_s)]=\frac{1}{2M A^{(\omega_s) 2}}\sum_{i\in\{x,y,z\}}\!\!\!\sigma_{\uparrow_i}^2\equiv\frac{\sigma^2}{M A^{(\omega_s) 2}}, 
\end{align}
where $A^{(\omega_s)}$ is defined in Eq. \eqref{eq::BandpassEst}. Together with 
the expected value of $\hat{S}_\Omega(\omega_s)$ in \erf{eq::BandpassExpect}, this yields 
the probability of finding $\hat{S}_\Omega(\omega_s)=s$ conditioned on $S_\Omega(\omega)$ in the 
explicit form
\begin{align}
\label{eq::condprob}
{\mathbb P}&\left[\hat{S}_\Omega(\omega_s)=s \,\big| \,S_\Omega(\omega)\right]\propto\\\notag
&\text{exp}\bigg\{\!\!-\frac{\left[\,s-\frac{1}{\pi A^{(\omega_s)\,2}}\!\int_0^\infty\text{d}\omega \, S_\Omega(\omega) F_\Omega(\omega,T)\,\, \right]^2} {{2\sigma^2/[M A^{(\omega_s)\,2}]}}\bigg\}.
\end{align}

\subsection{Band-shifted DPSS amplitude filters}
\label{sec::DPSSFF}

The eigenestimates in classical multitaper estimation are created by tapering discrete-time samples with a DPSS. For sensing amplitude noise in the quantum setting, we take a similar approach by using the continuous taper $\Omega(t)$, which creates a corresponding DPSS filter in the frequency domain.  Let the piecewise-constant amplitude control modulation of Eq. \eqref{eq::piecwiseOmega} be chosen so that
\begin{equation}
\Omega_n^{\text{DPSS}} \equiv\Omega\, v_n^{(k)}(N,W), \quad n=0,\ldots,N-1,
\label{eq::basic}
\end{equation}
where the constant $\Omega$ has units of angular frequency. From \erf{eq::amplitudeFF}, the resulting amplitude filter is proportional to the square of a DPSWF,
\begin{align}
\label{eq::kFF}
F_{\Omega}^{\text{DPSS}}(\omega,T) = \Omega^2\,\frac{\text{sin}^2(\omega\Delta t/2)}{\omega^2}\,U^{(k)}(N,W;\omega)^2.
\end{align} 
As noted in Sec. \ref{sec::FFs}, the envelope term $s(\omega,\Delta t) = \text{sin}^2(\omega\Delta t/2)/\omega^2$ arises 
from the finite duration $\Delta t$ of the piecewise constant intervals that comprise $\Omega(t)$ and is absent in the classical setting.  
Since QNS requires narrow band-pass filters with $k< K$ and $4\pi W/\Delta t \ll\omega_\text{N}$, $s(0,\Delta t)$ is locally flat in 
the region where $U^{(k)}(N,W;\omega)^2$ is spectrally concentrated for typical parameters. Consequently, $F_{\Omega}^{\text{DPSS}}(\omega,T)$ largely retains the shape of the corresponding DPSWF in the parameter regime relevant to QNS.  

\subsubsection{Shifting the passband}

\begin{figure}[t]
\centering
\hspace*{-3mm}
\includegraphics[scale=.42]{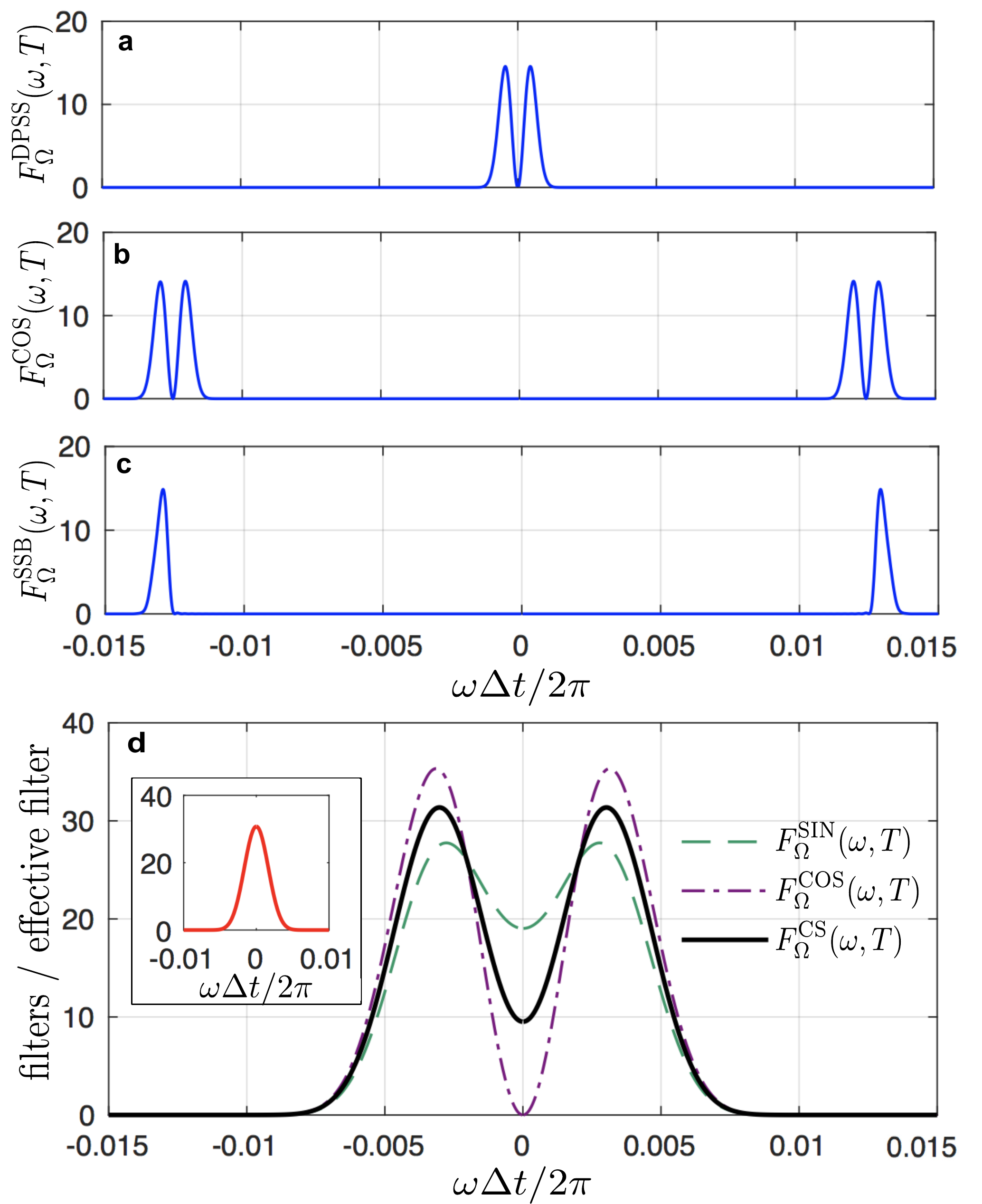}
\vspace*{-4mm}
\caption{(Color online) Shifted DPSS amplitude filters. The unshifted DPSS FF of order $k=1$ with $N=500$ and $W=10/N$ (a) has a passband centered at $\omega\Delta t/(2\pi)=0$. In (b),  the $k=1$ DPSS filter is shifted by $\omega_s\Delta t/(2\pi)=0.013$ with COS modulation, producing a filter with two duplicate passbands centered at 
$\omega\Delta t/(2\pi)=\pm 0.013$. In (c), SSB modulation with $\omega_s\Delta t/(2\pi)=0.013$ halves the passband of the $k=1$ DPSS filter, shifting the lower half by $-0.013$ and the upper half by $0.013$.  (d) The $k=0$ DPSS filter with  $N=500$ and $W=4/N$ (inset) is shifted by $\omega_s\Delta t/(2\pi)=0.003$ with SIN modulation (green dashed line), COS modulation (purple dash-dotted line) and CS modulation (black solid line).}
\label{fig::shift}
\end{figure}

For $k<K$, the DPSS filter in \erf{eq::kFF} is spectrally concentrated in a passband $B_0$ centered at $\omega=0$ 
[Fig. \ref{fig::shift}(a)]. 
To characterize the PSD at multiple frequencies, we need the ability to shift this passband along the frequency axis. 
The required shifted band-pass filters can be generated through simple modifications of the applied DPSS control in Eq. 
\eqref{eq::basic}, by using analogue modulation techniques similar to those commonly employed in radio and communication engineering.

Two such modulation techniques, co-sinusoidal (COS) and single-sideband (SSB) modulation, were introduced and 
demonstrated in \cite{Frey2017}. In a COS modulation protocol, the piecewise-constant control amplitude takes the form   
\begin{align}
&\Omega_n^{\text{COS}} \equiv \Omega\,v_n^{(k)}(N,W) \,\text{cos}(n\omega_s\Delta t), \quad \omega_s >0.
\label{eq::COS}
\end{align}
From \erf{eq::amplitudeFF}, the filter produced by this waveform is
\begin{align}
\label{eq::COSFF}
F_{\Omega}^{\text{COS}}&(\omega,T)=\frac{\Omega^2\,s(\omega,\Delta t)}{4}\\\notag
&\times\big[U^{(k)}(N,\!W;\omega\!-\!\omega_s)^2+
\;U^{(k)}(N,\!W;\omega\!+\!\omega_s)^2\\
&\;\;\;\;\;\;- 2U^{(k)}(N,\!W;\omega\!+\!\omega_s)U^{(k)}(N,\!W;\omega\!-\!\omega_s)\big]. \notag
\end{align} 
The final ``cross-term" in this expression is negligible when $\omega_s>2\pi W/\Delta t$ and $k< K$. In this regime, the filter is an even function proportional to the sum of two DPSS filters displaced by $\pm\omega_s$ and spectrally concentrated in passbands $B_{\pm\omega_s}$ [Fig. \ref{fig::shift}(b)].  Note that, on account of the periodicity of 
$\text{cos}(n\omega_s\Delta t)$, COS modulation cannot displace the passband by $\omega_s\geq 2\pi/\Delta t$.
Since both the PSD and the amplitude FFs are symmetric about $\omega=0$, we can restrict the  estimation problem 
to the positive frequency axis and treat $F_{\Omega}^{\text{COS}}(\omega,T)$ as a DPSS filter centered at $\omega_s$. 
This enables us to estimate the PSD at $\omega_s$ using the passband estimator of Sec. \ref{sec::QNS1}.

In a SSB band-shifting protocol, the amplitude control waveform is instead modified as  
\begin{align*}
\Omega_n^{\text{SSB}}\! \equiv \Omega[ v_n^{(k)}\!(N,\!W)\text{cos}(n\omega_s\Delta t)\!-\hat{v}_n^{(k)}\!(N,\!W)
\text{sin}(n\omega_s\Delta t)],
\end{align*}  
where $\{\hat{v}_n^{(k)}(N,W)\}$ denotes the discrete Hilbert transform of $\{v_n^{(k)}(N,W)\}$. SSB halves the original passband $B_0$, 
shifting the lower half by $-\omega_s$ and the upper half by $\omega_s$, thereby creating a filter spectrally concentrated in 
the ``half-bands" $(-\omega_s-2\pi W/\Delta t, -\omega_s)$ and $(\omega_s, \omega_s+2\pi W/\Delta t)$ [Fig. \ref{fig::shift}(c)].
While the reduced bandwidth provided by SSB can, in principle, allow for higher resolution estimates of the PSD at fixed $W$, this comes at the expense of the orthogonality and approximate completeness of the FFs in the passband. In contrast, in the regime where $W$ is small enough for $s(\omega,\Delta t)$ to be locally flat in $B_{\omega_s}$ and $\omega_s>2\pi W/\Delta t$, modulation by COS preserves the shape of the DPSWF in the translated passband. In particular, the Fourier transforms of the COS-modulated waveforms in \erf{eq::COS} for $k<K$ form an approximately complete 
functional basis in $B_{\omega_s}$. As a result, the sum of the filters in \erf{eq::COSFF} still approximates an ideal band-pass filter when $k<K$, analogous to \erf{eq::complete}, which makes COS modulation the appropriate choice for multitaper estimation.

Despite the above advantages, a shortcoming of COS modulation is filter distortion due to the cross-terms in 
Eq. \eqref{eq::COSFF} when $\omega_s \lesssim 2\pi W/\Delta t$. This effect can be removed, however, incorporating sinusoidal (SIN) modulation in a procedure we term {\em CS modulation}. In analogy to Eq. \eqref{eq::COS}, let SIN modulation be characterized 
by a control amplitude of the form
\begin{align*}
\Omega_n^{\text{SIN}} \equiv \Omega\,v_n^{(k)}(N,W) \,\text{sin}(n\omega_s\Delta t), \quad \omega_s >0, 
\end{align*}
which produces the filter
\begin{align*}
&\,F_{\Omega}^{\text{SIN}}(\omega,T) =\frac{\Omega^2\,s(\omega,\Delta t)}{4}\\\notag
&\times\big[U^{(k)}(N,\!W;\omega\!-\!\omega_s)^2+
\;U^{(k)}(N,\!W;\omega\!+\!\omega_s)^2\\
&\;\;\;\;\;\;+2U^{(k)}(N,\!W;\omega\!+\!\omega_s)U^{(k)}(N,\!W;\omega\!-\!\omega_s)\big]. \notag
\end{align*} 
Suppose the qubit is measured after COS (SIN) modulation is applied, producing the estimates $\hat{\mathcal{S}}^{\text{COS}}(T)$ $\big(\hat{\mathcal{S}}^{\text{SIN}}(T)\big)$. The expected value of the sum of these estimates is then 
\begin{align}
&\expect{\hat{\mathcal{S}}^{\text{COS}}(T)}_\mathcal{M}+\expect{\hat{\mathcal{S}}^{\text{SIN}}(T)}_\mathcal{M}\notag\\
&\;\;\;\;\;\;=\frac{1}{\pi}\! \int_{0}^{\infty}\!\!\!\! \mathrm{d}\omega\, S_{\Omega}(\omega)
\big[F_{\Omega}^{\text{COS}}(\omega,T) + F_{\Omega}^{\text{SIN}}(\omega,T) \big].\label{eq::cossin}
\end{align} 
That is, we obtain an overlap integral between the target PSD and an effective filter,
\begin{align}
&F_\Omega^{\text{CS}}(\omega,T)=\,F_{\Omega}^{\text{COS}}(\omega,T) + F_{\Omega}^{\text{SIN}}(\omega,T)\label{eq::Fcossin}
\\&\;\;\;\propto\,s(\omega,\Delta t)\big[U^{(k)}(N,\!W;\omega\!-\!\omega_s)^2
+U^{(k)}(N,\!W;\omega\!+\!\omega_s)^2\big].\notag
\end{align} 
This effective filter is free of the cross-term, while maintaining the desired 
orthogonality and (approximate) completeness properties of the DPSWF. 
The analogy between the effective filter $F_\Omega^{\text{CS}}(\omega,T)$ 
and the actual filter $F_\Omega(\omega,T)$ defines a similar passband estimator for CS modulation,
\begin{align*}
\hat{S}^{\text{CS}}_\Omega(t) \equiv \frac{\hat{\mathcal{S}}^{\text{COS}}(T)+\hat{\mathcal{S}}^{\text{SIN}}(T)}{A^{\text{CS}} },
\end{align*} 
where
\begin{align}
A^{\text{CS}} &\equiv \frac{1}{\pi}\int_{a}^{b}d\omega\big[F_{\Omega}^{\text{COS}}(\omega,T) + 
F_{\Omega}^{\text{SIN}}(\omega,T) \big], 
\nonumber \\ 
a&=
\begin{cases} 
\;0,&\omega_s\leq2\pi W/\Delta t\\
\;\omega_s-2\pi W/\Delta t,&\omega_s>2\pi W/\Delta t, \\
\end{cases} 
\label{eq::avalue}\\
b&=\omega_s+2\pi W/\Delta t. 
\label{eq::bvalue}
\end{align}
This estimate is unaffected by bias due to cross-terms when $\omega_s \lesssim 2\pi W/\Delta t$.

\subsubsection{Filter aliasing}

Regardless of the modulation strategy we employ, the amplitude filter takes the form given in \erf{eq::amplitudeFF}, namely, 
it is a product of $s(\omega,\Delta t)$ and the magnitude of a DTFT, $|\tilde{\Omega}(\omega)\big|^2$. As noted, 
$|\tilde{\Omega}(\omega)\big|^2$ is both symmetric about $\omega=0$ and periodic over $2\omega_\text{N}$, which implies 
that $|\tilde{\Omega}(\omega)\big|^2$ has an additional reflection symmetry about the Nyquist frequency, $\omega_\text{N}$. 
As the sum of two amplitude filters, the effective filter $F_\Omega^{\text{CS}}(\omega,T)$ also inherits this property.
In the classical setting, reflection symmetry about $\omega_\text{N}$ is responsible for {\em aliasing}.  For example, 
the expected value of the tapered estimate in \erf{eq::expecttaper} depends on $|\tilde{t}(\omega-\omega_s)|^2$.
If the latter is peaked at $\omega_s>0$, the reflection symmetry about $\omega_\text{N}$ implies that another peak, the ``alias'', 
lies at $\omega_s'=2\omega_\text{N}-\omega_s$. The presence of these dual peaks means that the estimator cannot distinguish between the values of the PSD at $\omega_s$ and $\omega_s'$. For spectral estimation strategies based on discrete samples of a signal that are \emph{evenly spaced in time}, aliasing prevents the PSD from being estimated at frequencies greater than 
$\omega_\text{N}$. 

\begin{figure}[t]
\centering
\hspace*{-1.5mm}
\includegraphics[scale=.75]{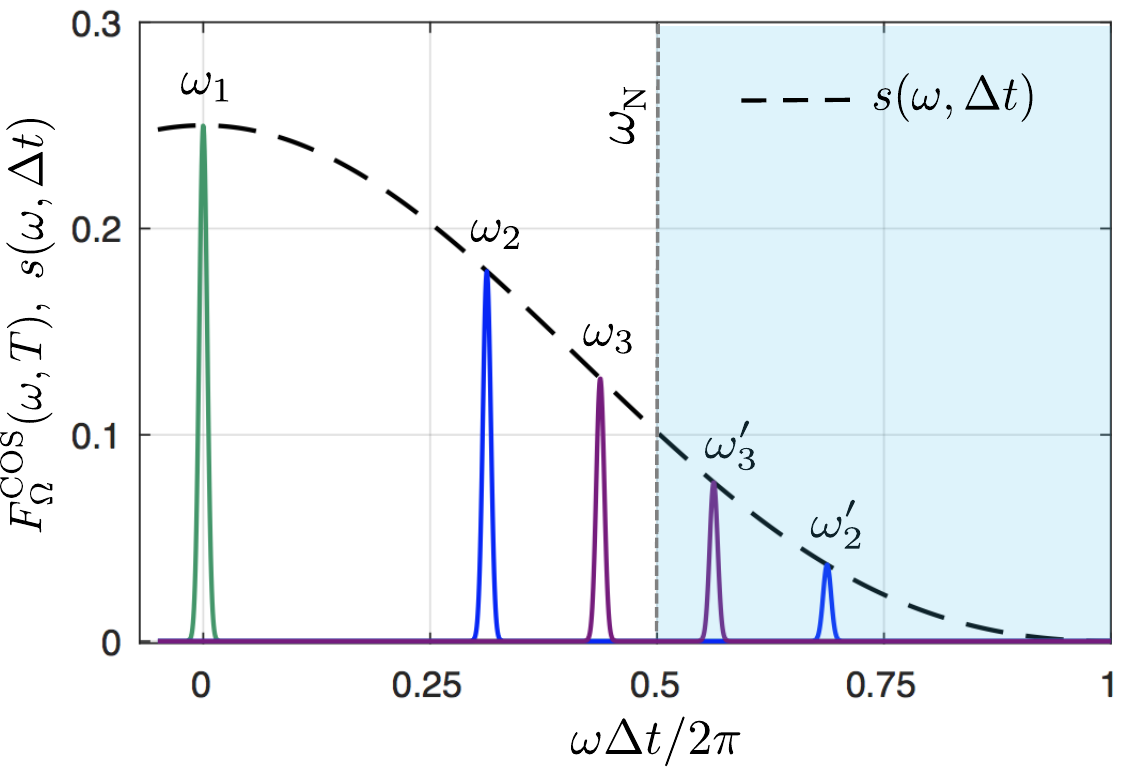}
\vspace*{-3mm}
\caption{(Color online) Aliasing of filters with spectral concentration near the Nyquist frequency $\omega_\text{N}$. The unshifted $k=0$ DPSS filter, centered at $\omega_1=0$ (green), is plotted along with $k=0$ DPSS filters COS-shifted by $\omega_2=5\pi/(8\Delta t)$ (blue) and $\omega_3=7\pi/(8\Delta t)$ (purple). The $k=0$ DPSS filters shifted by $\omega_2$ and $\omega_3$ have alias peaks beyond $\omega_\text{N}$ (shaded area), centered at $\omega_2'=11\pi/(8\Delta t)$ and $\omega_3'=9\pi/(8\Delta t)$. The magnitude of the peaks decreases with $\omega$ according to the envelope function $s(\omega,\Delta t)$ (black dashed line). }
\label{fig::alias}
\end{figure}

For the amplitude filter arising in the quantum setting, the symmetry of $|\tilde{\Omega}(\omega)\big|^2$ about $\omega_\text{N}$ causes 
an effect similar to aliasing. If, through appropriate band-shifting, we produce a DPSS filter with a passband containing $\omega_s$, 
an alias passband containing $\omega_s'$ will exist beyond 
$\omega_\text{N}$, as shown in Fig. \ref{fig::alias}. If the PSD is nonzero for $\omega>\omega_\text{N}$, \erf{eq::BandpassExpect} implies that the PSD in \emph{both} passbands will contribute to 
$\mathcal{S}(T)$, biasing any estimate of the PSD in the lower passband. Unlike in the classical case, however, the alias will not be the exact mirror image of the passband below $\omega_\text{N}$. Due to the presence of the envelope function $s(\omega,\Delta t)$, the magnitude of the amplitude FF is suppressed at frequencies beyond $\omega_\text{N}$ by a factor 
$$ \!F_{\Omega}(\omega_s',T)/ F_{\Omega}(\omega_s,T) =
\omega_s^2/(2\omega_\text{N}-\omega_s)^2, \quad 0<\omega_s\leq\omega_\text{N}.$$
A passband contained in $[0,\pi/(2\Delta t)]$, for example, will have an alias smaller in magnitude by at least a factor of about 1/9.  Despite the reduction of aliasing effects granted by the decay of the amplitude FF in the quantum setting,
shifting the passband of the FF beyond $\omega_\text{N}$ with the goal of estimating the PSD at some $\omega_s>\omega_\text{N}$ inevitably produces an additional passband below $\omega_\text{N}$.  Since the latter will always make the dominant contribution to 
$\mathcal{S}(T)$, the Nyquist frequency sets the upper bound for reconstructing the PSD, as in the classical case. 

In fact, this bound holds for {\em any} DD-QNS protocols composed of uniform, piecewise-constant intervals $\Delta t$ in the time domain.  This includes QNS based on pulsed DD when the control ``switching function" in the toggling-frame alternates between $\pm 1$ at multiples of some uniform $\Delta t$. As shown in Ref. \cite{Norris_Spectroscopy}, surpassing the Nyquist bound in spectral estimation is possible by employing control sequences composed of {\em non-uniform} intervals, effectively generating {\em incommensurate} periodicities.  In this case, the range of reconstruction is ultimately limited 
by the {\em timing resolution} of the control sequence, 
say, $\delta$ and, remarkably, the upper bound of spectral reconstruction, $\omega_{\text{max}}=\pi/\delta\geq\pi/\Delta t$, is 
achievable even when all piecewise-constant intervals of the control 	are larger in duration than $\Delta t$.
The subject of aliasing and its effect on spectral estimation is revisited in Sec. \ref{sec::comb}.

\subsection{Quantum eigenestimates}
\label{sec::eigen}

The passband estimator associated with a DPSS filter serves as the quantum analogue to Thomson's eigenestimate. In order to distinguish between different DPSS amplitude modulations, let $F^{(k,\omega_s)}(\omega,T)$ denote the amplitude filter produced by the control waveform in \erf{eq::COS}, which combines the DPSS of order $k$ with COS modulation to shift the passband by $\omega_s>0$. For a qubit that has evolved under the corresponding control, we shall define $\hat{\mathcal{S}}^{(k,\omega_s)}(T)$ as the estimate of $\mathcal{S}(T)$ given in \erf{eq::AxEst}. The $k$th-order quantum eigenestimate of the PSD is then given by
\begin{align}
\label{eq::eigenest}
\hat{S}_\Omega^{(k)}(\omega_s)&=\frac{\hat{\mathcal{S}}^{(k,\omega_s)}(T)}{A^{(k,\omega_s)}},
\;\; A^{(k,\omega_s)}=\frac{1}{\pi}\int_{a}^{b}\!\!d\omega \, F^{(k,\omega_s)}(\omega,T) , 
\end{align}
where $a$ and $b$ are same as in Eqs. \eqref{eq::avalue}-\eqref{eq::bvalue}.
Note that the eigenestimate differs slightly from \erf{eq::BandpassEst}, in that it allows the possibility of a passband containing $\omega=0$. As
$\hat{S}_\Omega^{(k)}(\omega_s)$ is a specialized form of the passband estimator, it is also normally distributed. From Eqs. 
\eqref{eq::BandpassExpect}-\eqref{eq::VarEst}, it follows that its expected value and variance are, respectively, given by 
\begin{eqnarray}
\hspace*{-5mm}\expect{\hat{S}_\Omega^{(k)}\!(\omega_s)}_{\mathcal{M}}&\!=\!& \frac{1}{\pi A^{(k,\omega_s)}} \!\int_{0}^{\infty}\!\!\!\!\! \mathrm{d}\omega\, 
F^{(k,\omega_s)}(\omega,T) S_{\Omega}(\omega), 
\label{eq::ExpectEigenQ} \\
\hspace*{-5mm}\text{var}[\hat{S}_\Omega^{(k)}(\omega_s)] & =& \frac{\sigma^2 }{ M A^{(k,\omega_s)\,2}}. 
\label{eq::ExpectEigenQ2}
\end{eqnarray}
Although we have specialized to COS modulation for concreteness, we can obtain analogous eigenestimates for CS modulation by equating $F^{(k,\omega_s)}(\omega,T)$ with the effective filter in \erf{eq::Fcossin} and $\hat{\mathcal{S}}^{(k,\omega_s)}(T)$ with $\hat{\mathcal{S}}^{\text{COS}}(T)+\hat{\mathcal{S}}^{\text{SIN}}(T)$. A similar eigenestimate for SSB is given in Ref. \cite{Frey2017}.

Like the classical case, each quantum eigenestimate is itself an estimate of the PSD prior to being combined into a multitaper estimate. The eigenestimates are biased estimators, however, stemming from both leakage of the filters ({\em broadband bias}, BB) and curvature of the PSD within the passband ({\em local bias}, LB). Following the methodology of Thomson, we separate the integral in \erf{eq::ExpectEigenQ} into three regions, namely:
\begin{align}
\expect{\hat{S}_\Omega^{(k)}\!(\omega_s)}_\mathcal{M}& \equiv \!\frac{1}{\pi A^{(k,\omega_s)}}\Big[\int_{a}^{b}\! \mathrm{d}\omega\,
F^{(k,\omega_s)}(\omega,T) S_{\Omega}(\omega)\notag\\
&\;\;\;\;+\int_{0}^{a}\! \mathrm{d}\omega\, F^{(k,\omega_s)}(\omega,T) S_{\Omega}(\omega) \notag \\
&\;\;\;\;+\int_{b}^{\infty}\! \mathrm{d}\omega\,F^{(k,\omega_s)}(\omega,T) S_{\Omega}(\omega)\,\Big]. \label{eq::eigenexpect2}
\end{align}
The last two terms in this equation, which depend on integrals below and above the passband, 
constitute the BB,
\begin{align}
\label{eq::BB}
{\cal B}_{\text{BB}}^{(k,\omega_s)}\!\equiv\!\frac{1}{\pi A^{(k,\omega_s)}}\,\ddashint_{0}^{\infty}\!\!\!\text{d}\omega\, 
F^{(k,\omega_s)}\!(\omega,T) S_\Omega(\omega), 
\end{align}
where $\ddashint_{0}^{\infty}\!\text{d}\omega\,\,\cdot \equiv \int_0^\infty \text{d}\omega\cdot-\int_{a}^{b}\text{d}\omega\,\,\cdot\,$ is a shorthand for integration over all frequencies outside $[a,b]$. ${\cal B}_{\text{BB}}^{(k,\omega_s)}$ increases with the leakage present in the FF and vanishes in the idealized limit where $F^{(k,\omega_s)}\!(\omega,T)$ is exactly band-limited. 

Even when no BB is present, $\hat{S}_\Omega^{(k)}\!(\omega_s)$ will still possess LB,  unless the PSD is flat within the passband, as seen from \erf{eq::ExpectEigenQ}. The LB is contained in the first term of \erf{eq::eigenexpect2} which, upon 
Taylor-expanding $S_{\Omega}(\omega)$  about $\omega_s$, becomes
\begin{align*}
&\!\int_{a}^{b}\!\!\! \mathrm{d}\omega\, F^{(k,\omega_s)}(\omega,T) \sum_{\ell=0}^\infty 
\frac{S_\Omega^{(\ell)}\!(\omega_s)(\omega-\omega_s)^\ell}{\pi {A}^{(k,\omega_s)}\ell!}=\\
&S_\Omega(\omega_s)+\!\!\int_{a}^{b}\!\!\! \mathrm{d}\omega\, F^{(k,\omega_s)}(\omega,T) 
\sum_{\ell=1}^\infty\!\frac{S_\Omega^{(\ell)}\!(\omega_s)(\omega-\omega_s)^\ell}{\pi {A}^{(k,\omega_s)}\ell!}.
\end{align*}
The final term in this expression, which depends on higher-order derivatives of $S_{\Omega}(\omega)$ at $\omega_s$, is the LB:
\begin{align}
\label{eq::LO}
{\cal B}_{\text{LB}}^{(k,\omega_s)}\!\equiv\!\!\int_{a}^{b}\!\!\! \mathrm{d}\omega\,F^{(k,\omega_s)}(\omega,T) \sum_{\ell=1}^\infty 
\frac{S_\Omega^{(\ell)}\!(\omega_s)(\omega-\omega_s)^\ell}{\pi {A}^{(k,\omega_s)}\,\ell!}.
\end{align}
This expression vanishes when the PSD is flat in $[a,b]$, in which case an estimate of the PSD at a single frequency $\omega_s$ accurately characterizes the PSD in the passband. The LB is particularly pronounced when the PSD exhibits peaks or other sharp spectral features. In terms of its bias contributions, the expected value of the eigenestimate may then be written as 
\begin{align*}
\expect{\hat{S}_\Omega^{(k)}\!(\omega_s)}_\mathcal{M}=S_\Omega(\omega_s)+{\cal B}_{\text{BB}}^{(k,\omega_s)}+{\cal B}_{\text{LB}}^{(k,\omega_s)}.
\end{align*}

\subsection{Adaptive quantum multitaper approach}
\label{sec::adaptive}

As in the classical case, the quantum multitaper estimate is formed by a weighted average of the quantum eigenestimates,
\begin{align}
\label{eq::QMultiEst}
\hat{S}_\Omega^\text{m}(\omega_s)=\sum_kd_k(\omega_s)\hat{S}_\Omega^{(k)}(\omega_s), \quad 
 \sum_kd_k(\omega_s)=1.
\end{align}
Recall that Thomson selected the weights in the classical multitaper estimate using an adaptive method designed to ensure that the eigenestimates with the least estimated bias had the highest weight in the final multitaper estimate. Here, we adapt this procedure to the quantum setting.

The adaptive method begins with a set of unnormalized weights, $\{\tilde{d}_k(\omega_s)\}$. Each $\tilde{d}_k(\omega_s)$ is defined as the weight minimizing the square of the bias, namely, 
\begin{align*}
E_k(\omega_s) \equiv \big[S_\Omega(\omega_s)-\tilde{d}_k(\omega_s)\expect{\hat{S}_\Omega^\text{(k)}(\omega_s)}_\mathcal{M}\big]^2.
\end{align*}
We can solve for the unnormalized weights through differentiation with respect to $\tilde{d}_k(\omega_s)$,  yielding 
$$\tilde{d}_k(\omega_s)={S_\Omega(\omega_s)} / {[ S_\Omega(\omega_s)+{\cal B}_{\text{BB}}^{(k,\omega_s)}+
{\cal B}_{\text{LB}}^{(k,\omega_s)} ]}.$$ 
As this expression depends on the unknown PSD explicitly and implicitly through the bias terms, we replace $S_\Omega(\omega_s)$ with a prior estimate of the PSD, $\hat{S}_{\Omega}^{[0]}(\omega_s)$. The estimate $\hat{S}_{\Omega}^{[0]}(\omega_s)$ is also used to estimate the broadband and local biases, which we detail below. The 0th-order estimates of the PSD and biases produce a 1st-order solution for the unnormalized weighting coefficients, say, $\tilde{d}_k^{[1]}(\omega_s)$. Substituting $d_k^{[1]}(\omega_s)\equiv \tilde{d}_k^{[1]}(\omega_s)/[\sum_k\tilde{d}_k^{[1]}(\omega_s)]$ as the weights in \erf{eq::QMultiEst} produces a new estimate of the spectrum, $\hat{S}_{\Omega}^{[1]}(\omega_s)$. This concludes the first cycle of a recursion. At iteration $n$, the unnormalized weighting coefficients and the spectral estimate are
\begin{align}
\tilde{d}_k^{[n]}(\omega_s) &=\frac{\hat{S}^{[n-1]}(\omega_s)}{\hat{S}^{[n-1]}(\omega_s)+\hat{\cal B}_\text{BB}^{k,\omega_s[n-1]}
+\hat{\cal B}_\text{LB}^{k,\omega_s[n-1]}}\;, \notag\\
\hat{S}^{[n]}(\omega_s) &=\frac{\sum_k\tilde{d}_k^{[n]}(\omega_s)\hat{S}_\Omega^{(k)}(\omega_s)}{\sum_k\tilde{d}_k^{[n]}(\omega_s)},\label{eq::Sn}
\end{align}
respectively. This process is repeated until $\hat{S}^{[n]}(\omega_s)$ converges at some $n \equiv n_c$. The final multitaper estimate is then 
$\hat{S}_\Omega^{\text{m}}(\omega_s) \equiv \hat{S}^{[n_c]}(\omega_s)$, with the weight on the $k$-th eigenestimate given by $d_k(\omega_s)=\tilde{d}_k^{[n_c]}(\omega_s)/[\sum_k\tilde{d}_k^{[n_c]}(\omega_s)]$. 

Each iteration of the adaptive procedure relies on the spectral estimate from the previous iteration. Beginning this process at iteration 
$n=1$ requires a prior 0th-order estimate of the PSD. If the ultimate goal is estimating the PSD at a set of shift frequencies 
$\{\omega_{s,p}|p=1, \ldots , P\}$, the 0th-order estimate takes the form $\hat{S}^{[0]}\equiv\{\hat{S}^{[0]}(\omega_{s,p})|p=1,\ldots ,P\}$, where each $\hat{S}^{[0]}(\omega_{s,p})$ is an estimate of the PSD at $\omega_{s,p}$. Thomson originally used the 
$k=0$ eigenestimates as the 0th-order estimate. Other options include a multitaper estimate of the form in \erf{eq::QMultiEst} with the weighting coefficients assumed equal {\em a priori}. At iteration $n$, computing the spectral estimate at each shift frequency via \erf{eq::Sn} produces another set of estimates, $\hat{S}^{[n]}\equiv\{\hat{S}^{[n]}(\omega_{s,p})|p=1,\ldots, P\}$. Given $\hat{S}^{[n]}$, the BB can be approximated from \erf{eq::BB} via numerical integration.  Approximating the LB depends on higher-order derivatives of the PSD through the Taylor expansion in \erf{eq::LO}. If COS or CS modulation is used to shift the DPSS filter and $s(\omega,\Delta t)$ is locally flat in the passband, the $\ell=1$ term in the expansion vanishes due to symmetry of the filter about $\omega_{s,p}$.  For SSB, the $\ell=1$ term is nonzero since the DPSS filters are no longer symmetric. In this case, the first order derivative of the PSD at $\omega_{s,p}$ can be approximated as a finite difference of $\hat{S}^{[n]}$, 
\begin{align}
\hat{S}^{[n]'}(\omega_{s,p})=\frac{\hat{S}^{[n]}(\omega_{s,p+1})-\hat{S}^{[n]}(\omega_{s,p})}{\omega_{s,p+1}-\omega_{s,p}}.
\end{align}
The $\ell=1$ term in \erf{eq::LO} can then be computed via numerical integration. For the $\ell>1$ terms, approximating higher-order derivatives of the PSD requires successive finite differences of $\hat{S}^{[n]}$. This approximation becomes progressively more sensitive to noise in $\hat{S}^{[n]}$ as $\ell$ increases, however. Following Thomson and Ref. \cite{Frey2017}, we truncate the Taylor expansion after 
$\ell=1$ in our simulations in Sec. \ref{sec::results}.

The approximate completeness of the DPSWF has important implications for the quantum multitaper estimate. 
Being a weighted sum of the normally distributed eigenestimates, such an estimate is also normally distributed, 
with expected value and variance 
following from \erf{eq::QMultiEst} and Eqs. \eqref{eq::ExpectEigenQ}-\eqref{eq::ExpectEigenQ2}:
\begin{eqnarray}
\notag
\expect{\hat{S}_\Omega^{\text{m}}(\omega_s)}_\mathcal{M} & \!=\! & \frac{1}{\pi}\!\int_{0}^\infty\!\!\!\!\!\text{d}\omega\!\Big[\sum_kd_k(\omega_s)
\frac{F^{(k,\omega_s)}(\omega,T)}{A^{(k,\omega_s)}}\Big]\!S_\Omega(\omega), \\\label{eq::MultiVar}
\text{var}[\hat{S}_\Omega^{\text{m}}(\omega_s)] & = &\sum_k d_k(\omega_s)^2\,\text{var}[\hat{S}_\Omega^{(k)}(\omega_s)].
\end{eqnarray}
Consider the effective filter in $\expect{\hat{S}_\Omega^{\text{m}}(\omega_s)}_\mathcal{M}$, namely, 
\begin{align*}
\rho^{(\text{m},\omega_s)}(\omega,T) \equiv \sum_k d_k(\omega_s)\frac{ F^{(k,\omega_s)}(\omega,T)}{A^{(k,\omega_s)}}.
\end{align*}
If the above sum extends over $k\in\{0,\ldots,K-1\}$ and $d_k(\omega_s)\approx1/K$ for all $k$, the effective filter takes a form similar to the near ideal band-pass filter in \erf{eq::complete}, namely, 
\begin{align}
\label{eq::EffFF}
\rho^{(\text{m},\omega_s)}(\omega,T)\approx\frac{\Delta t}{4\pi W}\,\mathds{1}_{B_{\omega_s}}(\omega).
\end{align}
If LB is ignored in the adaptive procedure, this correspondence becomes exact as $N\rightarrow\infty$. As a consequence of the uniformity 
of $\rho^{(\text{m},\omega_s)}(\omega,T)$, all spectral components of the PSD within the passband then contribute approximately equally to the estimate, 
implying that 
\begin{align}
\expect{\hat{S}_\Omega^{\text{m}}(\omega_s)}_\mathcal{M}\approx\frac{\Delta t}{4\pi W}\int_{B_{\omega_s}}\text{d}\omega S_\Omega(\omega).
\end{align}
Quantum multitaper thus estimates the average value of the PSD in the passband, which is particularly useful for the detection of fine spectral features, as we will explicitly demonstrate in Sec. \ref{sec::detect}. For instance, since $F^{(0,\omega_s)}(\omega,T)$ is preferentially concentrated at $\omega_s$, the center of the passband, the $k=0$ eigenestimate will be relatively unaffected by fine spectral features near the boundary of the passband. On the other hand, owing to the approximately uniform concentration of $\rho^{(\text{m},\omega_s)}(\omega,T)$, the quantum multitaper estimate has the capability to detect features at \emph{any} location in the passband.

\subsection{Non-adaptive DPSS-based spectral estimation}
\label{sec::DPSSFF}

\subsubsection{Single-setting quantum multitaper approach}

In practice, a disadvantage of the adaptive quantum multitaper (AQM) approach we just described is that it requires a different measurement setting for each order DPSS. If DPSS of orders up to $K-1$ are used, an experimenter must acquire $K$ different estimates of the survival probabilities for each shift frequency $\omega_s$. A single-taper eigenestimate, on the other hand, requires just one estimate of each survival probability for a given $\omega_s$. To offer an easier experimental implementation while retaining advantages of the AQM, we can form a {\em single-setting quantum multitaper} (SSQM) estimate by using an amplitude control waveform which is still piecewise-constant in time but consists of a linear combination of DPSS, 
\begin{equation}
\Omega_n^{\text{SSQM}} \equiv\Omega \sum_{k=0}^{K-1} c_k v_n^{(k)} (N,W), \quad c_k\in\mathbb{R}, \sum_{k=0}^{K-1}c_k^2=1.
\label{eq::SSC}
\end{equation}
By choosing the coefficients $c_k$ so that the resulting filter approximates an ideal one, we can achieve a result similar to the AQM using a single measurement setting and the standard passband estimator. In fact, this approach closely parallels classical multitaper in the sense that a single noise realization is tapered with multiple DPSS. 

\begin{figure}[t]
\centering
\hspace*{-.1cm}\includegraphics[scale=.62]{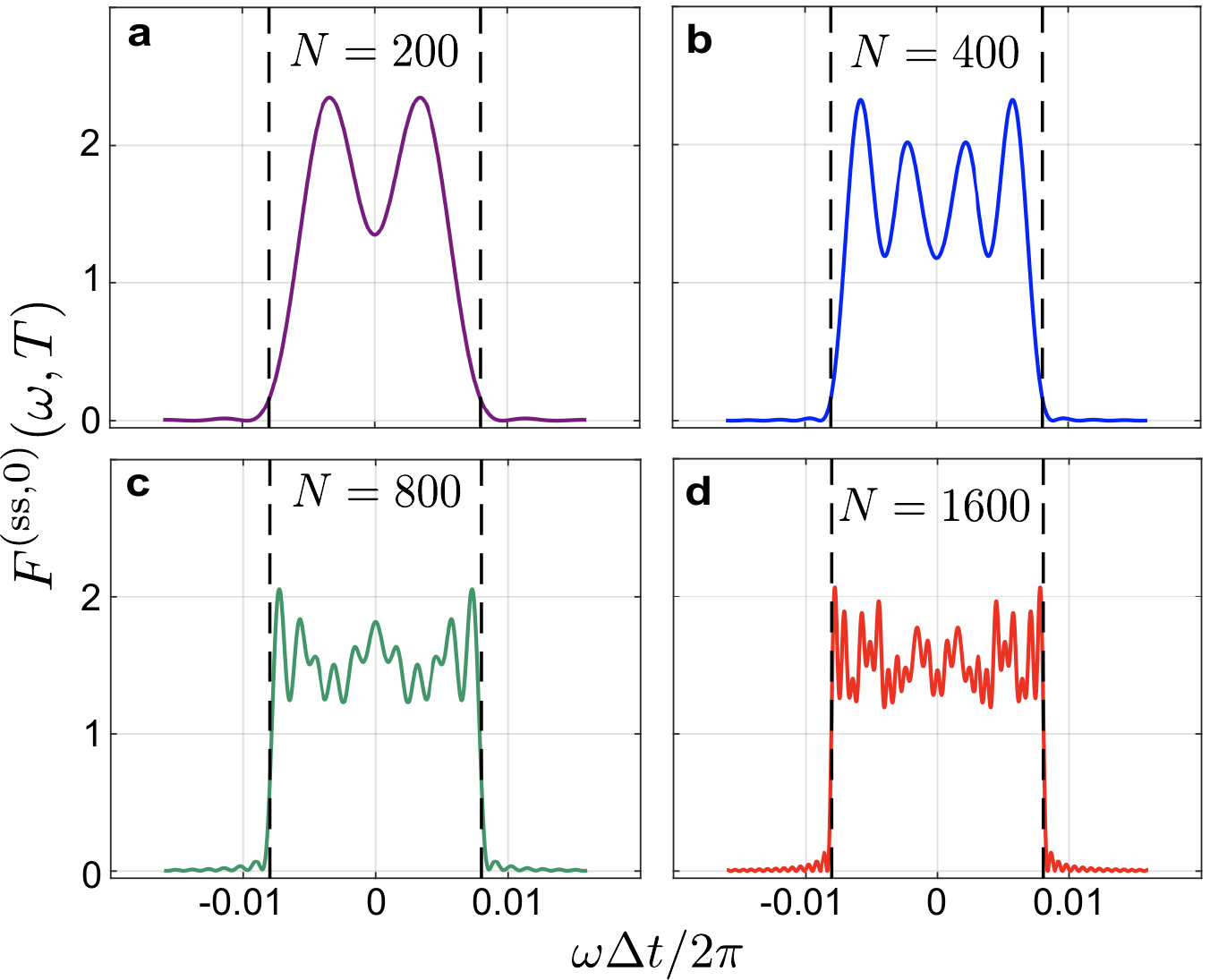}
\vspace*{-3mm}
\caption{(Color online) Filters produced by the SSQM optimization procedure for $N=200$ (a), $N=400$ (b), $N=800$ (c), and $N=1600$ (d). In the time domain, the filters are linear combinations of the first $K-1$ DPSS with $W=0.008$ and $\Delta t=1\,\mu$s. The boundaries of the filter passband at $\omega\Delta t/(2\pi)=\pm W$ are marked with vertical dashed lines.}\label{fig::ss}
\end{figure}

Direct calculation shows that the control waveform in Eq. \eqref{eq::SSC} yields an amplitude filter centered at $\omega_s=0$ given by
\begin{align*}
F^{(\text{ss},0)}(\omega,T)=&\, \Omega\, s(\omega,\Delta t)\Big|\sum_{k=0}^{K-1}\frac{c_k}{\epsilon_k}\,U^{(k)}(N,W;\omega)\Big|^2\\
=& \, \Omega\,s(\omega,\Delta t)\Big[\sum_{k=0}^{K-1}c_k^2\,U^{(k)}(N,W;\omega)^2\;+ \\
&\!\!\!\!\!\!\sum_{\substack{k\neq k'\\ \text{same parity}}}\!\!\!\!\!\!c_kc_{k'}\,U^{(k)}\!(N,W;\omega)U^{(k')}\!(N,W;\omega)\Big],
\end{align*}
where ``same parity" means that $k$ and $k'$ are either both even or both odd. From \erf{eq::complete} it follows that the sum 
in the second line converges to an ideal filter with increasing $N$ when $c_k^2=1/K$. The sum in the third line 
causes $F^{(\text{ss},0)}(\omega,T)$ to deviate from the ideal filter shape, however. At a given frequency, this deviation can be 
quantified by the error
\begin{align*}
\varepsilon[\omega,\{c_k\}] \equiv &\frac{1}{2W}\mathds{1}_{B_0}(\omega)-\!\Big[\sum_{k=0}^{K-1}c_k^2\,U^{(k)}\!(N,W;\omega)^2\\
&+\!\!\!\!\!\sum_{\substack{k\neq k'\\ \text{same parity}}}\!\!\!\!\!\!c_kc_{k'}\,U^{(k)}\!(N,W;\omega)U^{(k')}\!(N,W;\omega) \Big].
\end{align*}
To approximate an ideal band-pass filter, we numerically solve for $c_k$ that minimize a cost function consisting of the integrated squared error in the 
DPSS passband. Representative single-setting filters so constructed are 
are plotted in Fig. \ref{fig::ss}. At a fixed $N$, these filters do not approximate a square wave as closely as the $\rho_K(N,W;\omega)$ plotted in Fig. \ref{fig::IdealFilter}. Additionally, we cannot make any rigorous statements regarding their optimality or convergence properties as $N\rightarrow\infty$. 

In the same manner as ordinary DPSS filters, the single-setting filters can be shifted by COS or COS/SIN to produce a filter $F^{(\text{ss},\omega_s)}(\omega,T)$ with passband centered at $\omega_s$. The single-setting estimate of the PSD at $\omega_s$ then follows from the passband estimator of Sec. \ref{sec::QNS1},
\begin{eqnarray*}
\hat{S}_\Omega^{\text{ss}}(\omega_s)=\frac{\hat{\mathcal{S}}^{(\text{ss},\omega_s)}(T)}{A^{(\text{ss},\omega_s)}},
\; A^{(\text{ss},\omega_s)}=\frac{1}{\pi}\int_{a}^{b}d\omega F^{(\text{ss},\omega_s)}(\omega,T), 
\end{eqnarray*}
where $a$ and $b$ are given in Eqs. \eqref{eq::avalue}-\eqref{eq::bvalue}.
As in the AQM case, the relative uniformity of the filters makes the single setting estimate useful for the detection 
of fine spectral features.

\subsubsection{Interpolated estimate via Fisher information}
\label{sec::Fisher}

Regardless of the estimator we employ, characterizing the spectrum over a range of frequencies, say, $[0,\omega_{\text{max}})$, requires estimating the PSD at multiple shift frequencies. Consider a set of spectral estimates $\hat{S}(\omega_{s,1}),\ldots, \hat{S}(\omega_{s,P})$ at shift frequencies $\omega_{s,1},\ldots,\omega_{s,P}\in[0,\omega_{\text{max}})$. Here, $\hat{S}$ can refer to a passband estimate, as in the single-taper DPSS or SSQM approaches, or an estimate that depends on multiple measurement settings like the AQM. Each $\hat{S}(\omega_{s,p})$ is associated with a filter (or effective filter) with passband $B_{\omega_{s,p}}$ and, therefore, depends on the value of the PSD in $B_{\omega_{s,p}}$. To ensure that every $\omega\in[0,\omega_{\text{max}})$ is contained in at least one passband, the $B_{\omega_{s,p}}$ will generally overlap.  Consequently, some frequencies will be contained in multiple passbands. 
If $\omega$ belongs to the intersection of the overlapping passbands $B_{\omega_{s,p}}$ and $B_{\omega_{s,p'}}$, each of the associated estimates, $\hat{S}(\omega_{s,p})$ and $\hat{S}(\omega_{s,p'})$, carry information about $S_\Omega(\omega)$. How, 
then, should we combine $\hat{S}(\omega_{s,p})$ and $\hat{S}(\omega_{s,p'})$ in order to estimate $S_\Omega(\omega)$? 

Rather than considering an arbitrary $\omega\in[0,\omega_{\text{max}})$, we simplify the problem slightly by discretizing the relevant frequency range into small segments $\big[(q-1)\Delta\omega,q\Delta\omega\big)$ for $q=1,\ldots,Q$ and $Q\Delta\omega=\omega_{\text{max}}$. Let $\vec{S} \equiv (S_1,\ldots,S_Q)^T$ denote the ``discretized PSD",  where $S_q$ is the value of the PSD at the center of segment $q$. Given a set of estimates $\hat{S}(\omega_{s,1}),\ldots, \hat{S}(\omega_{s,P})$, we seek a way of using the information contained in each in order to estimate $S_q$. By quantifying the information that these estimates carry about $S_q$ using the classical Fisher information, we now show how to define an {\em interpolated estimate} for $S_q$ in which the $\hat{S}(\omega_{s,p})$ containing the most information are weighted preferentially. 

We first consider the Fisher information when $\hat{S}(\omega_{s,p})=\hat{S}_\Omega(\omega_{s,p})$ is an arbitrary passband estimator associated with an amplitude filter $F_\Omega^{(\omega_{s,p})}(\omega,T)$. This result can be specialized to a DPSS eigenestimate or a SSQM estimate by replacing $F_\Omega^{(\omega_s)}(\omega,T)$ with $F^{(k,\omega_s)}(\omega,T)$ or $F^{(\text{ss},\omega_s)}(\omega,T)$. 
Denote the integrated value of the amplitude filter in segment $q$ by
\begin{align}
\label{eq::Fq}
A_q^{(\omega_{s,p})}\equiv\frac{1}{\pi}\int_{(q-1)\Delta\omega}^{q\Delta\omega}\text{d}\omega F_\Omega^{(\omega_{s,p})}(\omega,T).
\end{align}
If $\mathbf{F}$ is the $P\times Q$ dimensional {\em filter matrix} with elements
\begin{align}\label{eq::Fpq}
\mathbf{F}_{pq} \equiv A_q^{(\omega_{s,p})}\big[A^{(\omega_{s,p})}\big]^{-1}, 
\end{align}
taking $\int_0^\infty\text{d}\omega S_\Omega(\omega) F_\Omega^{(\omega_{s,p})}(\omega,T)/[\pi A^{(\omega_{s,p})}]\approx 
\big(\mathbf{F}\vec{S}\big)_p$ in \erf{eq::condprob} determines the conditional probability that $\hat{S}_\Omega(\omega_{s,p})=s_p$ given $\vec{S}$ as follows:
\begin{align}
\label{eq::likelihood1}
{\mathbb P}\big[\,s_p\,\big|\,\vec{S}\;\big]\!\approx\!
\frac{1}{\sqrt{2\pi\Sigma_{pp}}}\;\text{exp} \bigg[\frac{\big(s_p-(\mathbf{F}\vec{S})_p \big)^2}{2\Sigma_{pp}}\bigg].
\end{align}
Here, $\Sigma_{pp'}=\delta_{pp'}\,\text{var}[\hat{S}_\Omega(\omega_{s,p})]$ is an element of the $P\times P$ covariance matrix $\Sigma$, which is diagonal thanks to the independence of the passband estimates. 
In the simplest case where measurement noise dominates over the statistical fluctuations of the amplitude noise, the covariance matrix is independent of the PSD. Owing to the Gaussianity of the conditional probability for $M\gg 1$, the Fisher information then reads 
\begin{align}
\mathcal{I}_{\hat{S}_\Omega(\omega_{s,p})}[S_q]
=\bigg[\frac{\partial (\mathbf{F}\vec{S})_p}{\partial S_q}\bigg]^2 \!\Sigma_{pp}^{-1}
=M\bigg[\frac{{A_q^{(\omega_{s,p})}}}{\sigma}\bigg]^2 \!.
\label{eq::FisherSk}
\end{align}
In Appendix \ref{app::fisher}, we examine the more general case in which the covariance matrix is permitted to depend on the PSD. This dependence results in an additive correction which is typically negligible for $M\gg1$. From Eq. \eqref{eq::FisherSk}, we see that for fixed $M$ and $\sigma$ the information the estimate carries about $S_q$ increases with the area underneath the filter in segment $q$.   

For the AQM approach, which relies on eigenestimates obtained over multiple measurement settings, a similar form of the Fisher information can be derived using the covariance matrix $\Sigma_{pp'}=\delta_{pp'}\text{var}[\hat{S}_\Omega^{\text{m}}(\omega_{s,p})]$ and the filter matrix
\begin{align}
\label{eq::Rq}
\mathbf{F}_{pq}=\mathcal{R}_q^{(\omega_{s,p})}, \quad 
\mathcal{R}_q^{(\omega_{s,p})}\equiv\frac{1}{\pi}\int_{(q-1)\Delta\omega}^{q\Delta\omega} \!\!\text{d}\omega\,\rho^{(\text{m},\omega_{s,p})}(\omega,T). 
\end{align}
In this case, the Fisher information becomes
\begin{align}
\label{eq::FisherSm}
\mathcal{I}_{\hat{S}_\Omega^{\text{m}}\!(\omega_{s,p})}[S_q]=&\frac{M}{\sum_{k}d_k(\omega_{s,p})^2/
A^{(k,\omega_{s,p})\,2}} \bigg[\frac{\mathcal{R}_q^{(\omega_{s,p})}}{\sigma}\bigg]^2.
\end{align}
Analogous to \erf{eq::FisherSk}, the information that the AQM estimate carries about $S_q$ grows with
the area underneath the effective filter in segment $q$.

Using Eq. (\ref{eq::FisherSk}) or Eq. (\ref{eq::FisherSm}),  we can estimate each $S_q$ as a linear combination of the $\hat{S}(\omega_{s,1}),\ldots , \hat{S}(\omega_{s,P})$ weighted by the Fisher information, 
\begin{align}
\label{eq::FisherEst}
\hat{S}_q^\mathcal{I}=\frac{\sum_{p=1}^P\mathcal{I}_{\hat{S}(\omega_{s,p})}[S_q]\,\hat{S}(\omega_{s,p})}{\sum_{p=1}^P\mathcal{I}_{\hat{S}(\omega_{s,p})}[S_q]}=\sum_{p=1}^P\mathcal{I}_{qp}\,\hat{S}(\omega_{s,p}).
\end{align}
Irrespective of the type of initial estimate, the more information that $\hat{S}(\omega_{s,p})$ contains about ${S}_q$, the more it contributes to $\hat{S}_q^\mathcal{I}$, as anticipated. In classical multitaper, Thomson proposed a similar 
``high resolution estimate" as a free parameter expansion in the 
$\omega_{s,p}$ \cite{thomson_multitaper}. Since the shift frequencies are chosen prior to measurement in the quantum case, 
however, they cannot be treated as free parameters in the analysis. 
Nonetheless, $\hat{S}_q^\mathcal{I}$ enables us to interpolate between existing estimates of the PSD, offering a higher 
apparent resolution in a manner similar to Thomson's free parameter expansion.

\subsubsection{Bayesian spectral estimation}
\label{sub::Bayes}

The Gaussian distribution of the conditional probability in \erf{eq::condprob} makes the spectral estimation problem particularly amenable to Bayesian techniques \cite{Frey2017,BayesianQNS}. If $\mathbf{d}$ is a vector of measurement data obtained from the qubit sensor and $\vec{S}=(S_1,\ldots,S_Q)^T$ is the discretized PSD of the previous section, the Bayesian spectral estimate is derived from the posterior distribution ${\mathbb P}\big[\vec{S}\big|\mathbf{d}\big]$, which is the probability that a particular PSD generated the observed data. In Ref. \cite{Frey2017}, the measurement data consisted of estimates of the qubit fidelity, $\hat{\mathcal{F}}_{\text{av}}\approx 1-\hat{\mathcal{S}}(T)$. Here, in order to make contact with the single and multitaper strategies of the previous sections, we employ DPSS eigenestimates as measurement data. Formally, these approaches are equivalent since both fidelity and the eigenestimates are statistics of $\hat{\mathcal{S}}(T)$.  

Up to normalization, the posterior distribution conditioned on the DPSS eigenestimates 
$\hat{\mathbf{S}}_\Omega^{(k)}=(s_1,\ldots,s_P)^T$ is given by
\begin{align*}
{\mathbb P}\big[\vec{S}\big|\hat{\mathbf{S}}_\Omega^{(k)}\big]\propto {\mathbb P}\big[\hat{\mathbf{S}}_\Omega^{(k)}\big|\,\vec{S}\big]\,
{\mathbb P}\big[\vec{S}\big],
\end{align*}
where ${\mathbb P}\big[\hat{\mathbf{S}}_\Omega^{(k)}\big|\,\vec{S}\big]$ is the likelihood function and ${\mathbb P}\big[\vec{S}\big]$ is the prior probability distribution of the PSD. Since the eigenestimates are uncorrelated, they are described by a $P\times P$ diagonal covariance matrix, with elements 
$\Sigma_{pp'}=\delta_{pp'}\,\text{var}[\hat{S}_\Omega^{(k)}(\omega_{s,p})]$. The likelihood, which is multivariate Gaussian, then follows from \erf{eq::likelihood1}, namely, 
\begin{align}\label{eq::LiklihoodBayes}
{\mathbb P}\Big[\hat{\mathbf{S}}_\Omega^{(k)}\,\big|\,\vec{S}\Big]=&\prod_{p=1}^P {\mathbb P}\Big[s_p\big|\vec{S}\,\Big]\\\notag
=&\frac{1}{\sqrt{(2\pi)^p\,\text{det}\,\Sigma}}\;e^{-\frac{1}{2}(\hat{\mathbf{S}}_\Omega^{(k)}-\mathbf{F}\vec{S})^T\Sigma^{-1}(\hat{\mathbf{S}}_\Omega^{(k)}-\mathbf{F}\vec{S})}.
\end{align}
We consider a prior distribution of the PSD described by a multivariate Gaussian with a $Q\times Q$ covariance matrix $\Sigma_0$,
\begin{align}
{\mathbb P}\big[\vec{S}\big]=
\frac{1}{\sqrt{(2\pi)^p\,\text{det}\,\Sigma_0}}\;e^{-\frac{1}{2}(\vec{S}-\vec{\mu}_0)^T\Sigma_0^{-1}(\vec{S}-\vec{\mu}_0)}.
\label{eq::priorG}
\end{align}
If the experimenter lacks any prior knowledge of the PSD, ${\mathbb P}\big[\vec{S}\big]$ can be made ``diffuse" or effectively flat by taking $\Sigma_0=\delta_{q,q'}\sigma_0^2$, where $\sigma_0\gg\text{max}[\hat{\mathbf{S}}_\Omega^{(k)}]$. In the Bayesian analysis of Sec. \ref{sec::results}, we employ the interpolated estimate in \erf{eq::FisherEst} based on previous qubit measurements as a prior, taking $\vec{\mu}_0=(\hat{S}_1^\mathcal{I},...,\hat{S}_Q^\mathcal{I})^T$ and $(\Sigma_0)_{qq'}=\sum_{p=1}^P\mathcal{I}_{qp}\mathcal{I}_{q'p}\text{var}[\hat{S}(\omega_{s,p})]$. 

The Gaussianity of the prior and the likelihood function ensure that the posterior is also multivariate Gaussian. From the Gaussian posterior, there are multiple approaches to obtain the final estimate of the PSD. In Ref. \cite{Frey2017}, we
employed DPSS of different orders that shared the same set of passbands. The PSD was estimated by the {\em maximum of the posterior} distribution of the discretized PSD within each passband, and the estimates in each overlapping passband were then combined via \erf{eq::FisherEst}. In the analysis of Sec. \ref{sec::results}, we use only $k=0$ DPSS with no two having the same passband, eliminating the complication of passbands shared by multiple FFs. The PSD is then estimated by the {\em mean of the posterior}, which is the approach taken in Ref. \cite{BayesianQNS} in the context of dephasing noise.

\section{Illustrative applications}
\label{sec::results}

In this section, we illustrate the main advantages that Slepian QNS protocols bestow in concrete spectral estimation 
scenarios involving both single- and multi- taper protocols, as compared to conventional protocols involving DD or RSE.  
For computational simplicity and in order to focus attention on the essential features, we assume in our simulations that dephasing noise is negligible. Taking $\beta_z(t)\equiv 0$ in Eq. \eqref{eq::Ham} implies that ${\mathbb P}(\uparrow_z,T)={\mathbb P}(\uparrow_y,T)$ and ${\mathbb P}(\uparrow_x,T)=1$, reducing the estimated signal in \erf{eq::AxEst} to $\hat{\cal S}(T) = 1-{ \hat{\mathbb P}}(\uparrow_z,T)$. Likewise, our simulations do not account for noise introduced by a measurement apparatus or other sources that could be present in an experimental implementation. Consequently, the variance of the spectral estimates results solely from statistical fluctuations of the measurement outcomes $\{i_m\}$ and its expected value can be determined via Eq. (\ref{eq::VarEst}), (\ref{eq::ExpectEigenQ2}) or (\ref{eq::MultiVar}) with $\sigma^2= {\mathbb P}(\uparrow_z,T)[1-{\mathbb P}(\uparrow_z,T)]$.

\subsection{Comparison to square-wave modulation techniques and leakage bias}

\begin{figure}
\centering
\includegraphics[scale=.84]{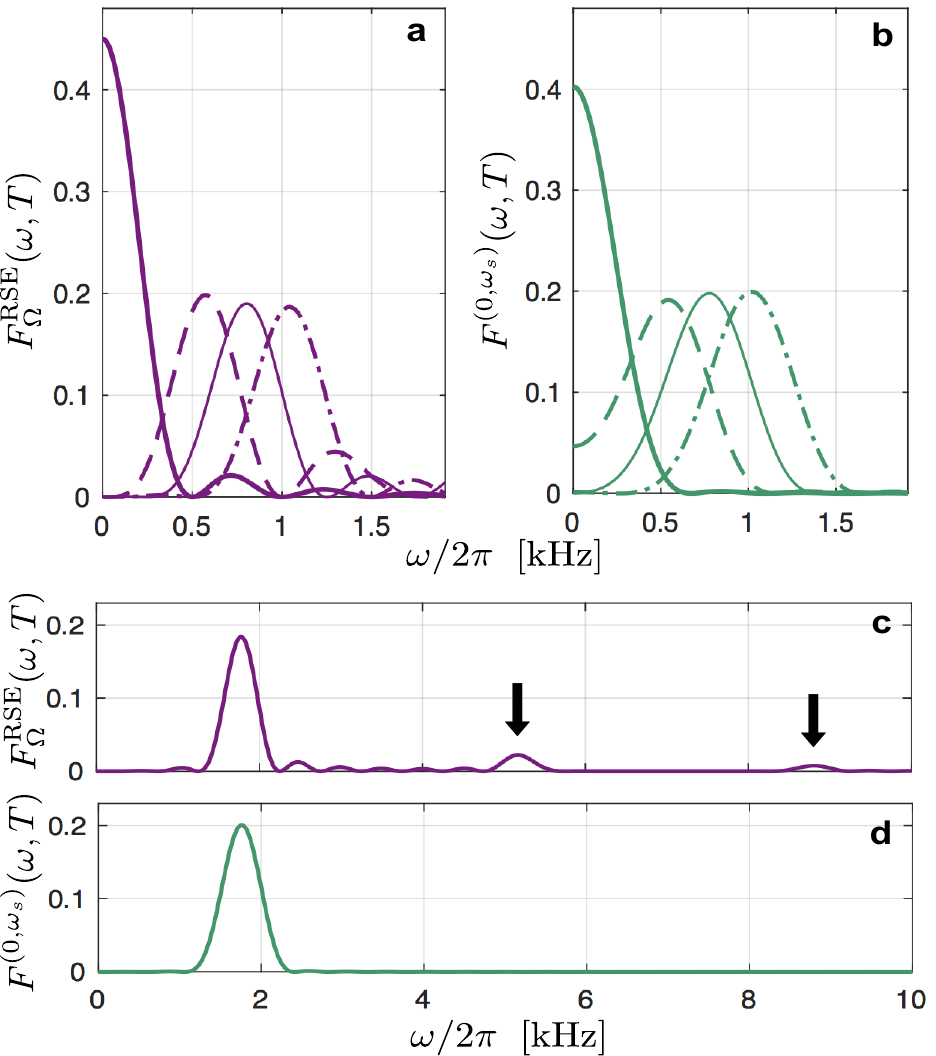}
\vspace*{-2mm}
\caption{Sample FFs used to reconstruct the Lorentzian PSDs. Filters corresponding to the first four RSE sequences 
are plotted in (a), which are (from left to right) a constant amplitude control waveform and CPMG with $n=2, 3, 4$ sign changes. Plotted in (b) are the first four DPSS filters used in the reconstruction, which are generated by the $k=0$ DPSS shifted by $\omega_s=0, 2\pi/T, 3\pi/T, 4\pi/T$ (left to right) with $N=500$, 
$\Delta t=4\,\mu$s and $W=1/N$. For both the RSE and DPSS, the control waveforms have total duration $T=2$ ms and are normalized so that $\int_0^T\text{d}t\,\Omega(t)^2=900$ rad$^2$/s, ensuring that the areas enclosed by the filters is equal by Eq. \eqref{eq::parseval}.
The leakage of the RSE filter for CPMG with $n=7$ is shown in (c) with arrows highlighting the largest lobes outside the main passband.  For comparison, the $k=0$ DPSS shifted by $\omega_s=7\pi/T$ is plotted in (d).  }
\label{fig:LorentzianFFs}
\end{figure}

Leakage bias can have a severe effect on QNS, though the extent of the error depends sensitively on the particular PSD and filters used in the reconstruction. To quantitatively illustrate this, we compare the performance of RSE and $k=0$ DPSS sequences in two sample spectral reconstructions. Sample filters produced by the RSE sequences are shown in Fig. \ref{fig:LorentzianFFs}(a). Constant amplitude modulation produces a filter peaked at $\omega=0$, while CPMG RSE with sign switches (or $\pi$-phase advances) numbering  $n=2, 3,\ldots,40$ generates the remaining filters with passbands centered at $\omega_n\approx n\pi/T$. For all of the RSE sequences, the approximate width of the passbands is $4\pi/T$. The RSE filters have leakage in the form of peaks or lobes occurring at frequencies above the passband: the $n=7$ CPMG RSE filter, plotted in Fig. \ref{fig:LorentzianFFs}(c), typifies this pattern. 

In contrast, the filters produced by the $k=0$ DPSS sequences in Fig. \ref{fig:LorentzianFFs}(b) and (d) exhibit no obvious lobes above the passband. To ensure a fair comparison with the RSE sequences, 
we chose the bandwidth parameter of the DPSS to be $W=1/N$, making the width of all passbands $4\pi W/\Delta t=4\pi/T$ for both the DPSS and RSE filters. COS modulation is used to shift the DPSS filters,
in such a way that the locations of the passbands correspond with those of the RSE filters. 
All of the sequences are normalized in the time domain using Eq. \eqref{eq::parseval}, which ensures that the area enclosed by all filters is uniform. Thus, the filters centered at $\omega=0$ in Figs. \ref{fig:LorentzianFFs}(a) and (b) have twice the magnitude of the filters centered at $n\pi/T>0$, since the latter also have passbands centered at $-n\pi/T$ on account of symmetry about $\omega=0$.

\begin{figure*}[t]
\centering
\hspace*{-1mm}
\includegraphics[scale=.87]{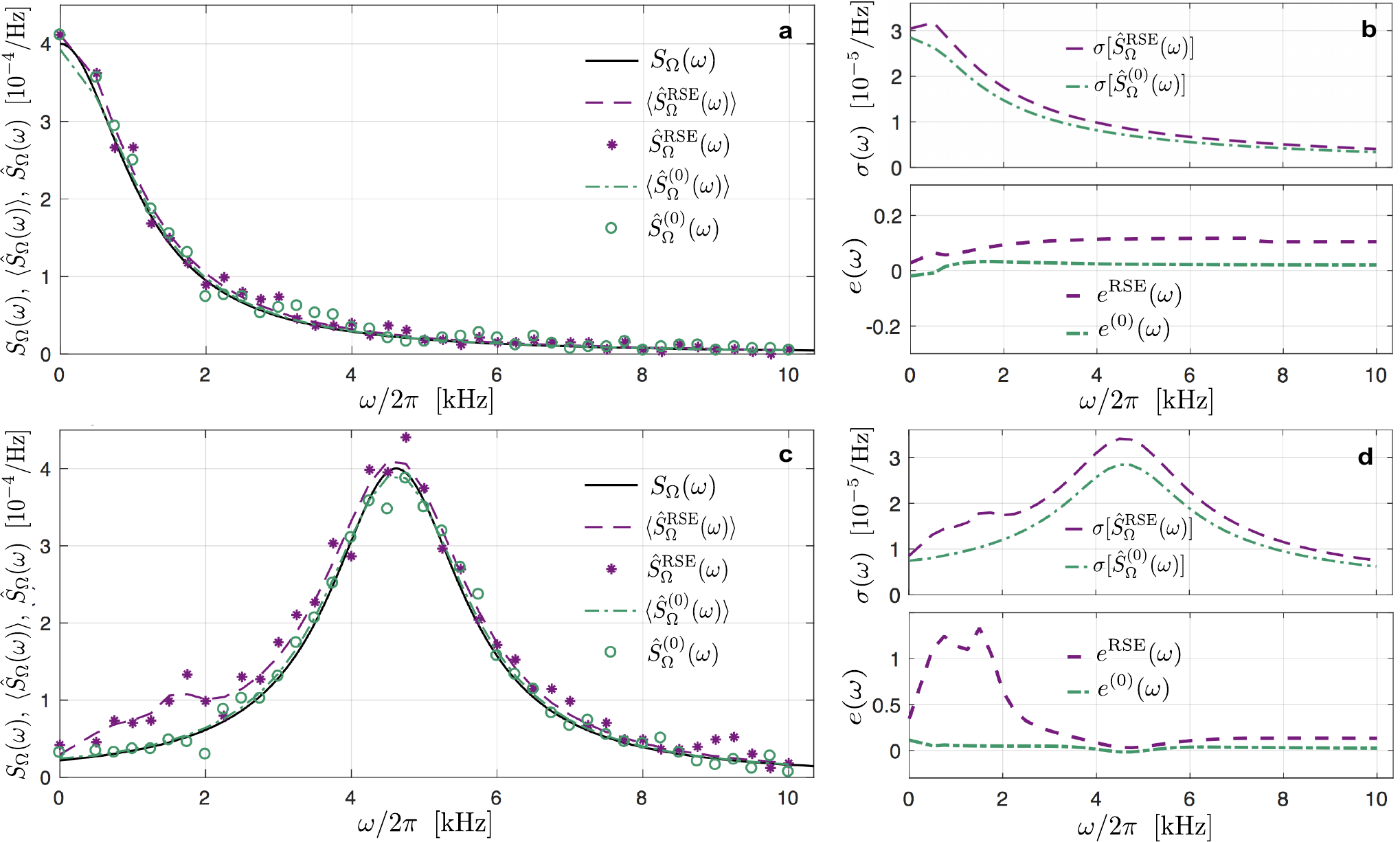}
\vspace*{-4mm}
\caption{(Color online) Effects of leakage bias on spectral reconstructions. Two Lorentzian PSDs as in Eq. \eqref{eq::lor}, 
with $A=4\times10^{-4}$ Hz$^{-1}$, $w_p/2\pi=1.11$ kHz and peaks centered at $p/2\pi=0$ kHz (a) and $p/2\pi=4.62$ kHz (c), are estimated using the RSE and $k=0$ DPSS FFs depicted in Fig. \ref{fig:LorentzianFFs}.  Plots (a) and (c) compare numerically simulated RSE (purple asterisks) and $k=0$ DPSS eigenestimates (green circles) of the target PSD (black solid line). The numerical estimates are produced by $M=2000$ simulated qubit measurements in the $z$ basis per control sequence. Also shown are the expected values of RSE (purple dashed line) vs. DPSS eigenestimates (green dash dotted line). For $p/2\pi=0$ kHz, the uncertainty and relative error $e(\omega)\equiv [\expect{\hat{S}_\Omega(\omega)}-S_\Omega(\omega)]/S_\Omega(\omega)$ of the RSE estimates (purple dashed line) and the DPSS eigenestimates (green dash dotted line) are plotted in (b). The same quantities for $p/2\pi=4.62$ kHz are shown in (d).}
\label{fig:Lorentzian}
\end{figure*}

We employ the above RSE and DPSS sequences to reconstruct two representative Lorentzian PSDs of the form
\begin{align}
S_\Omega(\omega)=\frac{C}{[(|\omega|-p)/w_p]^2+1},\quad C>0.
\label{eq::lor}
\end{align}
Figure \ref{fig:Lorentzian}(a) shows the reconstruction of the above spectrum when centered at $p=0$ kHz. Plotted along with the RSE and the DPSS eigenestimates are their expected values, which deviate little from the actual PSD.  Though the performance of the RSE and DPSS sequences are comparable, Fig. \ref{fig:Lorentzian}(b) shows that the estimates produced by the RSE sequences have a higher expected relative error.  The expected uncertainties of the RSE and DPSS eigenestimates
are also plotted in Fig. \ref{fig:Lorentzian}(b). The RSE filters have less concentration in the passband due to leakage, 
which results in a smaller integrated area in the passband and an uncertainty that is larger than the DPSS.

While the reconstruction of the Lorentzian PSD peaked at zero kHz is relatively unaffected by the leakage of the RSE filters, this is not the case for the reconstruction depicted in Fig. \ref{fig:Lorentzian}(c). In this case, the Lorentzian is peaked at $4.62$ kHz, implying that the PSD has a large magnitude in the region where the first nine RSE filters suffer considerable leakage (note the locations of the high frequency lobes belonging to the $n=7$ CPMG RSE filter in Fig. \ref{fig:LorentzianFFs}(c)). As a result, the RSE estimates in the $0$ - $2$ kHz region have significant error, which is further illustrated by the expected relative error in Fig. \ref{fig:Lorentzian}(d). In contrast, the DPSS eigenestimates agree closely with the actual PSD, thanks to the spectral concentration of the DPSS filters. As in the case of the first Lorentzian, the RSE estimates also exhibit a larger expected standard deviation. 

\subsection{Comparison to comb-based approaches and aliasing effects}
\label{sec::comb}

\begin{figure}[t]
\centering
\hspace*{-2mm}
\includegraphics[scale=.73]{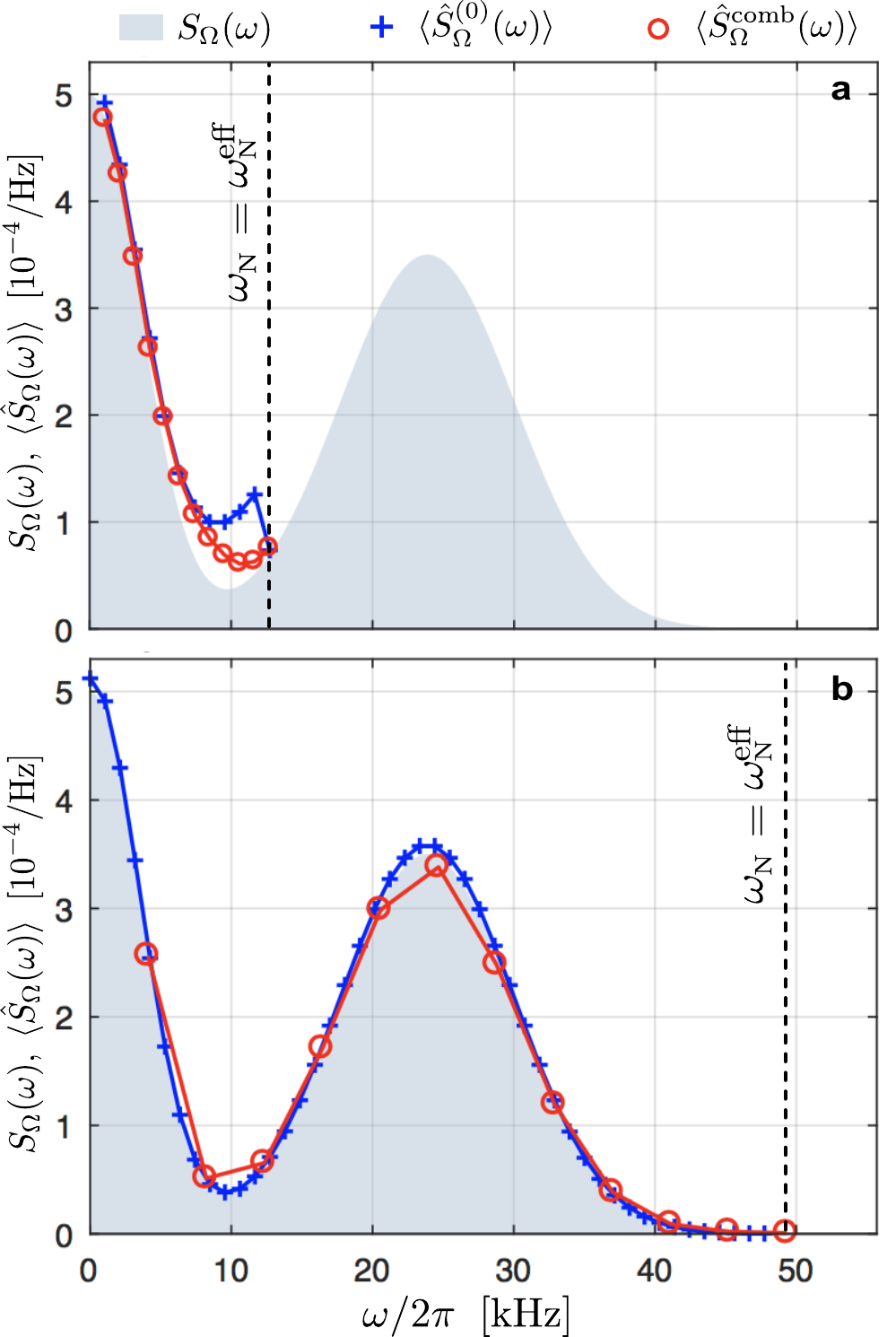}
\vspace*{-2mm}
\caption{(Color online) 
Aliasing and sampling resolution for DPSS vs. comb-based approaches. The shaded region is the area underneath the target PSD,
a multi-peaked Gaussian spectrum given by \erf{eq::multiGauss}, with $C_1=0.5$ ms, $\omega_1/2\pi=0$ kHz, $\sigma_1/2\pi=3.50$ 
kHz, $C_2=0.35$ ms, $\omega_2/2\pi=23.9$ kHz, $\sigma_1/2\pi=6.21$ kHz. Both (a) and (b) show the expected $k=0$ eigenestimates 
(blue crosses) and the expected frequency comb estimates (red hollow circles). In (a), the parameters of the sequences used are chosen so that $\omega_\text{N}/2\pi=\omega_\text{N}^{\text{eff}}/2\pi=12.7$ kHz (dashed vertical line), whereas $\omega_\text{N}/2\pi=\omega_\text{N}^{\text{eff}}/2\pi=49.0$ kHz (dashed vertical line) in (b). In the comb approach, extending the Nyquist frequency comes at the cost of degrading the sampling resolution, impacting the visibility of the first peak. }
\label{fig::comb}
\end{figure}

As mentioned in the Introduction, frequency-domain comb-based approaches to QNS
have been widely used across QIP. Although in their original formulation \cite{Alvarez_Spectroscopy}
and applications to date such protocols consider additive dephasing noise, the basic idea of frequency sampling via 
sequence repetition may also be adapted to the characterization of multiplicative amplitude noise, our present focus. 
We now use this setting to illustrate some limitations that are intrinsic to how comb-based QNS operates, 
when implemented in their simplest form involving repetition of a {\em fixed} control sequence.

Let $\Omega_B(t)$ denote such a ``base'' amplitude control waveform,
of duration $T_B$. Then applying $R$ repetitions of $\Omega_B(t)$ over total time $T=RT_B$ creates the filter
\begin{align*}
F_\Omega(\omega,RT_B)=\frac{\text{sin}^2(\omega R T_B/2)}{\text{sin}^2(\omega T_B/2)} F_{\Omega_B}(\omega,T_B),
\end{align*}
where $F_{\Omega_B}(\omega,T_B) = | \frac{1}{2}  \int_0^{T_B}\text{d}t\,e^{i\omega t}\Omega_B(t)|^2$ is the filter 
produced by a single repetition.  For $R\gg1$, a frequency comb emerges  in $ F_\Omega(\omega,RT_B)$, since
\begin{align*}
\frac{\text{sin}^2(\omega R T_B/2)}{\text{sin}^2(\omega T_B/2)}\approx\frac{2\pi R}{T_B}\sum_{h=-\infty}^\infty \!
\delta\!\left(\omega-\frac{2\pi h}{T_B}\right), 
\end{align*}
consisting of a sum of delta functions centered at $\omega_h \equiv h (2\pi/T_B)$, 
$h \in {\mathbb Z}$. If the comb is truncated at some maximum harmonic $h_{\text{max}}$, the expected squared 
magnitude of $a^{(1)}_x(RT_B)$ is determined by a linear equation, 
\begin{align*}
\mathcal{S}(RT_B)\approx \frac{2R}{T_B}\, \sum_{h=1}^{ h_{\text{max}} }  
F_{\Omega_B}\! \left( \omega_h,T_B \right)S_{\Omega}\! \left(\omega_h \right).  
\end{align*}
Estimating $\mathcal{S}$ for at least $h_{\text{max}}$ different base sequences, $\Omega_B(t)$, yields a system of linear equations, solving which produces an estimate of the target PSD at frequencies $\{ \omega_h| \, h=1,\ldots, h_{\text{max}} \}$.  In the original protocol for dephasing noise \cite{Alvarez_Spectroscopy}, CPMG sequences were used, with two instantaneous $\pi$-pulses and different durations, $T_B$, $T_B/2$, \ldots, $T_B/h_{\text{max}}$. In the amplitude noise setting, a similar protocol can be implemented by RSE CPMG with $n=2$ sign changes.

An intrinsic feature of  such a frequency comb approach is that the {\em sampling resolution}, 
namely, the minimum achievable distance between reconstructed points of the PSD, is determined by the duration of the base control waveform. Since the above method estimates the PSD at the harmonic frequencies, the resolution is necessarily $2\pi/T_B$. Indeed, the flat-top RSE estimates shown in Fig. \ref{fig:Lorentzian} suffer from a similar problem, as the sampling resolution is $\pi/T$.  In contrast, the sampling resolution of DPSS eigenestimates or multitaper estimates is determined by the spacing of the shift frequencies $\omega_s$, which are independently tunable parameters of the amplitude control waveforms. Although the shift frequencies have an upper bound set by the Nyquist frequency, their spacing has no dependence on the duration of the DPSS waveforms. This distinction is crucial, as the only way to improve the sampling resolution in comb-based approaches is by increasing $T_B$. Practically, this is not always possible as the qubit will approach a fully mixed state for sufficiently long evolution times, at which point no useful information about the noise may be extracted. Even when the evolution time is short enough to ensure the qubit does not completely decohere, aliasing can pose a barrier when $T_B$ is increased. 

As shown in Ref. \cite{Norris_Spectroscopy}, the largest harmonic that frequency comb protocols can estimate is upper-bounded by an \emph{effective Nyquist frequency} $\omega_\text{N}^{\text{eff}}$, which is determined by the durations of the 
piecewise-constant segments of the control waveforms. If $n_{\text{seq}}$ control sequences are used in the reconstruction, 
with piecewise-constant time-domain control waveforms 
having interior segments of duration $\Delta t_1,\ldots,\Delta t_L$, we have
\begin{align*}
&\omega_\text{N}^{\text{eff}}=\text{min}\Big\{ \frac{2 n_\text{seq}\pi}{T_B},\,\frac{\pi}{\Delta t_{\text{gcf}}}\Big\},\\
&\Delta t_{\text{gcf}} \equiv \text{max}\Big\{\Delta t_j\in\mathbb{R}\Big|\Delta t_\ell/\Delta t_j\in\mathbb{Z}^+, \ell=1,\ldots,L\Big\}.
\end{align*}
For a base sequence that consists of a RSE CPMG with $n=2$ sign changes, and variable duration ranging from $T_B$ to $T_B/h_{\text{max}}$, the interior pulse separations are $T_B/2,\ldots,T_B/2h_{\text{max}}$, implying that $\Delta t_{\text{gcf}}=(T_B/2)[\text{lcm}(2,...,h_{\text{max}})]^{-1}$, $n_\text{seq}=h_\text{max}$ and $\omega_\text{N}^{\text{eff}} = 2 h_{\text{max}}\pi/T_B$. This scaling of 
$\omega_\text{N}^{\text{eff}}$ with $1/T_B$ is not unique to CPMG, however. In fact, the same scaling holds for 
all implementations of the frequency comb approach that involve repetitions of a fixed base sequence.  In such cases, 
increasing $T_B$ to improve the resolution necessarily limits the range of the spectral reconstruction.

We quantitatively demonstrate the pitfalls of having both the sampling resolution and Nyquist frequency dependent on $T_B$ by estimating a multi-peaked Gaussian PSD. Specifically, we consider a noise PSD consisting of two Gaussian peaks, 
\begin{align}
\label{eq::multiGauss}
S_\Omega(\omega)=\sum_{i=1,2} C_i\,\text{exp}\bigg[\!-\frac{(\omega-\omega_i)^2}{2\sigma_i^2}\bigg], \quad C_i >0.
\end{align}
In Fig. \ref{fig::comb}(a), we plot the expected estimate of the PSD using the frequency comb method with $h_{\text{max}}=12$ CPMG sequences of durations $T_B$, $T_B/2$,\ldots, $T_B/12$ for $T_B=942\,\mu$s and $R=20$. In this way, the reconstructed PSD spans $12$ harmonics separated by a sampling resolution of $\omega_\text{res}/2\pi=1.06$ kHz. For the chosen $T_B$, we have 
$\omega_\text{N}^{\text{eff}}/2\pi=12.7$ kHz. Since the PSD has significant magnitude beyond $\omega_\text{N}^{\text{eff}}$, the expected comb estimates show error due to aliasing in the 5-12.7 kHz range. For comparison, we also plot the expected eigenestimates of the PSD at the exact same harmonic frequencies using $k=0$ DPSS with $\Delta t=39.3\,\mu$s, $N=260$, $W=1/N$ and shift frequencies 
$\omega_s=2\pi/T_B,\ldots,12\cdot 2\pi/T_B$. The Nyquist frequency of the DPSS control waveforms, $\omega_\text{N}=\pi/\Delta t=12.7$ kHz,  coincides by construction with $\omega_\text{N}^{\text{eff}}$. 
Figure \ref{fig::comb}(a) shows that the eigenestimates are more sensitive to error from aliasing than the comb estimates. In principle, the error in both of these 
estimates could be mitigated through the use of anti-aliasing filters, which  are designed to suppress components of the FF beyond the Nyquist frequency and are built into many commercial waveform generators. Nonetheless, the range of spectral reconstruction is limited by $\omega_\text{N}$ and $\omega_\text{N}^{\text{eff}}$ even when anti-aliasing filters are employed.  In Fig. \ref{fig::comb}(a), the locations of $\omega_\text{N}$ and $\omega_\text{N}^{\text{eff}}$ effectively prevent us from reconstructing the second peak of the PSD.

Resolving the second peak requires us to extend the range of spectral reconstruction by tuning 
$\omega_\text{N}$ and $\omega_\text{N}^{\text{eff}}$.
Unlike the standard comb method, DPSS allow us to accomplish this without sacrificing sampling resolution.  In Fig. \ref{fig::comb}(b), we plot the expected eigenestimates of the PSD using the $k=0$ DPSS with $\Delta t=10.2\,\mu$s, $N=1000$ and $W=1/N$. Though the total duration of the DPSS amplitude control waveforms, $T=N\Delta t$, is the {\em same} in both Fig. \ref{fig::comb}(a) and \ref{fig::comb}(b), the reduced $\Delta t$ results in $\omega_\text{N}/2\pi=49$ kHz in the second reconstruction. The eigenestimates resolve the second peak and show no signs of aliasing error since the PSD does not have appreciable magnitude beyond $\omega_\text{N}$.  In Fig. \ref{fig::comb}(b), the effective Nyquist frequency of the comb is moved to $\omega_\text{N}^{\text{eff}}/2\pi=49$ kHz by taking $T_B=245\,\mu$s with the same choice of CPMG sequences. While this choice of $T_B$ enables us to resolve the second peak of the PSD, the sampling resolution swells to  $\omega_\text{res}/2\pi=4.14$ kHz, making the first peak essentially unresolvable.

\begin{figure*}[t]
\centering
\hspace*{-2mm}
\includegraphics[scale=.88]{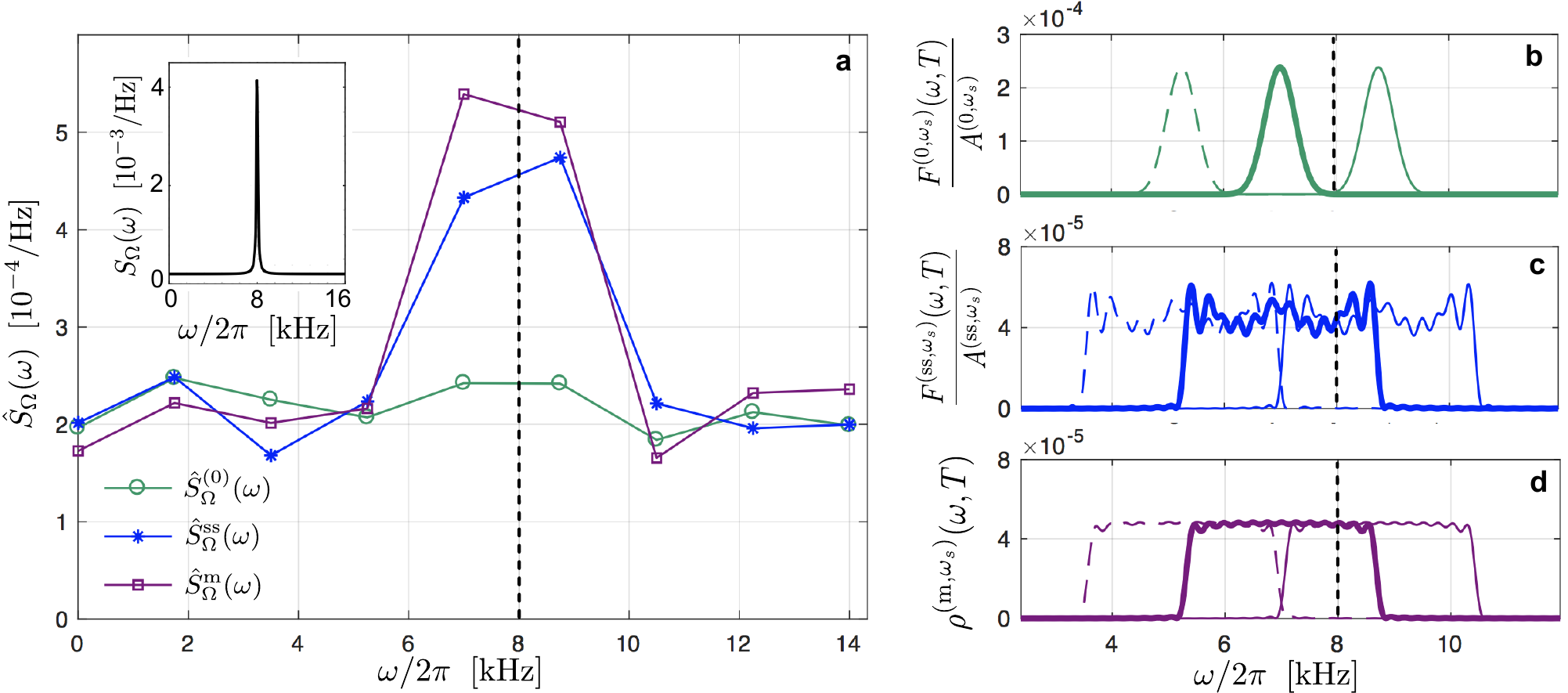}
\vspace*{-3mm}
\caption{Detection of fine spectral features by different Slepian-based estimation protocols.  The inset of (a) shows the target PSD, 
$S_\Omega(\omega)=C/\{[(|\omega|-p)/w_p]^2+1\}+s_0$, which consists of a narrow Lorentzian centered at $p/2\pi=7.96$ kHz, with amplitude $C=4$ ms and width $w_p/2\pi=0.08$ kHz, along with an additive white noise component of magnitude $s_0=2\times 10^{-4}\;\text{Hz}^{-1}$ and a cutoff frequency of $\omega_c/2\pi=17.5$ kHz.  Three different estimates used to detect the line component are plotted in (a): the  $k=0$ eigenestimate $\hat{S}_\Omega^{(0)}(\omega)$ (green hollow circles), the SSQM estimate $\hat{S}_\Omega^{\text{ss}}(\omega)$ (blue asterisks), and the AQM estimate $\hat{S}_\Omega^{\text{m}}(\omega)$ (purple hollow squares). The filters corresponding to the $k=0$ eigenestimates and SSQM estimates at $\omega_{s,3}/2\pi=5.25$ kHz, $\omega_{s,4}/2\pi=7.00$ kHz and $\omega_{s,5}/2\pi=8.75$ kHz are plotted in (b) and (c), respectively, whereas the effective filters of the AQM at the same frequencies are shown in (d). The location of the peak is represented by the black dashed line. As before, all DPSS amplitude control waveforms used are normalized so that $\int_0^T\text{d}t\,\Omega(t)^2=900$ rad$^2$/s. For each shift frequency, $M=2600$ simulated qubit measurements along $\sigma_z$ are used for estimation. In the AQM protocol, these measurements are divided between 13 measurement settings, corresponding to each DPSS.}
\label{fig::Line1}
\end{figure*}

\subsection{Detection and reconstruction of fine spectral features}
\label{sec::detect}

\subsubsection{Peak identification}

As discussed in Sec. \ref{sec::MultiQNS}, 
while the primary purpose of classical multitaper is to produce a statistically consistent estimator, in the quantum setting 
multitaper spectral estimation is most valuable for detecting spectral features and providing reliable prior estimates. In the detection scenario, the non-parametric quantum multitaper estimate is used to gain information about the general shape and structure of a completely unknown PSD. In particular, the estimate can successfully identify regions where fine spectral features, such as narrow peaks or abrupt spectral cutoffs, are likely present. With this knowledge, an experimentalist can then perform additional measurements targeted at these features, in order to refine the initial estimate of the underlying PSD. 

We demonstrate this method on the PSD shown in the inset of Fig. \ref{fig::Line1}(a), consisting of a white noise floor 
along with a sharp Lorentzian peak. 
The objective is to detect the presence of the peak and its rough location, so that this information may be used to select control sequences for subsequent measurements, aimed at achieving an accurate reconstruction. For this task, we compare the performance of a single-taper protocol using the $k=0$ DPSS to both the SSQM and the 
AQM protocols we described in Sec. \ref{sec::MultiQNS}. The DPSS used in all three have $W=7/N$ and $N=500$, which for $\Delta t=8\,\mu$s translates into a passband of width $ W/\Delta t\approx 3.5$ kHz. Note that this width is given in units of frequency, rather than angular frequency. Using COS modulation, we shift the passbands of the DPSS by $\omega_{s,0}/2\pi=0$ kHz,  $\omega_{s,1}/2\pi=1.75$ kHz, up to $\omega_{s,8}/2\pi=8\cdot1.75$ kHz, so that each passband overlaps its neighbor (or neighbors) by half of its width. 
Both types of quantum multitaper use the first $K-1=13$ DPSS. The weights on the different eigenestimates in the AQM were determined 
by the adaptive procedure taking into account BB, as LO vanishes to leading order. Convergence of the adaptive procedure occurred after 3 iterations. Figure \ref{fig::Line1}(b) shows the $k=0$ DPSS filters 
at shift frequencies $\omega_{s,3}/2\pi$, $\omega_{s,4}/2\pi$ and $\omega_{s,5}/2\pi$, which are highly concentrated at the centers of the passbands. In contrast, the corresponding filters produced by the SSQM optimization and effective filters generated by the AQM are distributed more uniformly in the passbands, as shown in Fig. \ref{fig::Line1}(c) and (d).

\begin{figure*}[t]
\centering
\includegraphics[scale=.84]{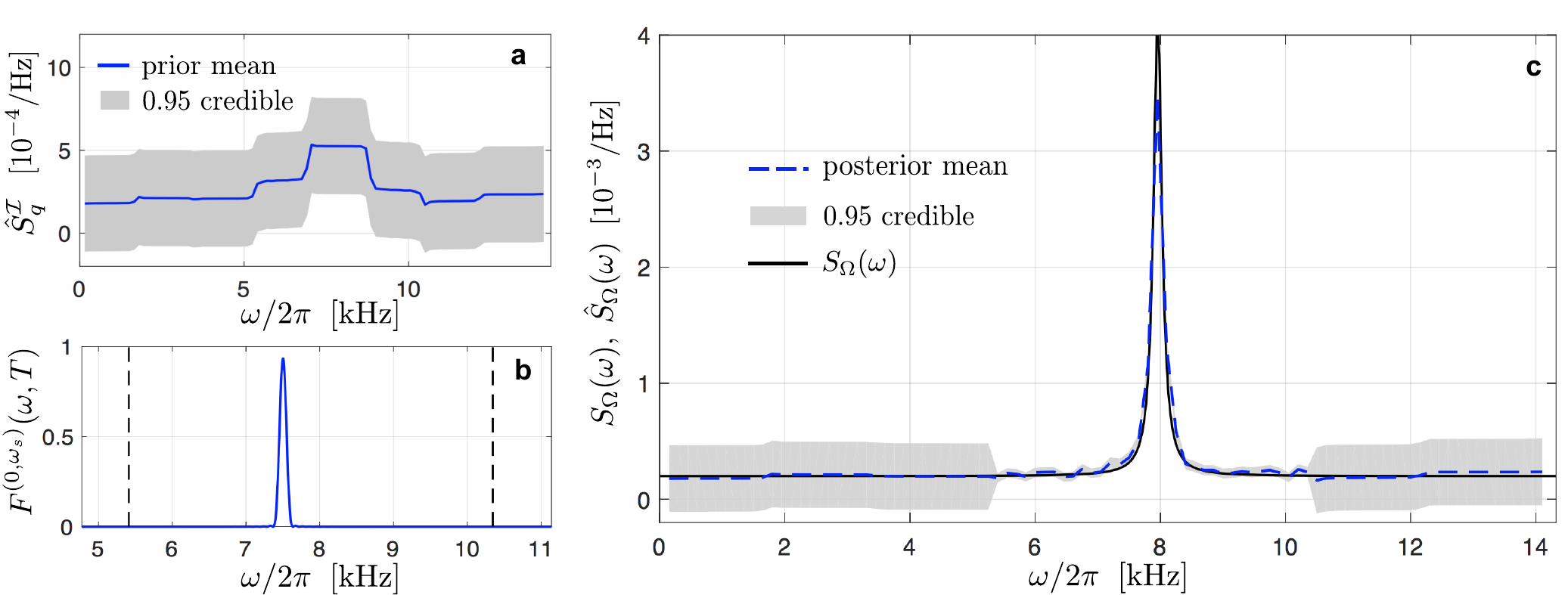}
\vspace*{-4mm}
\caption{(Color online) 
Reconstruction of a narrow Lorentzian peak via Bayesian estimation with multitaper prior. In (a), the prior mean of the discretized PSD, given by the renormalized interpolated estimate $\hat{S}_q^{\mathcal{I}}$, is plotted (blue line) along with the $0.95$ credible interval (shaded in grey).  To obtain a high resolution estimate of the fine spectral features in the 5.4-10.3 kHz region, additional measurements are performed using $k=0$ DPSS control waveforms with $N=500$, $\Delta t=20\;\mu s$, $W=1/N$ and same the time-domain normalization used previously. Plotted in (b) is a representative ``high-resolution" filter centered at 7.5 kHz (blue line) with black dashed lines at 5.4 and 10.3 kHz shown for scale. After $M=2600$ simulated qubit measurements for each control sequence, we obtain the posterior distribution of the PSD shown in (c) with mean (blue dashed line) and $0.95$ credible interval (shaded in grey). The mean shows excellent agreement with the actual PSD (black line).  The width of the $0.95$ credible interval in the 5.4-10.3 kHz region is substantially reduced due to the additional measurements.}
\label{fig::LineBayes}
\end{figure*}

The spectral reconstructions produced by the three approaches are plotted in the main panel of Fig. \ref{fig::Line1}(a). The large concentration at the center of the passband is clearly a detriment for the $k=0$ DPSS filter when it comes to detection, as sensitivity is lost at the boundary where the spectral concentration is low. Note from Fig. \ref{fig::Line1}(b) that the Lorentzian at $7.96$ kHz is located in a region of low concentration, which explains why the $k=0$ DPSS eigenestimates show no recognizable peak. Though the situation would be different if the Lorentzian were positioned closer to the center of the passband, there is no way to ensure this {\em a priori} without additional control sequences and measurements to ``fill in" the areas of low spectral concentration.  On account of the relative uniformity of the SSQM filter and the AQM effective filter, both of these estimates are instead capable of detecting fine features at any location with nearly equal likelihood: in Fig. \ref{fig::Line1}(a), both the AQM and SSQM reconstructions do exhibit a peak at $\omega_{s,4}$ and $\omega_{s,5}$, centered about  7.96 kHz. 

In an actual implementation of the detection strategy, we would need to verify whether the peaks in the estimates were truly the result of spectral structure or whether they could be explained by white noise. This can be accomplished with a simple significance test \cite{montgomery}.  As a null hypothesis, suppose that the actual PSD is flat. The expected values of the AQM and SSQM 
estimates are then given by their means over $\omega_{s,0}$,\ldots, $\omega_{s,8}$, which are, respectively, 
$\overline{S}_\Omega^{\text{m}}=27.7\times10^{-5}$ $1/$Hz and $\overline{S}_\Omega^{\text{ss}}=26.3\times10^{-5}$ $1/$Hz. 
Since, as mentioned, the simulation takes into account only statistical fluctuations of the measurement outcomes,
it follows from Eq. \ref{eq::VarEst} that the standard deviation of any passband estimator like the SSQM has the upper bound 
\begin{align*}
\sigma[\hat{S}_\Omega(\omega_s)] =
\bigg\{ \frac{{\mathbb P}(\uparrow_z,T)[1-{\mathbb P}(\uparrow_z,T)]}{M A^{(\omega_s)2}} \bigg\}^{\!1/2} 
\!\!\!\leq \frac{1}{ \sqrt{ 4MA^{(\omega_s)2}} }.
\end{align*}
For the AQM, we can find an upper bound for the standard deviation by applying the above formula to the eigenestimates. For both the SSQM and the AQM, the standard deviation of the peak estimates at $\omega_{s,4}$ and $\omega_{s,5}$ are upper-bounded by 
$\sigma[\hat{S}_\Omega^{\text{ss}}(\omega_{s,4})],\sigma[\hat{S}_\Omega^{\text{m}}(\omega_{s,4})],\sigma[\hat{S}_\Omega^{\text{m}}(\omega_{s,5})]\leq 6.3\times10^{-5}$ $1/$Hz and $\sigma[\hat{S}_\Omega^{\text{ss}}(\omega_{s,5})]\leq 6.4\times10^{-5}$ $1/$Hz. This implies that $\hat{S}_\Omega^{\text{ss}}(\omega_{s,4})$ and $\hat{S}_\Omega^{\text{ss}}(\omega_{s,5})$ are at least 2.7 and 3.3 standard deviations above the mean, while $\hat{S}_\Omega^{\text{m}}(\omega_{s,4})$ and $\hat{S}_\Omega^{\text{m}}(\omega_{s,5})$ are at least 4.2 and 3.7 standard deviations above the mean. The peaks in the multitaper reconstructions are therefore unlikely to be the result of a flat PSD, signifying spectral structure. In an actual experiment, more sophisticated significance tests may be performed, incorporating available knowledge of the measurement noise.

\subsubsection{High-resolution Bayesian estimation}

Although the SSQM and AQM approaches succeed in detecting the presence of spectral structure, the estimates of the PSD they provide are poor. Indeed, any DPSS estimate will have significant LB when the PSD contains features with widths much smaller than the widths of the filter passbands. To more accurately reconstruct the fine features of the PSD, we proceed to make additional estimates using DPSS of order $k=0$ with $N=500$, $\Delta t=20\,\mu$s, $W=1/N$ and the same time-domain normalization.
For these new parameters, the width of the passband is $W/\Delta t=0.21$ kHz, much narrower than the 3.5 kHz passbands of the filters  formerly used for detection. The $P=34$ filters total are shifted by $\omega_{s,p}/2\pi=(5.3+0.15p)$ kHz for $p\in\{1,\ldots,P\}$, so that they are spectrally concentrated in the 5.4-10.3 kHz region. A representative filter is plotted in Fig. \ref{fig::LineBayes}(b). Using the narrow passband filters, we obtained new eigenestimates $\hat{\mathbf{S}}_\Omega^{(0)}\equiv[\hat{S}_\Omega^{(0)}(\omega_{s,1}),\ldots, \hat{S}_\Omega^{(0)}(\omega_{s,P})]^T$ by making $M=2600$ numerically simulated qubit measurements along $z$ for each DPSS, as before. Note that reconstructing the entire PSD with the same sampling resolution would require 60 additional shifted DPSS, hence 156,000 additional measurements. The advantage of the detection strategy is that it enables us to target a specific region where spectral structure is present, thereby avoiding costly and unnecessary measurements.

Using the Bayesian approach described in Sec. \ref{sub::Bayes}, we can combine the new information about the fine features of the 
PSD contained in the eigenestimates $\hat{\mathbf{S}}_\Omega^{(0)}$ with the prior knowledge obtained in the detection stage.  We illustrate this in Fig. \ref{fig::LineBayes}. As a first step, we discretize the spectral estimation region from 0 to 14 kHz into $Q=94$ 
intervals of width $\Delta\omega/2\pi=0.15$ kHz, and correspondingly discretize the PSD, with $S_q$ denoting its value in interval $q\in\{1,\ldots,Q\}$. 
For the AQM estimate obtained in the detection stage, the interpolated estimate that follows from \erf{eq::FisherEst} is
$\hat{S}_q^\mathcal{I}=\sum_{p=0}^8\mathcal{I}_{qp}\hat{S}_\Omega^{\text{m}}(\omega_{s,p})$. As the mean of the prior distribution ${\mathbb P}\big[\vec{S}\big]$ described in \erf{eq::priorG}, we take $\mu_0=(\hat{S}_1^\mathcal{I},...,\hat{S}_Q^\mathcal{I})^T$. On account of the uniformity of the AQM effective filters, the weights $\mathcal{I}_{qp}$ and $\mathcal{I}_{q+1\,p}$ corresponding to neighboring frequency intervals $q$ and $q+1$ are often nearly equal, which produces an ill-conditioned 
prior covariance matrix $(\Sigma_0)_{qq'}=\sum_{p=0}^8\mathcal{I}_{qp}\mathcal{I}_{q'p}\text{var}[\hat{S}_\Omega^{\text{m}}(\omega_{s,p})]$. To ensure numerical stability, we employ Tikhonov regularization,
which involves substituting for $\Sigma_0$ a regularized covariance matrix, $\Sigma_0^\text{reg}\equiv\Sigma_0+\lambda\,I$, with $\lambda>0$ \cite{Neumaier1998}. Increasing the diagonal entries via Tikhonov regularization improves the conditioning of the covariance matrix, though it also expands the uncertainty of the prior. This is not a detriment in our case, however. The interpolated AQM estimate can be expected to have significant LB in the 5.4-10.3 kHz region, where fine features of the PSD were detected. Increasing the prior uncertainty ensures that the less biased narrow passband eigenestimates are weighted preferentially in the final estimate. The mean and $0.95$ credible region of the regularized prior are shown in Fig. \ref{fig::LineBayes}(a).  The likelihood ${\mathbb P}\big[\hat{\mathbf{S}}_\Omega^{(0)}|\vec{S}\big]$ follows from \erf{eq::LiklihoodBayes}. The final estimate of the PSD, namely, the mean of the posterior distribution ${\mathbb P}\big[\vec{S}\,\big|\,\hat{\mathbf{S}}_\Omega^{(0)}\big]\propto {\mathbb P}\big[\vec{S}\big]\,{\mathbb P}\big[\hat{\mathbf{S}}_\Omega^{(0)}\,\big|\,\vec{S}\big]$, is plotted in Fig. \ref{fig::LineBayes}(c). The posterior mean, which incorporates information gathered at each stage of the estimation procedure, is seen to agree very closely with the actual PSD.

\section{Conclusion and outlook} 
\label{sec::outlook}

In this work, we have provided a complete theoretical treatment of the optimally band-limited spectral estimation protocols for multiplicative control noise introduced and experimentally validated in Ref. \cite{Frey2017}. In addition to detailing the synthesis of these protocols via DPSS-defined amplitude control waveforms, we numerically demonstrated the applicability and performance of DPSS in a number of realistic spectral estimation scenarios.  DPSS controls clearly outperform established QNS methods based on DD in terms of suppressing leakage bias. Moreover, when combined with analogue modulation techniques, DPSS controls produce filter passbands with independently tunable positions. This affords the important advantage that the sampling resolution of a spectral reconstruction is independent of the total evolution time, unlike CPMG or comb-based approaches. As such, DPSS-based QNS can produce higher-quality spectral reconstructions even when sequence parameters must be chosen to compensate for 
limitations stemming from aliasing. The illustrative scenarios we have quantitatively analyzed further demonstrate how DPSS controls provide a very versatile and powerful tool for spectral estimation in the quantum setting. Conceptually, the problem of QNS can be divided into two interconnected tasks: detection and reconstruction. In a truly non-parametric scenario, we have shown how multitaper approaches are able to generate uniform filters ideally suited to identify regions of the frequency domain where fine spectral structure is present. These regions can then be targeted for high-resolution reconstruction using single-taper approaches and Bayesian inference techniques at a substantially reduced measurement cost.

This work can be extended in several directions. Presently, we have considered the application of DPSS controls to characterize multiplicative amplitude noise described by a Gaussian, zero-mean, stationary process. The multiplicative nature of this noise model, 
in which the amplitude control variable $\Omega(t)$ is coupled {\em directly} to the amplitude noise $\beta_\Omega(t)$ in the Hamiltonian of \erf{eq::Ham}, permits a straightforward analogy between the classical and quantum cases -- with $\Omega(t)$ serving as a ``continuous" taper in a quantum version of the classical tapered spectral estimate. This analogy is far less clear for the {\em additive} dephasing noise $\beta_z(t)$, which does not couple directly to the control variables.  While the qubit Hamiltonian in a suitably transformed frame can exhibit coupling between dephasing noise and the applied control [as evidenced by the toggling frame Hamiltonian of \erf{eq::toggle}], the control variables $\Omega(t)$ and $\phi(t)$ couple to $\beta_z(t)$ in an highly nonlinear manner.  For these reasons, qualitatively new methods are needed to generate DPSS filters for sensing additive dephasing noise, which will be the subject of future work.

Another key assumption in this work (indeed, in all QNS work to date we are aware of), is that the noise sources are {\em stationary}, meaning that their statistical properties are invariant under time translation. Yet, non-stationary, {\em deterministic} signals oscillating at an unknown frequency $\overline{\omega}$ are often encountered even when stationary noise is present, forming a composite noise process which may be described by $\tilde{\beta}_\Omega(t)=\beta_\Omega(t)+A\sin(\overline{\omega}t+\alpha)$, in terms 
of parameters $A, \overline{\omega}, \alpha$, on top of the stationary zero-mean background noise. A process such as 
$\tilde{\beta}_\Omega(t)$ has {\it mixed spectra}, exhibiting line components at $\pm\overline{\omega}$, which can be confused 
with peaks and other fine spectral features, complicating the interpretation of the reconstructed PSD. Using a two-stage multitaper approach, Thomson detailed a method for detecting line components and selectively reconstructing the underlying PSD \cite{thomson_multitaper}. Generalizing these ideas to the quantum regime, by accessing the {\em first moment} of the error vector components, will extend the applicability of QNS to common non-stationary processes.

Finally, recent work has extended comb-based DD-QNS to non-Gaussian dephasing noise \cite{Norris_Spectroscopy}, often encountered in superconducting qubits and other solid-state devices. The spectral properties of non-Gaussian noise sources are characterized by higher dimensional analogues of the PSD, known as polyspectra. The effect of leakage bias will likely be compounded in this higher dimensional setting, necessitating the extension of DPSS control techniques to non-Gaussian spectral estimation. Again, classical estimation techniques \cite{thompson1990} may provide a valuable starting point for exploration.

\section*{acknowledgements} 

It is a pleasure to thank F\'{e}lix Beaudoin and Gerardo A. Paz-Silva for a critical reading of the manuscript. 
Work partially supported by the US Army Research Office under Contract W911NF-12-R-0012 and the ARC Centre of 
Excellence for Engineered Quantum Systems CE110001013. D.L. acknowledges support from the 
Intelligence Advanced Research Projects Activity via Department of Interior National Business Center contract number 2012-12050800010.


\begin{appendix}

\section{Optimal concentration in the time domain}
\label{app::TimeConcen}

Though the converse of the spectral concentration problem we have discussed in the main text -- namely, 
the question of what band-limited sequence is maximally time-concentrated -- is not directly relevant to the time-limited 
signals we consider, a mention is nevertheless worthwhile in discussing notable properties of the DPSS and DPSWF. 

A discrete-time sequence $\{u_n\}$ whose representation in the frequency domain is band-limited 
is necessarily infinite in the time domain, as it cannot be simultaneously band-limited and time-limited. In analogy with 
\erf{eq::SpectralCon}, the extent to which $\{u_n\}$ can be approximately time-limited to an interval $[0,N\Delta t)$ is 
quantified by the ratio
\begin{align*}
E_{N}[\{u_n\}] \equiv \frac{\sum_{n=0}^{N-1}|u_n|^2}{\sum_{n=-\infty}^{\infty}|u_n|^2}.
\end{align*}
Slepian showed that the infinite sequence maximizing this expression solves a singular value problem 
equivalent to the Toeplitz eigenvalue problem discussed in Sec. \ref{sec::dpss}, 
albeit with $n\in(-\infty,\infty)$. The solutions are 
{\em doubly-infinite extensions} of the DPSS, which we still denote by 
$\{v_n^{(k)}(N,W)|n\in(-\infty,\infty)\}$. This extended sequence is identical to the DPSS of order $k$ 
under the restriction $n\in\{0,\ldots,N-1\}$. The time concentration of the extended DPSS is given by
\begin{align*}
E_{N}[\{v_n^{(k)}(N,W)|n\in(-\infty,\infty)\}]=\lambda_k(N,W).
\end{align*}
The infinite sequence with maximum time concentration is, thus, the doubly-infinite $k=0$ DPSS. 

To solve the time concentration problem, the extended DPSS must also be band-limited, however. Define the band-limiting operator 
$B_{\Delta b}$, acting on some frequency domain function $G(\omega)$, to be an operator that converts $G(\omega)$
into a function band-limited to $(-\Delta b,\Delta b)$, i.e., for $\omega\in[\omega_\text{N},\omega_\text{N})$,
\begin{align*}
B_{\Delta b} [G(\omega)] \equiv 
\begin{cases} 
\;G(\omega),&|\omega|< \Delta b\\
\;0,&|\omega|\geq \Delta b.
\end{cases}
\end{align*}
The DTFT of the extended DPSS are then the band-limited DPSWF,
\begin{align*}
B_{2\pi W/\Delta t}[U^{(k)}(N,W;\omega)] &= \\
\epsilon_k\!\sum_{n=-\infty}^{\infty}\! v_n^{(k)}(N,W)& e^{i\omega [n-(N-1)/2]\Delta t} , 
\end{align*}
which, similar to the DPSWF in Eq. \eqref{eq::DPSWF}, are also real functions of frequency. 
The extended $k=0$ DPSS, thus, is the optimal solution to the time concentration problem.

\vspace*{5mm}

\section{Fisher information}
\label{app::fisher}

The expressions for the Fisher information given in Eqs. (\ref{eq::FisherSk}) and (\ref{eq::FisherSm}) were derived under 
the assumption that measurement noise dominated over statistical fluctuations of the amplitude noise, which implies that 
the covariance matrix is independent of the amplitude noise PSD. Here, we give the more general form these expressions 
take in the case where $\Sigma$ does depend on the PSD. For $\hat{S}\in\{\hat{S}_\Omega(\omega_{s,p}), \hat{S}_\Omega^{\text{m}}(\omega_{s,p})\}$, let
\begin{align*}
\Sigma_{pp'} \equiv 
\begin{cases} 
\;\delta_{pp'}\text{var}[\hat{S}_\Omega(\omega_{s,p})],&\;\hat{S}=\hat{S}_\Omega(\omega_{s,p}),\\
\;\delta_{pp'}\text{var}[\hat{S}_\Omega^{\text{m}}(\omega_{s,p})],&\;\hat{S}=\hat{S}_\Omega^{\text{m}}(\omega_{s,p}),\\
\end{cases}
\end{align*}
\begin{align*}
\mathbf{F}_{pq} \equiv 
\begin{cases} 
\;A_q^{(\omega_{s,p})}/A^{(\omega_{s,p})},&\;\hat{S}=\hat{S}_\Omega(\omega_{s,p}),\\
\;\mathcal{R}_q^{(\omega_{s,p})},&\;\hat{S}=\hat{S}_\Omega^{\text{m}}(\omega_{s,p}),\\
\end{cases}
\end{align*}
and
\begin{align*}
\vec{S} \equiv (S_1,\ldots,S_Q)^T , 
\end{align*}
where $S_q=S_\Omega[(2q-1)\Delta\omega/2]$, with 
$q\in\{1,\ldots,Q\}$ and $p\in\{1,\ldots,P\}$, and $A_q^{(\omega_{s,p})}$ and  $\mathcal{R}_q^{(\omega_{s,p})}$ are defined in 
Eqs. (\ref{eq::Fq}) and (\ref{eq::Rq}), respectively. The Fisher information quantifying the information that $\hat{S}$ carries about $S_q$ is then given by
\begin{align*}
\mathcal{I}_{\hat{S}(\omega_{s,p})}[S_q]&=\bigg[\frac{\partial (\mathbf{F}\vec{S})_p}{\partial S_q}\bigg]^2\Sigma_{pp}^{-1}
+\frac{1}{2}\bigg(\frac{\partial \Sigma_{pp}}{\partial S_q}\bigg)^2\Sigma_{pp}^{-2}.
\end{align*}
The second term of this expression, which is absent in Eqs. (\ref{eq::FisherSk}) and  (\ref{eq::FisherSm}), results from dependence of $\Sigma$ on the PSD. Note that the first term scales with $M\propto\Sigma_{pp}^{-1}$, while the second term is independent of $M$. Accordingly, he second term is a small correction in the regime where $M\gg1$, as we consider.

In the numerical simulations in Sec. \ref{sec::results}, dephasing noise is absent and amplitude noise makes the only contribution to 
$\Sigma$, through  
$\sigma^2= {\mathbb P}(\uparrow_z,T)[1-{\mathbb P}(\uparrow_z,T)]$. The Fisher information of the eigenestimate has then 
the explicit expression
\begin{align*}
\mathcal{I}_{\hat{S}_\Omega(\omega_{s,p})}[S_q]= &\,M\frac{A_q^{(\omega_{s,p})\,2}}{{\mathbb P}(\uparrow_z,T)
[1-{\mathbb P}(\uparrow_z,T)]}\\\notag
&+2\bigg\{\frac{A_q^{(\omega_{s,p})}[2 {\mathbb P}(\uparrow_z,T)-1]}{{\mathbb P}(\uparrow_z,T)
[1-{\mathbb P}(\uparrow_z,T)]}\bigg\}^2.
\end{align*}
For $M\gtrsim1000$ as used in the simulations, the second term can indeed be verified to be negligible.

\end{appendix}

\bibliography{SlepianRefs}

\begin{thebibliography}{55}%
\makeatletter
\providecommand \@ifxundefined [1]{%
 \@ifx{#1\undefined}
}%
\providecommand \@ifnum [1]{%
 \ifnum #1\expandafter \@firstoftwo
 \else \expandafter \@secondoftwo
 \fi
}%
\providecommand \@ifx [1]{%
 \ifx #1\expandafter \@firstoftwo
 \else \expandafter \@secondoftwo
 \fi
}%
\providecommand \natexlab [1]{#1}%
\providecommand \enquote  [1]{``#1''}%
\providecommand \bibnamefont  [1]{#1}%
\providecommand \bibfnamefont [1]{#1}%
\providecommand \citenamefont [1]{#1}%
\providecommand \href@noop [0]{\@secondoftwo}%
\providecommand \href [0]{\begingroup \@sanitize@url \@href}%
\providecommand \@href[1]{\@@startlink{#1}\@@href}%
\providecommand \@@href[1]{\endgroup#1\@@endlink}%
\providecommand \@sanitize@url [0]{\catcode `\\12\catcode `\$12\catcode
  `\&12\catcode `\#12\catcode `\^12\catcode `\_12\catcode `\%12\relax}%
\providecommand \@@startlink[1]{}%
\providecommand \@@endlink[0]{}%
\providecommand \url  [0]{\begingroup\@sanitize@url \@url }%
\providecommand \@url [1]{\endgroup\@href {#1}{\urlprefix }}%
\providecommand \urlprefix  [0]{URL }%
\providecommand \Eprint [0]{\href }%
\providecommand \doibase [0]{http://dx.doi.org/}%
\providecommand \selectlanguage [0]{\@gobble}%
\providecommand \bibinfo  [0]{\@secondoftwo}%
\providecommand \bibfield  [0]{\@secondoftwo}%
\providecommand \translation [1]{[#1]}%
\providecommand \BibitemOpen [0]{}%
\providecommand \bibitemStop [0]{}%
\providecommand \bibitemNoStop [0]{.\EOS\space}%
\providecommand \EOS [0]{\spacefactor3000\relax}%
\providecommand \BibitemShut  [1]{\csname bibitem#1\endcsname}%
\let\auto@bib@innerbib\@empty
\bibitem [{\citenamefont {Percival}\ and\ \citenamefont
  {Walden}(1993)}]{percival1993spectral}%
  \BibitemOpen
  \bibfield  {author} {\bibinfo {author} {\bibfnamefont {D.~B.}\ \bibnamefont
  {Percival}}\ and\ \bibinfo {author} {\bibfnamefont {A.~T.}\ \bibnamefont
  {Walden}},\ }\href@noop {} {\emph {\bibinfo {title} {Spectral Analysis for
  Physical Applications}}}\ (\bibinfo  {publisher} {Cambridge University
  Press},\ \bibinfo {year} {1993})\BibitemShut {NoStop}%
\bibitem [{\citenamefont {Thomson}(1982)}]{thomson_multitaper}%
  \BibitemOpen
  \bibfield  {author} {\bibinfo {author} {\bibfnamefont {D.~J.}\ \bibnamefont
  {Thomson}},\ }\bibfield  {title} {\enquote {\bibinfo {title} {Spectrum
  estimation and harmonic analysis},}\ }\href@noop {} {\bibfield  {journal}
  {\bibinfo  {journal} {Proc. IEEE}\ }\textbf {\bibinfo {volume} {70}},\
  \bibinfo {pages} {1055} (\bibinfo {year} {1982})}\BibitemShut {NoStop}%
\bibitem [{\citenamefont {Slepian}\ and\ \citenamefont
  {Pollak}(1961)}]{Slepian1961}%
  \BibitemOpen
  \bibfield  {author} {\bibinfo {author} {\bibfnamefont {D.}~\bibnamefont
  {Slepian}}\ and\ \bibinfo {author} {\bibfnamefont {H.~O.}\ \bibnamefont
  {Pollak}},\ }\bibfield  {title} {\enquote {\bibinfo {title} {Prolate
  spheroidal wave functions, {F}ourier analysis, and uncertainty---{i}},}\
  }\href@noop {} {\bibfield  {journal} {\bibinfo  {journal} {Bell System Tech.
  J.}\ }\textbf {\bibinfo {volume} {40}},\ \bibinfo {pages} {43} (\bibinfo
  {year} {1961})}\BibitemShut {NoStop}%
\bibitem [{\citenamefont {Landau}\ and\ \citenamefont
  {Pollak}(1961{\natexlab{a}})}]{Landau1961a}%
  \BibitemOpen
  \bibfield  {author} {\bibinfo {author} {\bibfnamefont {H.~J.}\ \bibnamefont
  {Landau}}\ and\ \bibinfo {author} {\bibfnamefont {H.~O.}\ \bibnamefont
  {Pollak}},\ }\bibfield  {title} {\enquote {\bibinfo {title} {Prolate
  spheroidal wave functions, {F}ourier analysis and uncertainty---{ii}},}\
  }\href@noop {} {\bibfield  {journal} {\bibinfo  {journal} {Bell System Tech.
  J.}\ }\textbf {\bibinfo {volume} {41}},\ \bibinfo {pages} {65} (\bibinfo
  {year} {1961}{\natexlab{a}})}\BibitemShut {NoStop}%
\bibitem [{\citenamefont {Landau}\ and\ \citenamefont
  {Pollak}(1961{\natexlab{b}})}]{Landau1961b}%
  \BibitemOpen
  \bibfield  {author} {\bibinfo {author} {\bibfnamefont {H.~J.}\ \bibnamefont
  {Landau}}\ and\ \bibinfo {author} {\bibfnamefont {H.~O.}\ \bibnamefont
  {Pollak}},\ }\bibfield  {title} {\enquote {\bibinfo {title} {Prolate
  spheroidal wave functions, {F}ourier analysis and uncertainty---{iii}: The
  dimension of the space of essentially time- and band-limited signals},}\
  }\href@noop {} {\bibfield  {journal} {\bibinfo  {journal} {Bell System Tech.
  J.}\ }\textbf {\bibinfo {volume} {41}},\ \bibinfo {pages} {1295} (\bibinfo
  {year} {1961}{\natexlab{b}})}\BibitemShut {NoStop}%
\bibitem [{\citenamefont {Slepian}(1964)}]{Slepian1964}%
  \BibitemOpen
  \bibfield  {author} {\bibinfo {author} {\bibfnamefont {D.}~\bibnamefont
  {Slepian}},\ }\bibfield  {title} {\enquote {\bibinfo {title} {Prolate
  spheroidal wave functions, {F}ourier analysis and uncertainty ---{iv}:
  Extensions to many dimensions; generalized prolate spheroidal functions},}\
  }\href@noop {} {\bibfield  {journal} {\bibinfo  {journal} {Bell System Tech.
  J.}\ }\textbf {\bibinfo {volume} {43}},\ \bibinfo {pages} {3009} (\bibinfo
  {year} {1964})}\BibitemShut {NoStop}%
\bibitem [{\citenamefont {Papoulis}\ and\ \citenamefont
  {Bertran}(1972)}]{Papoulis1972}%
  \BibitemOpen
  \bibfield  {author} {\bibinfo {author} {\bibfnamefont {A.}~\bibnamefont
  {Papoulis}}\ and\ \bibinfo {author} {\bibfnamefont {M.~S.}\ \bibnamefont
  {Bertran}},\ }\bibfield  {title} {\enquote {\bibinfo {title} {Digital
  filtering and prolate functions},}\ }\href@noop {} {\bibfield  {journal}
  {\bibinfo  {journal} {IEEE Trans. Circuit Theory}\ }\textbf {\bibinfo
  {volume} {19}} (\bibinfo {year} {1972})}\BibitemShut {NoStop}%
\bibitem [{\citenamefont {Slepian}(1978)}]{Slepian1978}%
  \BibitemOpen
  \bibfield  {author} {\bibinfo {author} {\bibfnamefont {D.}~\bibnamefont
  {Slepian}},\ }\bibfield  {title} {\enquote {\bibinfo {title} {Prolate
  spheroidal wave functions, {F}ourier analysis, and uncertainty---{v}: The
  discrete case},}\ }\href@noop {} {\bibfield  {journal} {\bibinfo  {journal}
  {Bell System Tech. J.}\ }\textbf {\bibinfo {volume} {57}},\ \bibinfo {pages}
  {1371} (\bibinfo {year} {1978})}\BibitemShut {NoStop}%
\bibitem [{\citenamefont {Verd\'u}(1998)}]{verdu}%
  \BibitemOpen
  \bibfield  {author} {\bibinfo {author} {\bibfnamefont {S.}~\bibnamefont
  {Verd\'u}},\ }\bibfield  {title} {\enquote {\bibinfo {title} {Fifty years of
  {S}hannon theory},}\ }\href@noop {} {\bibfield  {journal} {\bibinfo
  {journal} {IEEE Trans. Inf. Theory}\ }\textbf {\bibinfo {volume} {44}},\
  \bibinfo {pages} {2057} (\bibinfo {year} {1998})}\BibitemShut {NoStop}%
\bibitem [{\citenamefont {Bronez}(1992)}]{Bronez}%
  \BibitemOpen
  \bibfield  {author} {\bibinfo {author} {\bibfnamefont {T.~P.}\ \bibnamefont
  {Bronez}},\ }\bibfield  {title} {\enquote {\bibinfo {title} {On the
  performance advantage of multitaper spectral analysis},}\ }\href@noop {}
  {\bibfield  {journal} {\bibinfo  {journal} {IEEE Trans. Signal Proc.}\
  }\textbf {\bibinfo {volume} {40}},\ \bibinfo {pages} {2941} (\bibinfo {year}
  {1992})}\BibitemShut {NoStop}%
\bibitem [{\citenamefont {Babadi}\ and\ \citenamefont {Brown}(2014)}]{babadi}%
  \BibitemOpen
  \bibfield  {author} {\bibinfo {author} {\bibfnamefont {B.}~\bibnamefont
  {Babadi}}\ and\ \bibinfo {author} {\bibfnamefont {E.~N.}\ \bibnamefont
  {Brown}},\ }\bibfield  {title} {\enquote {\bibinfo {title} {A review of
  multitaper spectral analysis},}\ }\href@noop {} {\bibfield  {journal}
  {\bibinfo  {journal} {IEEE Trans. Biomed. Eng.}\ }\textbf {\bibinfo {volume}
  {61}} (\bibinfo {year} {2014})}\BibitemShut {NoStop}%
\bibitem [{Aas()}]{Aash}%
  \BibitemOpen
  \href@noop {} {\ }\bibinfo {note} {R. J. Schoelkopf and A. A. Clerk and S. M.
  Girvin and K. W. Lehnert and M. H. Devoret, in {\em Quantum Noise in
  Mesoscopic Physics}, NATO Science Series {\bf 97}, 173 (2003).}\BibitemShut
  {Stop}%
\bibitem [{\citenamefont {Faoro}\ and\ \citenamefont {Viola}(2004)}]{Faoro}%
  \BibitemOpen
  \bibfield  {author} {\bibinfo {author} {\bibfnamefont {L.}~\bibnamefont
  {Faoro}}\ and\ \bibinfo {author} {\bibfnamefont {L.}~\bibnamefont {Viola}},\
  }\bibfield  {title} {\enquote {\bibinfo {title} {Dynamical suppression of
  $1/f$ noise processes in qubit systems},}\ }\href@noop {} {\bibfield
  {journal} {\bibinfo  {journal} {Phys. Rev. Lett.}\ }\textbf {\bibinfo
  {volume} {92}},\ \bibinfo {pages} {117905} (\bibinfo {year}
  {2004})}\BibitemShut {NoStop}%
\bibitem [{\citenamefont {Young}\ and\ \citenamefont
  {Whaley}(2012)}]{Young2012}%
  \BibitemOpen
  \bibfield  {author} {\bibinfo {author} {\bibfnamefont {K.~C.}\ \bibnamefont
  {Young}}\ and\ \bibinfo {author} {\bibfnamefont {K.~B.}\ \bibnamefont
  {Whaley}},\ }\bibfield  {title} {\enquote {\bibinfo {title} {Qubits as
  spectrometers of dephasing noise},}\ }\href@noop {} {\bibfield  {journal}
  {\bibinfo  {journal} {Phys. Rev. A}\ }\textbf {\bibinfo {volume} {86}}
  (\bibinfo {year} {2012})}\BibitemShut {NoStop}%
\bibitem [{\citenamefont {Gaebler}\ \emph {et~al.}(2016)\citenamefont
  {Gaebler}, \citenamefont {Tan}, \citenamefont {Lin}, \citenamefont {Wan},
  \citenamefont {Bowler}, \citenamefont {Keith}, \citenamefont {Glancy},
  \citenamefont {Coakley}, \citenamefont {Knill}, \citenamefont {Leibfried},\
  and\ \citenamefont {Wineland}}]{wineland}%
  \BibitemOpen
  \bibfield  {author} {\bibinfo {author} {\bibfnamefont {J.~P.}\ \bibnamefont
  {Gaebler}}, \bibinfo {author} {\bibfnamefont {T.~R.}\ \bibnamefont {Tan}},
  \bibinfo {author} {\bibfnamefont {Y.}~\bibnamefont {Lin}}, \bibinfo {author}
  {\bibfnamefont {Y.}~\bibnamefont {Wan}}, \bibinfo {author} {\bibfnamefont
  {R.}~\bibnamefont {Bowler}}, \bibinfo {author} {\bibfnamefont {A.~C.}\
  \bibnamefont {Keith}}, \bibinfo {author} {\bibfnamefont {S.}~\bibnamefont
  {Glancy}}, \bibinfo {author} {\bibfnamefont {K.}~\bibnamefont {Coakley}},
  \bibinfo {author} {\bibfnamefont {E.}~\bibnamefont {Knill}}, \bibinfo
  {author} {\bibfnamefont {D.}~\bibnamefont {Leibfried}}, \ and\ \bibinfo
  {author} {\bibfnamefont {D.~J.}\ \bibnamefont {Wineland}},\ }\bibfield
  {title} {\enquote {\bibinfo {title} {High-fidelity universal gate set for
  ${^{9}\mathrm{Be}}^{+}$ ion qubits},}\ }\href@noop {} {\bibfield  {journal}
  {\bibinfo  {journal} {Phys. Rev. Lett.}\ }\textbf {\bibinfo {volume} {117}},\
  \bibinfo {pages} {060505} (\bibinfo {year} {2016})}\BibitemShut {NoStop}%
\bibitem [{\citenamefont {Zajac}\ \emph {et~al.}(2017)\citenamefont {Zajac},
  \citenamefont {Sigillito}, \citenamefont {Russ}, \citenamefont {Borjans},
  \citenamefont {Taylor}, \citenamefont {Burkard},\ and\ \citenamefont
  {Petta}}]{petta}%
  \BibitemOpen
  \bibfield  {author} {\bibinfo {author} {\bibfnamefont {D.~M.}\ \bibnamefont
  {Zajac}}, \bibinfo {author} {\bibfnamefont {A.~J.}\ \bibnamefont
  {Sigillito}}, \bibinfo {author} {\bibfnamefont {M.}~\bibnamefont {Russ}},
  \bibinfo {author} {\bibfnamefont {F.}~\bibnamefont {Borjans}}, \bibinfo
  {author} {\bibfnamefont {J.~M.}\ \bibnamefont {Taylor}}, \bibinfo {author}
  {\bibfnamefont {G.}~\bibnamefont {Burkard}}, \ and\ \bibinfo {author}
  {\bibfnamefont {J.~R.}\ \bibnamefont {Petta}},\ }\bibfield  {title} {\enquote
  {\bibinfo {title} {Resonantly driven {\sc cnot} gate for electron spins},}\
  }\href@noop {} {\bibfield  {journal} {\bibinfo  {journal} {Science}\ ,\
  \bibinfo {pages} {10.1126/science.aao5965}} (\bibinfo {year}
  {2017})}\BibitemShut {NoStop}%
\bibitem [{\citenamefont {Ng}\ and\ \citenamefont
  {Preskill}(2009)}]{NgPreskill2009}%
  \BibitemOpen
  \bibfield  {author} {\bibinfo {author} {\bibfnamefont {H.~K.}\ \bibnamefont
  {Ng}}\ and\ \bibinfo {author} {\bibfnamefont {J.}~\bibnamefont {Preskill}},\
  }\bibfield  {title} {\enquote {\bibinfo {title} {Fault-tolerant quantum
  computation versus {G}aussian noise},}\ }\href@noop {} {\bibfield  {journal}
  {\bibinfo  {journal} {Phys. Rev. A}\ }\textbf {\bibinfo {volume} {79}},\
  \bibinfo {pages} {032318} (\bibinfo {year} {2009})}\BibitemShut {NoStop}%
\bibitem [{\citenamefont {Novais}\ and\ \citenamefont
  {Mucciolo}(2013)}]{Mucciolo2013}%
  \BibitemOpen
  \bibfield  {author} {\bibinfo {author} {\bibfnamefont {E.}~\bibnamefont
  {Novais}}\ and\ \bibinfo {author} {\bibfnamefont {E.~R.}\ \bibnamefont
  {Mucciolo}},\ }\bibfield  {title} {\enquote {\bibinfo {title} {Surface code
  threshold in the presence of correlated errors},}\ }\href@noop {} {\bibfield
  {journal} {\bibinfo  {journal} {Phys. Rev. Lett.}\ }\textbf {\bibinfo
  {volume} {110}},\ \bibinfo {pages} {010502} (\bibinfo {year}
  {2013})}\BibitemShut {NoStop}%
\bibitem [{\citenamefont {Boixo}\ \emph {et~al.}(2017)\citenamefont {Boixo},
  \citenamefont {Isakov}, \citenamefont {Smelyanskiy}, \citenamefont {Babbush},
  \citenamefont {Ding}, \citenamefont {Jiang}, \citenamefont {Bremner},
  \citenamefont {Martinis},\ and\ \citenamefont {Neven}}]{Sergio}%
  \BibitemOpen
  \bibfield  {author} {\bibinfo {author} {\bibfnamefont {S.}~\bibnamefont
  {Boixo}}, \bibinfo {author} {\bibfnamefont {S.~V.}\ \bibnamefont {Isakov}},
  \bibinfo {author} {\bibfnamefont {V.~N.}\ \bibnamefont {Smelyanskiy}},
  \bibinfo {author} {\bibfnamefont {R.}~\bibnamefont {Babbush}}, \bibinfo
  {author} {\bibfnamefont {N.}~\bibnamefont {Ding}}, \bibinfo {author}
  {\bibfnamefont {Z.}~\bibnamefont {Jiang}}, \bibinfo {author} {\bibfnamefont
  {M.~J.}\ \bibnamefont {Bremner}}, \bibinfo {author} {\bibfnamefont {J.~M.}\
  \bibnamefont {Martinis}}, \ and\ \bibinfo {author} {\bibfnamefont
  {H.}~\bibnamefont {Neven}},\ }\bibfield  {title} {\enquote {\bibinfo {title}
  {Characterizing quantum supremacy in near-term devices},}\ }\href@noop {}
  {\bibfield  {journal} {\bibinfo  {journal} {arXiv:1608.00263}\ } (\bibinfo
  {year} {2017})}\BibitemShut {NoStop}%
\bibitem [{\citenamefont {Johnson}\ \emph {et~al.}(2017)\citenamefont
  {Johnson}, \citenamefont {Romero}, \citenamefont {Olson}, \citenamefont
  {Cao},\ and\ \citenamefont {Aspuru-Guzik}}]{Peter}%
  \BibitemOpen
  \bibfield  {author} {\bibinfo {author} {\bibfnamefont {P.~D.}\ \bibnamefont
  {Johnson}}, \bibinfo {author} {\bibfnamefont {J.}~\bibnamefont {Romero}},
  \bibinfo {author} {\bibfnamefont {J.}~\bibnamefont {Olson}}, \bibinfo
  {author} {\bibfnamefont {Y.}~\bibnamefont {Cao}}, \ and\ \bibinfo {author}
  {\bibfnamefont {A.}~\bibnamefont {Aspuru-Guzik}},\ }\bibfield  {title}
  {\enquote {\bibinfo {title} {{\sc qvector}: an algorithm for device-tailored
  quantum error correction},}\ }\href@noop {} {\bibfield  {journal} {\bibinfo
  {journal} {arXiv:1711.02249}\ } (\bibinfo {year} {2017})}\BibitemShut
  {NoStop}%
\bibitem [{\citenamefont {Biercuk}\ \emph
  {et~al.}(2011{\natexlab{a}})\citenamefont {Biercuk}, \citenamefont
  {Doherty},\ and\ \citenamefont {Uys}}]{BiercukJPB2011}%
  \BibitemOpen
  \bibfield  {author} {\bibinfo {author} {\bibfnamefont {M.~J.}\ \bibnamefont
  {Biercuk}}, \bibinfo {author} {\bibfnamefont {A.~C.}\ \bibnamefont
  {Doherty}}, \ and\ \bibinfo {author} {\bibfnamefont {H.}~\bibnamefont
  {Uys}},\ }\bibfield  {title} {\enquote {\bibinfo {title} {Dynamical
  decoupling sequence construction as a filter-design problem},}\ }\href@noop
  {} {\bibfield  {journal} {\bibinfo  {journal} {J. Phys. B}\ }\textbf
  {\bibinfo {volume} {44}},\ \bibinfo {pages} {154002} (\bibinfo {year}
  {2011}{\natexlab{a}})}\BibitemShut {NoStop}%
\bibitem [{\citenamefont {Norris}\ \emph {et~al.}(2016)\citenamefont {Norris},
  \citenamefont {Paz-Silva},\ and\ \citenamefont
  {Viola}}]{Norris_Spectroscopy}%
  \BibitemOpen
  \bibfield  {author} {\bibinfo {author} {\bibfnamefont {L.~M.}\ \bibnamefont
  {Norris}}, \bibinfo {author} {\bibfnamefont {G.~A.}\ \bibnamefont
  {Paz-Silva}}, \ and\ \bibinfo {author} {\bibfnamefont {L.}~\bibnamefont
  {Viola}},\ }\bibfield  {title} {\enquote {\bibinfo {title} {Qubit noise
  spectroscopy for non-{G}aussian dephasing environments},}\ }\href@noop {}
  {\bibfield  {journal} {\bibinfo  {journal} {Phys. Rev. Lett.}\ }\textbf
  {\bibinfo {volume} {116}},\ \bibinfo {pages} {150503} (\bibinfo {year}
  {2016})}\BibitemShut {NoStop}%
\bibitem [{\citenamefont {Szankowski}\ \emph {et~al.}(2016)\citenamefont
  {Szankowski}, \citenamefont {Trippenbach},\ and\ \citenamefont
  {Cywinski}}]{Cywinski2016}%
  \BibitemOpen
  \bibfield  {author} {\bibinfo {author} {\bibfnamefont {P.}~\bibnamefont
  {Szankowski}}, \bibinfo {author} {\bibfnamefont {M.}~\bibnamefont
  {Trippenbach}}, \ and\ \bibinfo {author} {\bibfnamefont {L.}~\bibnamefont
  {Cywinski}},\ }\bibfield  {title} {\enquote {\bibinfo {title} {Spectroscopy
  of cross correlations of environmental noises with two qubits},}\ }\href@noop
  {} {\bibfield  {journal} {\bibinfo  {journal} {Phys. Rev. A}\ }\textbf
  {\bibinfo {volume} {94}},\ \bibinfo {pages} {012109} (\bibinfo {year}
  {2016})}\BibitemShut {NoStop}%
\bibitem [{\citenamefont {Paz-Silva}\ \emph {et~al.}(2017)\citenamefont
  {Paz-Silva}, \citenamefont {Norris},\ and\ \citenamefont {Viola}}]{Paz2017}%
  \BibitemOpen
  \bibfield  {author} {\bibinfo {author} {\bibfnamefont {G.~A.}\ \bibnamefont
  {Paz-Silva}}, \bibinfo {author} {\bibfnamefont {L.~M.}\ \bibnamefont
  {Norris}}, \ and\ \bibinfo {author} {\bibfnamefont {L.}~\bibnamefont
  {Viola}},\ }\bibfield  {title} {\enquote {\bibinfo {title} {Multiqubit
  spectroscopy of {G}aussian quantum noise},}\ }\href@noop {} {\bibfield
  {journal} {\bibinfo  {journal} {Phys. Rev. A}\ }\textbf {\bibinfo {volume}
  {95}},\ \bibinfo {pages} {022121} (\bibinfo {year} {2017})}\BibitemShut
  {NoStop}%
\bibitem [{\citenamefont {\'Alvarez}\ and\ \citenamefont
  {Suter}(2011)}]{Alvarez_Spectroscopy}%
  \BibitemOpen
  \bibfield  {author} {\bibinfo {author} {\bibfnamefont {G.~A.}\ \bibnamefont
  {\'Alvarez}}\ and\ \bibinfo {author} {\bibfnamefont {D.}~\bibnamefont
  {Suter}},\ }\bibfield  {title} {\enquote {\bibinfo {title} {Measuring the
  spectrum of colored noise by dynamical decoupling},}\ }\href@noop {}
  {\bibfield  {journal} {\bibinfo  {journal} {Phys. Rev. Lett.}\ }\textbf
  {\bibinfo {volume} {107}},\ \bibinfo {pages} {230501} (\bibinfo {year}
  {2011})}\BibitemShut {NoStop}%
\bibitem [{\citenamefont {Muhonen}\ \emph {et~al.}(2014)\citenamefont
  {Muhonen}, \citenamefont {Dehollain}, \citenamefont {Laucht}, \citenamefont
  {Hudson}, \citenamefont {Kalra}, \citenamefont {Sekiguchi}, \citenamefont
  {Itoh}, \citenamefont {Jamieson}, \citenamefont {McCallum}, \citenamefont
  {Dzurak},\ and\ \citenamefont {Morello}}]{Morello2014}%
  \BibitemOpen
  \bibfield  {author} {\bibinfo {author} {\bibfnamefont {J.~T.}\ \bibnamefont
  {Muhonen}}, \bibinfo {author} {\bibfnamefont {J.~P.}\ \bibnamefont
  {Dehollain}}, \bibinfo {author} {\bibfnamefont {A.}~\bibnamefont {Laucht}},
  \bibinfo {author} {\bibfnamefont {F.~E.}\ \bibnamefont {Hudson}}, \bibinfo
  {author} {\bibfnamefont {R.}~\bibnamefont {Kalra}}, \bibinfo {author}
  {\bibfnamefont {T.}~\bibnamefont {Sekiguchi}}, \bibinfo {author}
  {\bibfnamefont {K.~M.}\ \bibnamefont {Itoh}}, \bibinfo {author}
  {\bibfnamefont {D.~N.}\ \bibnamefont {Jamieson}}, \bibinfo {author}
  {\bibfnamefont {J.~C.}\ \bibnamefont {McCallum}}, \bibinfo {author}
  {\bibfnamefont {A.~S.}\ \bibnamefont {Dzurak}}, \ and\ \bibinfo {author}
  {\bibfnamefont {A.}~\bibnamefont {Morello}},\ }\bibfield  {title} {\enquote
  {\bibinfo {title} {Storing quantum information for 30 seconds in a
  nanoelectronic device},}\ }\href@noop {} {\bibfield  {journal} {\bibinfo
  {journal} {Nature Nanotech.}\ }\textbf {\bibinfo {volume} {9}},\ \bibinfo
  {pages} {986} (\bibinfo {year} {2014})}\BibitemShut {NoStop}%
\bibitem [{\citenamefont {Malinowski}\ \emph {et~al.}(2017)\citenamefont
  {Malinowski}, \citenamefont {Martins}, \citenamefont
  {Cywi\ifmmode~\acute{n}\else \'{n}\fi{}ski}, \citenamefont {Rudner},
  \citenamefont {Nissen}, \citenamefont {Fallahi}, \citenamefont {Gardner},
  \citenamefont {Manfra}, \citenamefont {Marcus},\ and\ \citenamefont
  {Kuemmeth}}]{Malinowski2017}%
  \BibitemOpen
  \bibfield  {author} {\bibinfo {author} {\bibfnamefont {F.~K.}\ \bibnamefont
  {Malinowski}}, \bibinfo {author} {\bibfnamefont {F.}~\bibnamefont {Martins}},
  \bibinfo {author} {\bibfnamefont {L.}~\bibnamefont
  {Cywi\ifmmode~\acute{n}\else \'{n}\fi{}ski}}, \bibinfo {author}
  {\bibfnamefont {M.~S.}\ \bibnamefont {Rudner}}, \bibinfo {author}
  {\bibfnamefont {P.~D.}\ \bibnamefont {Nissen}}, \bibinfo {author}
  {\bibfnamefont {S.}~\bibnamefont {Fallahi}}, \bibinfo {author} {\bibfnamefont
  {G.~C.}\ \bibnamefont {Gardner}}, \bibinfo {author} {\bibfnamefont {M.~J.}\
  \bibnamefont {Manfra}}, \bibinfo {author} {\bibfnamefont {C.~M.}\
  \bibnamefont {Marcus}}, \ and\ \bibinfo {author} {\bibfnamefont
  {F.}~\bibnamefont {Kuemmeth}},\ }\bibfield  {title} {\enquote {\bibinfo
  {title} {Spectrum of the nuclear environment for {G}a{A}s spin qubits},}\
  }\href@noop {} {\bibfield  {journal} {\bibinfo  {journal} {Phys. Rev. Lett.}\
  }\textbf {\bibinfo {volume} {118}},\ \bibinfo {pages} {177702} (\bibinfo
  {year} {2017})}\BibitemShut {NoStop}%
\bibitem [{\citenamefont {Bylander}\ \emph {et~al.}(2011)\citenamefont
  {Bylander}, \citenamefont {Gustavsson}, \citenamefont {Yan}, \citenamefont
  {Yoshihara}, \citenamefont {Harrabi}, \citenamefont {Fitch}, \citenamefont
  {Cory}, \citenamefont {Nakamura}, \citenamefont {Tsai},\ and\ \citenamefont
  {Oliver}}]{Bylander2011}%
  \BibitemOpen
  \bibfield  {author} {\bibinfo {author} {\bibfnamefont {J.}~\bibnamefont
  {Bylander}}, \bibinfo {author} {\bibfnamefont {S.}~\bibnamefont
  {Gustavsson}}, \bibinfo {author} {\bibfnamefont {F.}~\bibnamefont {Yan}},
  \bibinfo {author} {\bibfnamefont {F.}~\bibnamefont {Yoshihara}}, \bibinfo
  {author} {\bibfnamefont {K.}~\bibnamefont {Harrabi}}, \bibinfo {author}
  {\bibfnamefont {G.}~\bibnamefont {Fitch}}, \bibinfo {author} {\bibfnamefont
  {D.~G.}\ \bibnamefont {Cory}}, \bibinfo {author} {\bibfnamefont
  {Y.}~\bibnamefont {Nakamura}}, \bibinfo {author} {\bibfnamefont {J.-S.}\
  \bibnamefont {Tsai}}, \ and\ \bibinfo {author} {\bibfnamefont {W.~D.}\
  \bibnamefont {Oliver}},\ }\bibfield  {title} {\enquote {\bibinfo {title}
  {Noise spectroscopy through dynamical decoupling with a superconducting flux
  qubit},}\ }\href@noop {} {\bibfield  {journal} {\bibinfo  {journal} {Nature
  Phys.}\ }\textbf {\bibinfo {volume} {7}},\ \bibinfo {pages} {565} (\bibinfo
  {year} {2011})}\BibitemShut {NoStop}%
\bibitem [{\citenamefont {Romach}\ \emph {et~al.}(2015)\citenamefont {Romach},
  \citenamefont {M\"uller}, \citenamefont {Unden}, \citenamefont {Rogers},
  \citenamefont {Isoda}, \citenamefont {Itoh}, \citenamefont {Markham},
  \citenamefont {Stacey}, \citenamefont {Meijer}, \citenamefont {Pezzagna},
  \citenamefont {Naydenov}, \citenamefont {McGuinness}, \citenamefont
  {Bar-Gill},\ and\ \citenamefont {Jelezko}}]{Jelezko2015}%
  \BibitemOpen
  \bibfield  {author} {\bibinfo {author} {\bibfnamefont {Y.}~\bibnamefont
  {Romach}}, \bibinfo {author} {\bibfnamefont {C.}~\bibnamefont {M\"uller}},
  \bibinfo {author} {\bibfnamefont {T.}~\bibnamefont {Unden}}, \bibinfo
  {author} {\bibfnamefont {L.~J.}\ \bibnamefont {Rogers}}, \bibinfo {author}
  {\bibfnamefont {T.}~\bibnamefont {Isoda}}, \bibinfo {author} {\bibfnamefont
  {K.~M.}\ \bibnamefont {Itoh}}, \bibinfo {author} {\bibfnamefont
  {M.}~\bibnamefont {Markham}}, \bibinfo {author} {\bibfnamefont
  {A.}~\bibnamefont {Stacey}}, \bibinfo {author} {\bibfnamefont
  {J.}~\bibnamefont {Meijer}}, \bibinfo {author} {\bibfnamefont
  {S.}~\bibnamefont {Pezzagna}}, \bibinfo {author} {\bibfnamefont
  {B.}~\bibnamefont {Naydenov}}, \bibinfo {author} {\bibfnamefont {L.~P.}\
  \bibnamefont {McGuinness}}, \bibinfo {author} {\bibfnamefont
  {N.}~\bibnamefont {Bar-Gill}}, \ and\ \bibinfo {author} {\bibfnamefont
  {F.}~\bibnamefont {Jelezko}},\ }\bibfield  {title} {\enquote {\bibinfo
  {title} {Spectroscopy of surface-induced noise using shallow spins in
  diamond},}\ }\href@noop {} {\bibfield  {journal} {\bibinfo  {journal} {Phys.
  Rev. Lett.}\ }\textbf {\bibinfo {volume} {114}},\ \bibinfo {pages} {017601}
  (\bibinfo {year} {2015})}\BibitemShut {NoStop}%
\bibitem [{\citenamefont {Kotler}\ \emph {et~al.}(2013)\citenamefont {Kotler},
  \citenamefont {Akerman}, \citenamefont {Glickman},\ and\ \citenamefont
  {Ozeri}}]{Ozeri2013}%
  \BibitemOpen
  \bibfield  {author} {\bibinfo {author} {\bibfnamefont {S.}~\bibnamefont
  {Kotler}}, \bibinfo {author} {\bibfnamefont {N.}~\bibnamefont {Akerman}},
  \bibinfo {author} {\bibfnamefont {Y.}~\bibnamefont {Glickman}}, \ and\
  \bibinfo {author} {\bibfnamefont {R.}~\bibnamefont {Ozeri}},\ }\bibfield
  {title} {\enquote {\bibinfo {title} {Nonlinear single-spin spectrum
  analyzer},}\ }\href@noop {} {\bibfield  {journal} {\bibinfo  {journal} {Phys.
  Rev. Lett.}\ }\textbf {\bibinfo {volume} {110}},\ \bibinfo {pages} {110503}
  (\bibinfo {year} {2013})}\BibitemShut {NoStop}%
\bibitem [{\citenamefont {Almog}\ \emph {et~al.}(2016)\citenamefont {Almog},
  \citenamefont {Loewenthal}, \citenamefont {Coslovsky}, \citenamefont {Sagi},\
  and\ \citenamefont {Davidson}}]{Davidson2016}%
  \BibitemOpen
  \bibfield  {author} {\bibinfo {author} {\bibfnamefont {I.}~\bibnamefont
  {Almog}}, \bibinfo {author} {\bibfnamefont {G.}~\bibnamefont {Loewenthal}},
  \bibinfo {author} {\bibfnamefont {J.}~\bibnamefont {Coslovsky}}, \bibinfo
  {author} {\bibfnamefont {Y.}~\bibnamefont {Sagi}}, \ and\ \bibinfo {author}
  {\bibfnamefont {N.}~\bibnamefont {Davidson}},\ }\bibfield  {title} {\enquote
  {\bibinfo {title} {Dynamic decoupling in the presence of colored control
  noise},}\ }\href@noop {} {\bibfield  {journal} {\bibinfo  {journal} {Phys.
  Rev. A}\ }\textbf {\bibinfo {volume} {94}},\ \bibinfo {pages} {042317}
  (\bibinfo {year} {2016})}\BibitemShut {NoStop}%
\bibitem [{\citenamefont {Szankowski}\ and\ \citenamefont
  {Cywinski}(2018)}]{LukasAccuracy}%
  \BibitemOpen
  \bibfield  {author} {\bibinfo {author} {\bibfnamefont {P.}~\bibnamefont
  {Szankowski}}\ and\ \bibinfo {author} {\bibfnamefont {L.}~\bibnamefont
  {Cywinski}},\ }\bibfield  {title} {\enquote {\bibinfo {title} {On accuracy of
  dynamical decoupling based spectroscopy of {G}aussian noise},}\ }\href@noop
  {} {\bibfield  {journal} {\bibinfo  {journal} {Phys. Rev. A}\ }\textbf
  {\bibinfo {volume} {97}},\ \bibinfo {pages} {032101} (\bibinfo {year}
  {2018})}\BibitemShut {NoStop}%
\bibitem [{\citenamefont {Loretz}\ \emph {et~al.}(2014)\citenamefont {Loretz},
  \citenamefont {Rosskopf}, \citenamefont {Boss}, \citenamefont {Pezzagna},
  \citenamefont {Meijer},\ and\ \citenamefont {Degen}}]{degen_retract}%
  \BibitemOpen
  \bibfield  {author} {\bibinfo {author} {\bibfnamefont {M.}~\bibnamefont
  {Loretz}}, \bibinfo {author} {\bibfnamefont {T.}~\bibnamefont {Rosskopf}},
  \bibinfo {author} {\bibfnamefont {J.~M.}\ \bibnamefont {Boss}}, \bibinfo
  {author} {\bibfnamefont {S.}~\bibnamefont {Pezzagna}}, \bibinfo {author}
  {\bibfnamefont {J.}~\bibnamefont {Meijer}}, \ and\ \bibinfo {author}
  {\bibfnamefont {C.~L.}\ \bibnamefont {Degen}},\ }\bibfield  {title} {\enquote
  {\bibinfo {title} {Single-proton spin detection by diamond magnetometry},}\
  }\href@noop {} {\bibfield  {journal} {\bibinfo  {journal} {Science}\ }
  (\bibinfo {year} {2014})}\BibitemShut {NoStop}%
\bibitem [{\citenamefont {Loretz}\ \emph {et~al.}(2015)\citenamefont {Loretz},
  \citenamefont {Boss}, \citenamefont {Rosskopf}, \citenamefont {Mamin},
  \citenamefont {Rugar},\ and\ \citenamefont {Degen}}]{degen_followup}%
  \BibitemOpen
  \bibfield  {author} {\bibinfo {author} {\bibfnamefont {M.}~\bibnamefont
  {Loretz}}, \bibinfo {author} {\bibfnamefont {J.~M.}\ \bibnamefont {Boss}},
  \bibinfo {author} {\bibfnamefont {T.}~\bibnamefont {Rosskopf}}, \bibinfo
  {author} {\bibfnamefont {H.~J.}\ \bibnamefont {Mamin}}, \bibinfo {author}
  {\bibfnamefont {D.}~\bibnamefont {Rugar}}, \ and\ \bibinfo {author}
  {\bibfnamefont {C.~L.}\ \bibnamefont {Degen}},\ }\bibfield  {title} {\enquote
  {\bibinfo {title} {Spurious harmonic response of multipulse quantum sensing
  sequences},}\ }\href@noop {} {\bibfield  {journal} {\bibinfo  {journal}
  {Phys. Rev. X}\ }\textbf {\bibinfo {volume} {5}},\ \bibinfo {pages} {021009}
  (\bibinfo {year} {2015})}\BibitemShut {NoStop}%
\bibitem [{\citenamefont {Frey}\ \emph {et~al.}(2017)\citenamefont {Frey},
  \citenamefont {Mavadia}, \citenamefont {Norris}, \citenamefont {de~Ferranti},
  \citenamefont {Lucarelli}, \citenamefont {Viola},\ and\ \citenamefont
  {Biercuk}}]{Frey2017}%
  \BibitemOpen
  \bibfield  {author} {\bibinfo {author} {\bibfnamefont {V.~M.}\ \bibnamefont
  {Frey}}, \bibinfo {author} {\bibfnamefont {S.}~\bibnamefont {Mavadia}},
  \bibinfo {author} {\bibfnamefont {L.~M.}\ \bibnamefont {Norris}}, \bibinfo
  {author} {\bibfnamefont {W.}~\bibnamefont {de~Ferranti}}, \bibinfo {author}
  {\bibfnamefont {D.}~\bibnamefont {Lucarelli}}, \bibinfo {author}
  {\bibfnamefont {L.}~\bibnamefont {Viola}}, \ and\ \bibinfo {author}
  {\bibfnamefont {M.~J.}\ \bibnamefont {Biercuk}},\ }\bibfield  {title}
  {\enquote {\bibinfo {title} {Application of optimal band-limited control
  protocols to quantum noise sensing},}\ }\href@noop {} {\bibfield  {journal}
  {\bibinfo  {journal} {Nature Comms.}\ }\textbf {\bibinfo {volume} {8}},\
  \bibinfo {pages} {2189} (\bibinfo {year} {2017})}\BibitemShut {NoStop}%
\bibitem [{Bay()}]{BayesianQNS}%
  \BibitemOpen
  \href@noop {} {\ }\bibinfo {note} {C. Ferrie, C. Granade, G. A. Paz-Silva,
  and H. M. Wiseman, ``{B}ayesian quantum noise spectroscopy,''
  arXiv:1707.05088 (2017).}\BibitemShut {Stop}%
\bibitem [{\citenamefont {Soare}\ \emph
  {et~al.}(2014{\natexlab{a}})\citenamefont {Soare}, \citenamefont {Ball},
  \citenamefont {Hayes}, \citenamefont {Zhen}, \citenamefont {Jarratt},
  \citenamefont {Sastrawan}, \citenamefont {Uys},\ and\ \citenamefont
  {Biercuk}}]{Soare2014bath}%
  \BibitemOpen
  \bibfield  {author} {\bibinfo {author} {\bibfnamefont {A.}~\bibnamefont
  {Soare}}, \bibinfo {author} {\bibfnamefont {H.}~\bibnamefont {Ball}},
  \bibinfo {author} {\bibfnamefont {D.}~\bibnamefont {Hayes}}, \bibinfo
  {author} {\bibfnamefont {X.}~\bibnamefont {Zhen}}, \bibinfo {author}
  {\bibfnamefont {M.~C.}\ \bibnamefont {Jarratt}}, \bibinfo {author}
  {\bibfnamefont {J.}~\bibnamefont {Sastrawan}}, \bibinfo {author}
  {\bibfnamefont {H.}~\bibnamefont {Uys}}, \ and\ \bibinfo {author}
  {\bibfnamefont {M.~J.}\ \bibnamefont {Biercuk}},\ }\bibfield  {title}
  {\enquote {\bibinfo {title} {Experimental bath engineering for quantitative
  studies of quantum control},}\ }\href@noop {} {\bibfield  {journal} {\bibinfo
   {journal} {Phys. Rev. A}\ }\textbf {\bibinfo {volume} {\textbf{89}}},\
  \bibinfo {pages} {042329} (\bibinfo {year} {2014}{\natexlab{a}})}\BibitemShut
  {NoStop}%
\bibitem [{\citenamefont {Soare}\ \emph
  {et~al.}(2014{\natexlab{b}})\citenamefont {Soare}, \citenamefont {Ball},
  \citenamefont {Hayes}, \citenamefont {Sastrawan}, \citenamefont {Jarratt},
  \citenamefont {McLoughlin}, \citenamefont {Zhen}, \citenamefont {Green},\
  and\ \citenamefont {Biercuk}}]{SoareNatPhys2014}%
  \BibitemOpen
  \bibfield  {author} {\bibinfo {author} {\bibfnamefont {A.}~\bibnamefont
  {Soare}}, \bibinfo {author} {\bibfnamefont {H.}~\bibnamefont {Ball}},
  \bibinfo {author} {\bibfnamefont {D.}~\bibnamefont {Hayes}}, \bibinfo
  {author} {\bibfnamefont {J.}~\bibnamefont {Sastrawan}}, \bibinfo {author}
  {\bibfnamefont {M.~C.}\ \bibnamefont {Jarratt}}, \bibinfo {author}
  {\bibfnamefont {J.~J.}\ \bibnamefont {McLoughlin}}, \bibinfo {author}
  {\bibfnamefont {X.}~\bibnamefont {Zhen}}, \bibinfo {author} {\bibfnamefont
  {T.~J.}\ \bibnamefont {Green}}, \ and\ \bibinfo {author} {\bibfnamefont
  {M.~J.}\ \bibnamefont {Biercuk}},\ }\bibfield  {title} {\enquote {\bibinfo
  {title} {Experimental noise filtering by quantum control},}\ }\href@noop {}
  {\bibfield  {journal} {\bibinfo  {journal} {Nature Phys.}\ }\textbf {\bibinfo
  {volume} {10}},\ \bibinfo {pages} {825} (\bibinfo {year}
  {2014}{\natexlab{b}})}\BibitemShut {NoStop}%
\bibitem [{\citenamefont {Ball}\ and\ \citenamefont
  {Biercuk}(2015)}]{Ball2014}%
  \BibitemOpen
  \bibfield  {author} {\bibinfo {author} {\bibfnamefont {H.}~\bibnamefont
  {Ball}}\ and\ \bibinfo {author} {\bibfnamefont {M.~J.}\ \bibnamefont
  {Biercuk}},\ }\bibfield  {title} {\enquote {\bibinfo {title}
  {{W}alsh-synthesized noise filters for quantum logic},}\ }\href@noop {}
  {\bibfield  {journal} {\bibinfo  {journal} {EPJ Quantum Tech.}\ }\textbf
  {\bibinfo {volume} {2}} (\bibinfo {year} {2015})}\BibitemShut {NoStop}%
\bibitem [{\citenamefont {Khodjasteh}\ \emph {et~al.}(2012)\citenamefont
  {Khodjasteh}, \citenamefont {Bluhm},\ and\ \citenamefont {Viola}}]{aDCG}%
  \BibitemOpen
  \bibfield  {author} {\bibinfo {author} {\bibfnamefont {K.}~\bibnamefont
  {Khodjasteh}}, \bibinfo {author} {\bibfnamefont {H.}~\bibnamefont {Bluhm}}, \
  and\ \bibinfo {author} {\bibfnamefont {L.}~\bibnamefont {Viola}},\ }\bibfield
   {title} {\enquote {\bibinfo {title} {Automated synthesis of dynamically
  corrected quantum gates},}\ }\href@noop {} {\bibfield  {journal} {\bibinfo
  {journal} {Phys. Rev. A}\ }\textbf {\bibinfo {volume} {86}},\ \bibinfo
  {pages} {042329} (\bibinfo {year} {2012})}\BibitemShut {NoStop}%
\bibitem [{\citenamefont {Khodjasteh}\ and\ \citenamefont
  {Viola}(2009)}]{KavehDCG}%
  \BibitemOpen
  \bibfield  {author} {\bibinfo {author} {\bibfnamefont {K.}~\bibnamefont
  {Khodjasteh}}\ and\ \bibinfo {author} {\bibfnamefont {L.}~\bibnamefont
  {Viola}},\ }\bibfield  {title} {\enquote {\bibinfo {title} {Dynamically
  error-corrected gates for universal quantum computation},}\ }\href@noop {}
  {\bibfield  {journal} {\bibinfo  {journal} {Phys. Rev. Lett.}\ }\textbf
  {\bibinfo {volume} {102}},\ \bibinfo {pages} {080501} (\bibinfo {year}
  {2009})}\BibitemShut {NoStop}%
\bibitem [{\citenamefont {Paz-Silva}\ and\ \citenamefont
  {Viola}(2014)}]{PazFFF}%
  \BibitemOpen
  \bibfield  {author} {\bibinfo {author} {\bibfnamefont {G.~A.}\ \bibnamefont
  {Paz-Silva}}\ and\ \bibinfo {author} {\bibfnamefont {L.}~\bibnamefont
  {Viola}},\ }\bibfield  {title} {\enquote {\bibinfo {title} {General
  transfer-function approach to noise filtering in open-loop quantum
  control},}\ }\href@noop {} {\bibfield  {journal} {\bibinfo  {journal} {Phys.
  Rev. Lett.}\ }\textbf {\bibinfo {volume} {113}},\ \bibinfo {pages} {250501}
  (\bibinfo {year} {2014})}\BibitemShut {NoStop}%
\bibitem [{\citenamefont {Green}\ \emph {et~al.}(2013)\citenamefont {Green},
  \citenamefont {Sastrawan}, \citenamefont {Uys},\ and\ \citenamefont
  {Biercuk}}]{GreenNJP2013}%
  \BibitemOpen
  \bibfield  {author} {\bibinfo {author} {\bibfnamefont {T.~J.}\ \bibnamefont
  {Green}}, \bibinfo {author} {\bibfnamefont {J.}~\bibnamefont {Sastrawan}},
  \bibinfo {author} {\bibfnamefont {H.}~\bibnamefont {Uys}}, \ and\ \bibinfo
  {author} {\bibfnamefont {M.~J.}\ \bibnamefont {Biercuk}},\ }\bibfield
  {title} {\enquote {\bibinfo {title} {Arbitrary quantum control of qubits in
  the presence of universal noise},}\ }\href@noop {} {\bibfield  {journal}
  {\bibinfo  {journal} {New J. Phys.}\ }\textbf {\bibinfo {volume} {15}},\
  \bibinfo {pages} {095004} (\bibinfo {year} {2013})}\BibitemShut {NoStop}%
\bibitem [{\citenamefont {Clerk}\ \emph {et~al.}(2010)\citenamefont {Clerk},
  \citenamefont {Devoret}, \citenamefont {Girvin}, \citenamefont {Marquardt},\
  and\ \citenamefont {Schoelkopf}}]{ClerkRMP}%
  \BibitemOpen
  \bibfield  {author} {\bibinfo {author} {\bibfnamefont {A.~A.}\ \bibnamefont
  {Clerk}}, \bibinfo {author} {\bibfnamefont {M.~H.}\ \bibnamefont {Devoret}},
  \bibinfo {author} {\bibfnamefont {S.~M.}\ \bibnamefont {Girvin}}, \bibinfo
  {author} {\bibfnamefont {F.}~\bibnamefont {Marquardt}}, \ and\ \bibinfo
  {author} {\bibfnamefont {R.~J.}\ \bibnamefont {Schoelkopf}},\ }\bibfield
  {title} {\enquote {\bibinfo {title} {Introduction to quantum noise,
  measurement, and amplification},}\ }\href@noop {} {\bibfield  {journal}
  {\bibinfo  {journal} {Rev. Mod. Phys.}\ }\textbf {\bibinfo {volume} {82}},\
  \bibinfo {pages} {1155} (\bibinfo {year} {2010})}\BibitemShut {NoStop}%
\bibitem [{\citenamefont {Kofman}\ and\ \citenamefont
  {Kurizki}(2004)}]{Kofman2004}%
  \BibitemOpen
  \bibfield  {author} {\bibinfo {author} {\bibfnamefont {A.~G.}\ \bibnamefont
  {Kofman}}\ and\ \bibinfo {author} {\bibfnamefont {G.}~\bibnamefont
  {Kurizki}},\ }\bibfield  {title} {\enquote {\bibinfo {title} {Unified theory
  of dynamically suppressed qubit decoherence in thermal baths},}\ }\href@noop
  {} {\bibfield  {journal} {\bibinfo  {journal} {Phys. Rev. Lett.}\ }\textbf
  {\bibinfo {volume} {93}},\ \bibinfo {pages} {130406} (\bibinfo {year}
  {2004})}\BibitemShut {NoStop}%
\bibitem [{\citenamefont {Biercuk}\ \emph
  {et~al.}(2011{\natexlab{b}})\citenamefont {Biercuk}, \citenamefont
  {Doherty},\ and\ \citenamefont {Uys}}]{Biercuk_Filter}%
  \BibitemOpen
  \bibfield  {author} {\bibinfo {author} {\bibfnamefont {M.~J.}\ \bibnamefont
  {Biercuk}}, \bibinfo {author} {\bibfnamefont {A.~C.}\ \bibnamefont
  {Doherty}}, \ and\ \bibinfo {author} {\bibfnamefont {H.}~\bibnamefont
  {Uys}},\ }\bibfield  {title} {\enquote {\bibinfo {title} {Dynamical
  decoupling sequence construction as a filter-design problem},}\ }\href@noop
  {} {\bibfield  {journal} {\bibinfo  {journal} {J. Phys. B}\ }\textbf
  {\bibinfo {volume} {44}},\ \bibinfo {pages} {154002} (\bibinfo {year}
  {2011}{\natexlab{b}})}\BibitemShut {NoStop}%
\bibitem [{One()}]{One}%
  \BibitemOpen
  \href@noop {} {\ }\bibinfo {note} {Note that the expression of the amplitude
  filter $F_\Omega(\omega,t)$ differs from the one we used in Ref.
  \cite{Frey2017}. Here, we follow more closely Ref. \cite{PazFFF}, motivated
  by the use of fundamental FFs.}\BibitemShut {Stop}%
\bibitem [{Two()}]{Two}%
  \BibitemOpen
  \href@noop {} {\ }\bibinfo {note} {Physically, the required sign changes may
  be obtained through phase modulation, via $\pi$ phase-shifts in Eq.
  \eqref{eq::Hc}.}\BibitemShut {Stop}%
\bibitem [{\citenamefont {Bar-Gill}\ \emph {et~al.}(2012)\citenamefont
  {Bar-Gill}, \citenamefont {Pham}, \citenamefont {Belthangady}, \citenamefont
  {Le~Sage}, \citenamefont {Cappellaro}, \citenamefont {Maze}, \citenamefont
  {Lukin}, \citenamefont {Yacoby},\ and\ \citenamefont
  {Walsworth}}]{BarGill2012}%
  \BibitemOpen
  \bibfield  {author} {\bibinfo {author} {\bibfnamefont {N.}~\bibnamefont
  {Bar-Gill}}, \bibinfo {author} {\bibfnamefont {L.~M.}\ \bibnamefont {Pham}},
  \bibinfo {author} {\bibfnamefont {C.}~\bibnamefont {Belthangady}}, \bibinfo
  {author} {\bibfnamefont {D.}~\bibnamefont {Le~Sage}}, \bibinfo {author}
  {\bibfnamefont {P.}~\bibnamefont {Cappellaro}}, \bibinfo {author}
  {\bibfnamefont {J.~R.}\ \bibnamefont {Maze}}, \bibinfo {author}
  {\bibfnamefont {M.~D.}\ \bibnamefont {Lukin}}, \bibinfo {author}
  {\bibfnamefont {A.}~\bibnamefont {Yacoby}}, \ and\ \bibinfo {author}
  {\bibfnamefont {R.}~\bibnamefont {Walsworth}},\ }\bibfield  {title} {\enquote
  {\bibinfo {title} {Suppression of spin-bath dynamics for improved coherence
  of multi-spin-qubit systems},}\ }\href@noop {} {\bibfield  {journal}
  {\bibinfo  {journal} {Nature Comms.}\ }\textbf {\bibinfo {volume} {3}},\
  \bibinfo {pages} {858} (\bibinfo {year} {2012})}\BibitemShut {NoStop}%
\bibitem [{\citenamefont {Harris}(1978)}]{Harris1978}%
  \BibitemOpen
  \bibfield  {author} {\bibinfo {author} {\bibfnamefont {F.~J.}\ \bibnamefont
  {Harris}},\ }\bibfield  {title} {\enquote {\bibinfo {title} {On the use of
  windows for harmonic analysis with the discrete {F}ourier transform},}\
  }\href@noop {} {\bibfield  {journal} {\bibinfo  {journal} {Proc. IEEE}\
  }\textbf {\bibinfo {volume} {66}},\ \bibinfo {pages} {51} (\bibinfo {year}
  {1978})}\BibitemShut {NoStop}%
\bibitem [{\citenamefont {Abreu}\ and\ \citenamefont
  {Romero}(2015)}]{Abreu2016}%
  \BibitemOpen
  \bibfield  {author} {\bibinfo {author} {\bibfnamefont {L.~D.}\ \bibnamefont
  {Abreu}}\ and\ \bibinfo {author} {\bibfnamefont {J.~L.}\ \bibnamefont
  {Romero}},\ }\bibfield  {title} {\enquote {\bibinfo {title} {The
  bias-variance trade-off in {T}homson's multitaper estimator},}\ }\href@noop
  {} {\bibfield  {journal} {\bibinfo  {journal} {arXiv:1503.02991}\ } (\bibinfo
  {year} {2015})}\BibitemShut {NoStop}%
\bibitem [{\citenamefont {Brillinger}(1981)}]{Brillinger1981}%
  \BibitemOpen
  \bibfield  {author} {\bibinfo {author} {\bibfnamefont {D.~R.}\ \bibnamefont
  {Brillinger}},\ }\href@noop {} {\emph {\bibinfo {title} {Time Series: Data
  Analysis and Theory}}}\ (\bibinfo  {publisher} {SIAM},\ \bibinfo {year}
  {1981})\BibitemShut {NoStop}%
\bibitem [{\citenamefont {Montgomery}(2012)}]{montgomery}%
  \BibitemOpen
  \bibfield  {author} {\bibinfo {author} {\bibfnamefont {D.~C.}\ \bibnamefont
  {Montgomery}},\ }\href@noop {} {\emph {\bibinfo {title} {Design and Analysis
  of Experiments}}}\ (\bibinfo  {publisher} {Wiley \& Sons},\ \bibinfo {year}
  {2012})\BibitemShut {NoStop}%
\bibitem [{\citenamefont {Neumaier}(1998)}]{Neumaier1998}%
  \BibitemOpen
  \bibfield  {author} {\bibinfo {author} {\bibfnamefont {A.}~\bibnamefont
  {Neumaier}},\ }\bibfield  {title} {\enquote {\bibinfo {title} {Solving
  ill-conditioned and singular linear systems: a tutorial on regularization},}\
  }\href@noop {} {\bibfield  {journal} {\bibinfo  {journal} {SIAM}\ }\textbf
  {\bibinfo {volume} {40}},\ \bibinfo {pages} {636} (\bibinfo {year}
  {1998})}\BibitemShut {NoStop}%
\bibitem [{\citenamefont {Thomson}(1990)}]{thompson1990}%
  \BibitemOpen
  \bibfield  {author} {\bibinfo {author} {\bibfnamefont {D.~J.}\ \bibnamefont
  {Thomson}},\ }\bibfield  {title} {\enquote {\bibinfo {title} {Multi-window
  bispectrum estimates},}\ }\href@noop {} {\bibfield  {journal} {\bibinfo
  {journal} {in: IEEE Workshop on Higher-Order Spectral Analysis}\ ,\ \bibinfo
  {pages} {p. 19}} (\bibinfo {year} {1990})}\BibitemShut {NoStop}%
\end{thebibliography}%

\end{document}